\documentclass{article}
\usepackage[utf8]{inputenc}
\usepackage{rotating}
\usepackage{csquotes}
\usepackage{authblk}

\title{Forecasting and predicting stochastic agent-based model data with biologically-informed neural networks  }

\date{\today}

\author[1]{John T. Nardini}

\affil[1]{Department of Mathematics and Statistics, The College of New Jersey, Ewing, NJ, 08628, USA. \\
nardinij@tcnj.edu}

\usepackage[square,sort,comma,numbers]{natbib}
\usepackage{multirow}

\usepackage{graphicx}
\usepackage{amsmath}
\usepackage{geometry}
\usepackage{units}
\usepackage{amsfonts}
\usepackage{amssymb}
\usepackage{pifont}
\usepackage{multirow}
\usepackage{pdflscape}
\usepackage[dvipsnames]{xcolor}
\usepackage{bm}
\usepackage{hyperref}
\usepackage{verbatim}
\usepackage{ulem}
\usepackage[utf8]{inputenc}
\usepackage[T1]{fontenc}

\definecolor{green}{RGB}{0,128,0}
\definecolor{red}{RGB}{200,0,0}

\newcommand{\rmp}{r_m^{pull}}
\newcommand{\rmh}{r_m^{adh}}
\newcommand{\ppull}{p_{pull}}
\newcommand{\padh}{p_{adh}}
\newcommand{\DMLP}{\mathcal{D}^{MLP}}
\newcommand{\D}{\mathcal{D}}
\newcommand{\Dinterp}{\mathcal{D}^{interp}}
\newcommand{\TMLP}{T^{MLP}}
\newcommand{\PABM}{P^{ABM}}
\newcommand{\HABM}{H^{ABM}}
\newcommand{\TABM}{T^{ABM}}
\newcommand{\mean}[1]{\langle #1 \rangle} 

\newcommand{\ddt}[1]{\dfrac{\partial #1 }{\partial t}} 
\newcommand{\Pm}{\bm{p}}
\newcommand{\Pmold}{\bm{p}}
\newcommand{\Pmnew}{\bm{p}^{new}}
\newcommand{\Tftrain}{T_f^{train}}
\newcommand{\Tftest}{T_f^{test}}

\newcommand{\new}[1]{#1}
\newcommand{\old}[1]{}

\usepackage[ruled,vlined]{algorithm2e}

\usepackage{todonotes}

\renewcommand{\mean}[1]{\langle #1 \rangle} 
\usepackage{comment}

\begin{document}

\maketitle

\begin{abstract}
\old{Stochastic agent-based models (ABMs) allow researchers to understand how individual-level interactions scale into emergent behavior. These models are widely used in ecology, epidemiology, and cellular biology; however, thorough exploration of ABM behavior is challenging due to their heavy computational nature.}\new{Collective migration is an important component of many biological processes, including wound healing, tumorigenesis, and embryo development. Spatial agent-based models (ABMs) are often used to model collective migration, but it is challenging to thoroughly predict these models' behavior throughout parameter space due to their random and computationally intensive nature.} \old{Performing extensive ABM simulations is not feasible for complex ABMs due to their long simulation times. Instead, }Modelers often coarse-grain ABM rules into mean-field differential equation (DE) models. While these DE models are fast to simulate, they suffer from poor (or even ill-posed) ABM predictions in some regions of parameter space. \old{At present, we lack efficient algorithms to accurately predict how ABMs behave throughout parameter space.} \new{In this work, we describe how biologically-informed neural networks (BINNs) can be trained to learn interpretable BINN-guided DE models capable of accurately predicting ABM behavior. In particular, we show that BINN-guided partial DE (PDE) simulations can 1.) forecast future spatial ABM data not seen during model training, and 2.) predict ABM data at previously-unexplored parameter values.} This latter task is achieved by combining BINN-guided PDE simulations with multivariate interpolation. \old{Visualizing the learned terms of the BINN-guided PDE allow us to interpret how individual-scale ABMs rules scale into emergent population behavior.} We demonstrate our approach using three case study ABMs of collective migration that imitate cell biology experiments and find that BINN-guided PDEs accurately forecast and predict ABM data with a one-compartment PDE when the mean-field PDE is ill-posed or requires two compartments. \old{This work is broadly applicable to studying biological systems that exhibit collective behavior.} \new{This work suggests that BINN-guided PDEs allow modelers to efficiently explore parameter space, which may enable data-driven tasks for ABMs, such as estimating parameters from experimental data.}  All code and data from our study is available at \href{https://github.com/johnnardini/Forecasting_predicting_ABMs}{https://github.com/johnnardini/Forecasting\_predicting\_ABMs}.
\end{abstract}

\section{Introduction}

Many population-level patterns in biology arise from the actions of individuals. For example, predator-prey interactions determine ecological population dynamics; individuals' adherence to public health policies limit disease spread; and cellular interactions drive wound healing and tumor invasion. Mathematical modeling is a useful tool to understand and predict how such individual actions scale into collective behavior \cite{anguige_one-dimensional_2009,brauer_mathematical_2019,gibbs_coexistence_2022, huppert_mathematical_2013,   nardini_modeling_2016,   raja_noureen_swapping_2023, xiao_modeling_2001}. In particular, stochastic agent-based models (ABMs) are a widely-used modeling framework where autonomous agents mimic the individuals of a population \cite{baker_correcting_2010, grimm_pattern-oriented_2005, marshall_formalizing_2015, tracy_agent-based_2018}. ABMs are advantageous because they capture the discrete and stochastic nature of many biological processes \cite{chappelle_pulling_2019}. 
However, ABMs are computationally intensive, and their simulations can become time-consuming to perform when the population is comprised of many individuals \cite{simpson_reliable_2022, nardini_learning_2021}. This computational restraint prevents modelers from efficiently exploring how model parameters alter model outputs. As such, there is a need for the development of methods to \new{efficiently and accurately predict ABM behavior \cite{nardini_learning_2021, kieu_dealing_2020, larie_use_2021}.} \old{address ABMs' computational expenses by 1.) forecasting future model output from limited simulations, and 2.) predicting ABM data at previously-unexplored parameter values \cite{nardini_learning_2021, kieu_dealing_2020, larie_use_2021}.} %Such methods would be of particular importance to cell biology, where ABMs may be used to forecast the future results of expensive experiments or predict the experimental conditions most likely to lead to a successful outcome, such as successful wound closure or tumor eradication. 

Modelers often perform ABM prediction by coarse-graining ABM rules into continuous differential equation (DE) models \cite{baker_correcting_2010, simpson_reliable_2022}. Ordinary DEs (ODEs) describe how a quantity (e.g., agent density) changes over time, and Partial DEs (PDEs) describe how spatially-varying ABMs change with time \cite{simpson_reliable_2022}. Such DE models are useful surrogates for ABMs because they are cheap and efficient to simulate\old{ and are amenable to analytical methods, which modelers can use to precisely infer how model parameters impact their outputs. The most commonly used coarse-grained DE models are \emph{mean-field DE models}, which are derived by assuming agents respond to the average behavior of their neighbors \cite{nardini_learning_2021}}. Mean-field DE models\new{, which assume agents respond to the average behavior of their neighbors,} have been shown to accurately predict ABM behavior at some parameter values. Unfortunately, these models can poorly predict ABM outputs \new{when the mean-field assumption is violated} \cite{baker_correcting_2010, thompson_modelling_2012}. For example, Baker and Simpson 2010 \cite{baker_correcting_2010} demonstrated that the mean-field DE model for a population growth ABM only accurately predict ABM data when agents proliferate slowly. A further complication of mean-field DEs is that they may be ill-posed at certain parameter values. Anguige and Schmeiser 2009 \cite{anguige_one-dimensional_2009} used a stochastic space-jump model to study how cell adhesion impacts collective migration and found that the resulting mean-field PDE model is ill-posed (and thus cannot predict ABM behavior) for large adhesion values. \old{Modelers may improve DE models' predictive capability by implementing pair-wise interactions or moment closure approximations in lieu of the mean-field assumption, but the resulting models are often complicated and may significantly increase computation time \cite{baker_correcting_2010, johnston_mean-field_2012, simpson_distinguishing_2014}.}

Despite the inability of mean-field DE models to predict ABM behavior at all parameter values, ABM simulations do obey conservation laws (e.g., conservation of mass for spatial ABMs) \cite{vandenheuvel_pushing_2024}. \old{This observation suggests that} Alternative DE models may thus be capable of accurately describing ABM behavior. Equation learning (EQL) is a new area of research on the development and application of algorithms to discover the dynamical systems model that best describes a dataset \cite{brunton_discovering_2016, kaiser_sparse_2018, rudy_data-driven_2019, champion_data-driven_2019, mangan_inferring_2016, mangan_model_2017, messenger_weak_2021-1, messenger_weak_2021, lagergren_learning_2020, lagergren_biologically-informed_2020, nardini_learning_2020}. Brunton et al. 2016 \cite{brunton_discovering_2016} introduced a sparse regression-based EQL approach to learn DE models from data with a user-specified library of candidate terms. This method has proven very successful in recovering informative models from simulated and experimental data \cite{rudy_data-driven_2017}. There is a growing understanding that EQL methods can aid the prediction of ABM data \cite{nardini_learning_2021, messenger_learning_2022-1,messenger_learning_2022, supekar_learning_2023}. For example, we recently demonstrated that the least squares EQL approach learns ODE equations that accurately describe simulated ABM data, even when the collected data is incomplete or sparsely sampled \cite{nardini_learning_2021}. Supekar et al. 2023 \cite{supekar_learning_2023} coupled this method with spectral basis representation data to discover PDE models that capture the emergent behavior found in active matter ABMs. Another popular EQL approach includes physics-informed neural networks (PINNs), where modelers embed physical knowledge (in the form of a known PDE framework) into the training procedure for artificial neural networks (ANNs) \cite{cai_physics-informed_2021, kaplarevic-malisic_identifying_2023, linka_bayesian_2022,  raissi_physics-informed_2019, shin_convergence_2020}. Trained PINN models can predict complex, sparse, and noisy data while also obeying known physical principles. Lagergren et al. 2020 \cite{lagergren_biologically-informed_2020} extended the PINNs framework by replacing physics-based mechanistic terms with function-approximating multi-layer perceptions (MLPs) to develop the biologically-informed neural network (BINN) methodology. As a result, BINN models can learn PDE models from data with terms that depend on space, time, or agent density. Training the BINN to simulated ABM data ensures that a realization of this PDE that best matches the data is learned. Standard methods of DE analysis, including bifurcation analysis and pattern formation, can be used to understand the ABM's behavior. BINNs thus present a promising and interpretable tool for ABM forecasting and prediction. However, determining how BINNs can be used to learn predictive DE models for ABMs remains an open area of research.

In this work, we demonstrate how to combine BINNs and PDE model simulations to forecast and predict ABM behavior. Our approach leverages BINNs' vast data and modeling approximation capability with the computational efficiency of PDE models to develop a potent ABM surrogate modeling tool. In particular, we demonstrate how to use trained BINN models to \new{1.) forecast future ABM data at a fixed parameter value, and 2.) predict ABM data at previously-unexplored parameter values.} \old{forecast future ABM data at a fixed parameter, as well as how to predict ABM data at new parameter values by combining BINN modeling terms with multivariate interpolation.} \new{This latter task is achieved using multivariate interpolation, which} provides a straightforward approach for inferring PDE modeling terms. \old{though more complex methodologies, such as ANNs or Gaussian processes, could also be used \cite{kennedy_bayesian_2001}.} We demonstrate that visually inspecting the BINN modeling terms over a range of ABM parameter values allows us to interpret how ABM parameters impact model behavior.

We apply the BINNs methodology to three case study ABMs in this work. Each case study models collective migration in cell biological experiments, such as barrier and scratch assays \cite{nardini_modeling_2016, simpson_reliable_2022, johnston_mean-field_2012, lagergren_biologically-informed_2020,decaestecker_can_2007}. In a barrier assay, a two-dimensional layer of cells is cultured inside a physical boundary. Microscopy is used to image how the cell population migrates outwards once the barrier has been removed \cite{decaestecker_can_2007,das_ring_2015}. Cells are closely packed in these experiments and thus interact with their neighbors. Our case study ABMs simulate how two stimuli, namely, cell pulling and adhesion, impact collectively migrating cell populations. \new{These processes are ubiquitous in cell biology. For example, leader cells pull their followers into the wound area to heal wounded epithelial tissue, and cell adhesions in embryonic cells ensures the self organization of the different germ layers \cite{kashef_quantitative_2015,venhuizen_making_2017,vishwakarma_mechanobiology_2020}.} ABMs provide a promising avenue to model the impacts of these stimuli on collectively migrating cell populations.

We begin this work in Section \ref{sec:ABMs_coarse_graining} by presenting the case study ABMs and notation. In Section \ref{sec:data_analysis_methods}, we discuss our methodologies to forecast and predict ABM behavior. In Section \ref{sec:results}, we detail our results on using these methods to forecast and predict data from the three case study ABMs; this section concludes with a brief discussion on the computational expenses of each method. We conclude these results and suggest areas for future work in Section \ref{sec:discussion}.

\section{\old{Coarse-graining collectively migrating ABMs into PDE models}\new{The case study ABMs}}\label{sec:ABMs_coarse_graining}

 \old{In this study, we apply our BINNs methodology} \new{We consider three case study ABMs that imitate collective migration during cell biological experiments, including scratch and} barrier assays \cite{ nardini_modeling_2016, simpson_reliable_2022, johnston_mean-field_2012, lagergren_biologically-informed_2020, decaestecker_can_2007}. Each case study ABM models how cell pulling and adhesion impact collective cell migration during these experiments \cite{janiszewska_cell_2020,rothenberg_rap1_2023}. The ABMs are two-dimensional cellular automata with pulling agents that perform cell pulling rules and/or adhesive agents that perform rules on cell adhesion. Each model is an exclusion process, meaning that each agent can only occupy one lattice site at a time, and each lattice site is occupied by at most one agent. The first model is borrowed from \cite{chappelle_pulling_2019} and consists only of pulling agents; the second model is inspired by the stochastic space jump model from \cite{anguige_one-dimensional_2009} and consists only of adhesive agents; to the best of our knowledge, we are the first to study the third model, which consists of both pulling and adhesive agents. 

In this section, we briefly introduce our case study ABMs and their rules on agent pulling and adhesion in Section \ref{subsec:ABM_rules_MF}; we then detail our ABM notation and simulation in Section \ref{subsec:ABM_notation}. \old{and we present the mean-field PDE models for each ABM in Section \ref{subsec:ABM_MF_PDE}.} Additional details on the ABM rules and implementation can be found in electronic supplementary materials \ref{app:ABM_Rules} and \ref{app:ABM_implementation}, respectively.

\subsection{Brief introduction to the case study ABMs and their model rules}\label{subsec:ABM_rules_MF}

\begin{figure}
    \centering
    \includegraphics[width=.9\textwidth]{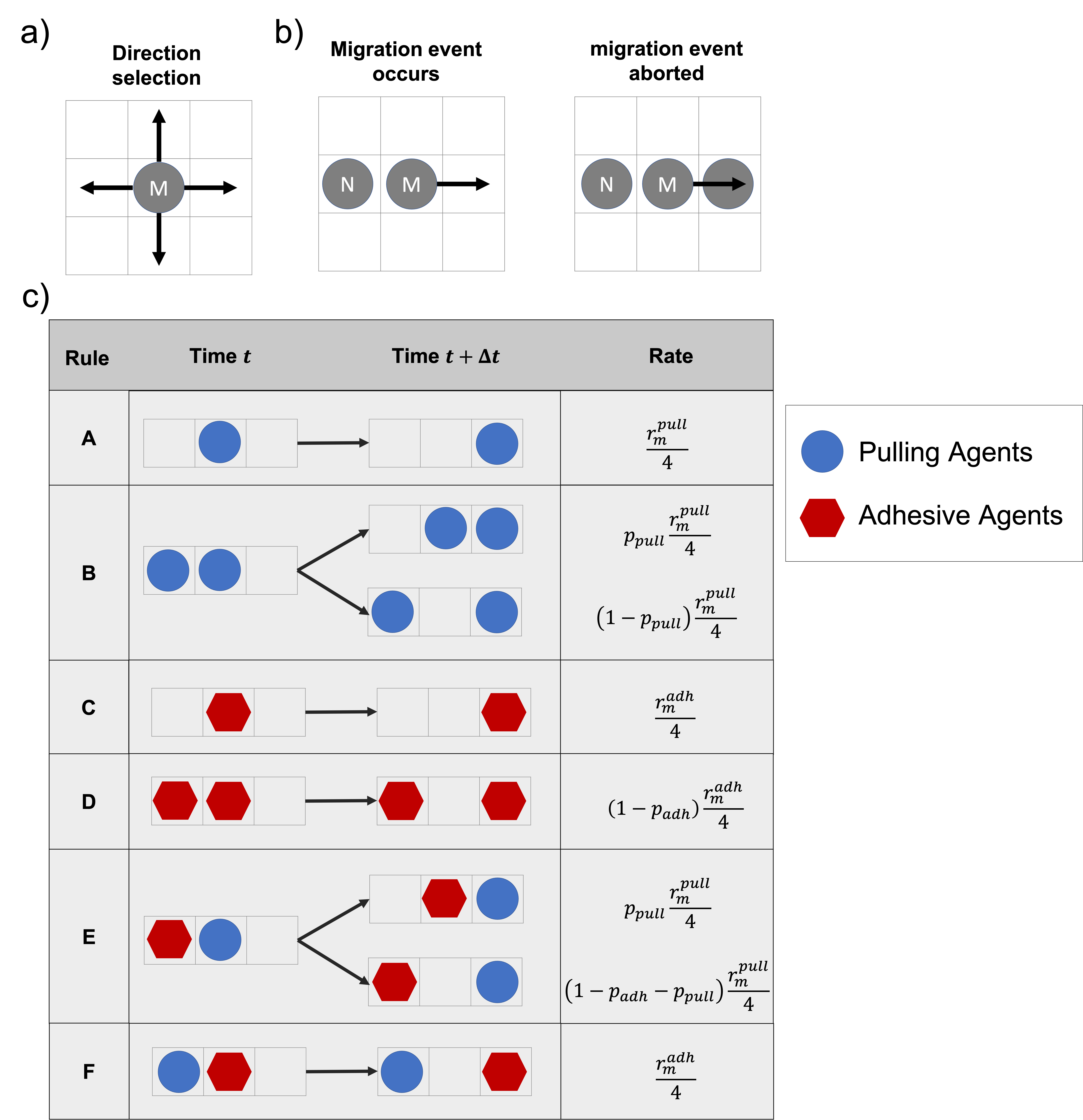}
    \caption{\linespread{1} \linespread{1} \small{ABM rules on migration, pulling, and adhesion.}  a) When an agent performs a migration event, it chooses one of the four cardinal directions to move towards with equal probability; \new{migration can also lead to a pulling or adhesion event in the chosen direction}. The migrating agent is referred to as a migrating agent (M) b) A migration event requires the lattice site in the chosen migration direction to be empty; otherwise, the migration event is aborted. A neighboring agent (N) is an agent located in the direction opposite the chosen migration direction. c) Rules A-F dictate the rules on agent migration, pulling, and adhesion. Here, we show each rule when an agent chooses to move rightwards. Rule A prescribes how a pulling agent (blue circle) migrates when there is no neighboring agent. Rule B prescribes how a pulling agent migrates and attempts to pull a neighboring pulling agent with it. Rule C prescribes how an adhesive agent (red hexagon) migrates when there is no neighboring agent. Rule D prescribes how a neighboring adhesive agent attempts to adhere to a migrating adhesive agent and abort its migration event. Rule E prescribes how a migrating pulling agent attempts to pull its neighboring adhesive agent, while the adhesive agent attempts to adhere to the pulling agent. Rule F prescribes how a migrating adhesive agent and neighboring pulling agent do not interact with each other. The last column documents the rate at which each lattice site configuration at time $t$ changes to the updated lattice site configuration at time $t+\Delta t$.}
    \label{fig:ABM_rules}
\end{figure}

Rules A-F governing agent pulling and adhesion are visually depicted in Figure \ref{fig:ABM_rules}, and the parameters for each rule are described in Table \ref{tab:ABM_params}. In all rules, a \emph{migrating agent} chooses one of its four neighboring lattice site to move into with equal probability (Figure \ref{fig:ABM_rules}(a)). A migration event is aborted if the lattice site in the chosen direction is already occupied (Figure \ref{fig:ABM_rules}(b)). We refer to a \emph{neighboring agent} as an agent located next to the migrating agent in the direction opposite of the chosen migration direction.  

\begin{table}
\centering
    \begin{tabular}{|c|c|c|}
    \hline
    Variable & Description & Range \\ \hline
    $\rmp$ & Pulling agent migration rate & $[0,\infty)$ \\    \hline
    $\rmh$ & Adhesive agent migration rate & $[0,\infty)$ \\    \hline
    $\ppull$ & Probability of successful pulling event & $[0,1]$ \\    \hline
    $\padh$ & Probability of successful adhesion event & $[0,1]$ \\    \hline
    $\alpha$ & Proportion of adhesive agents & $[0,1]$ \\    \hline   
    \end{tabular}
    \caption{ABM model parameters. We describe each model parameter and present their range of possible values.}
    \label{tab:ABM_params}
\end{table}

Rules A, B, and E are initiated when  a pulling agent attempts to migrate, which occurs with rate $\rmp$. \new{Migratory pulling agents pull their neighboring agents along with them with probability $\ppull$.} Rules C, D, and F are initiated when an adhesive agent attempts to migrate, which occurs with rate $\rmh$. \new{Neighboring adhesive agents adhere to migrating agents and abort the migration event with probability $\padh$. The parameter $\alpha$ corresponds to the proportion of adhesive agents in the simulation. Even though we eventually summarize each ABM simulation along the $x$-direction, all rules on migration, pulling, and adhesion occur in all four cardinal directions.} \old{Each rule can be described as follows:}

\hfill\begin{minipage}{\dimexpr\textwidth-.5cm}\indent \noindent
\end{minipage}

Our three case study ABMs are:
\begin{enumerate}
    \item \textbf{\uline{The Pulling ABM}}, which consists of rules A and B. \new{This model has parameters $\Pm = (\rmp,\ppull)^T$.}
    \item \textbf{\uline{The Adhesion ABM}}, which consists of rules C and D. \new{This model has parameters $\Pm = (\rmh,\padh)^T$.}
    \item \textbf{\uline{The Pulling \& Adhesion ABM}}, which consists of rules A-F. \new{This model has parameters $\Pm = (\rmp, \rmh, \ppull, \padh, \alpha)^T$.}
\end{enumerate}

\subsection{ABM notation} \label{subsec:ABM_notation}
\new{All parameters used to configure ABM simulations are summarized in Table \ref{tab:ABM_setup}.} Each model is simulated in the spatial domain $(x,y)\in[0,X]\times[0,Y]$. \old{We choose $X=200\text{ and } Y=40$ to represent a thin rectangle where collective migration primarily occurs along the $x$-dimension and is not affected by the boundary in this dimension.} We represent this space with a two-dimensional lattice with square lattice sites of length $\Delta=1$ to imitate a typical cell length.  \old{The variables $P_{i,j}(t)$, $H_{i,j}(t)$, and $0_{i,j}(t)$ denote the probabilities that lattice site $(i,j)$ is occupied at time $t$ by a pulling agent, adhesive agent, or empty, respectively for $i=1,\dots,X$ and $j=1,\dots,Y$. } Let $N^{(r)}_P(x_i,t_j)$ and $N^{(r)}_H(x_i,t_j)$ denote the number of pulling and adhesive agents, respectively, in the $i^\text{th}$ column \new{at the $j^\text{th}$ timepoint} for $i = 1,\dots,X$ \new{and $j = 1,\dots,T_f$} from the $r^{\text{th}}$ of $R$ identically prepared ABM simulations \new{ (the input model parameters are fixed but the $R$ model initializations and subsequent agent behaviors are stochastic)}. \new{Here, $X$ and $T_f$ denote the number of spatial columns and temporal grid points, respectively.} To estimate the \new{spatiotemporal} pulling and adhesive agent densities \old{in the $i^{\text{th}}$ column} from the $r^{\text{th}}$ simulation, we compute 
$$P^{(r)}(x_i,t_j) = \dfrac{ N^{(r)}_P(x_i,t_j)}{Y} \text{ and } H^{(r)}(x_i,t_j) = \dfrac{N^{(r)}_H(x_i,t_j)}{Y}, \text{ for } i = 1,\dots,X,\text{ and } j = 1,\dots,T_f, $$
respectively. The \emph{total} agent density in the $r^\text{th}$ simulation is then estimated by $$T^{(r)}(x_i,t_j) = P^{(r)}(x_i,t_j) + H^{(r)}(x_i,t_j).$$ To estimate the averaged pulling, adhesive, and total agent density in the $i^{\text{th}}$ column from $R$ identically prepared ABM simulations over time, we compute:
\begin{align*}
    \mean{\PABM(x_i,t_j)} &= \dfrac{1}{R}\sum_{r=1}^RP^{(r)}(x_i,t_j); \\
    \mean{\HABM(x_i,t_j)} &= \dfrac{1}{R}\sum_{r=1}^RH^{(r)}(x_i,t_j); \text{ and }\\
    \mean{\TABM(x_i,t_j)} &= \dfrac{1}{R}\sum_{r=1}^RT^{(r)}(x_i,t_j),\ \text{ for }  i = 1,\dots,X \text{ and } j = 1,\dots,T_f.
\end{align*}

\begin{table}
    \centering
    \begin{tabular}{|c|c|c|}
    \hline
    Variable & Description & Value \\ \hline
    $R$ & Number of averaged ABM simulations per dataset & 25 \\ \hline
    $t_f$ & Ending simulation time & 1000 \\ \hline    
    $\Delta t$ & Spacing between temporal gridpoints & 10  \\ \hline    
    $T_f$ & Number of total timepoints & 100  \\ \hline
    $T_f^{train}$ &Number of training timepoints & 75 \\ \hline
    $T_f^{test}$ &Number of testing timepoints & 25   \\  \hline
    $X$ & Number of horizontal lattice sites & 200 \\ \hline    
    $Y$ & Number of vertical lattice sites & 40 \\ \hline    
    $\Delta x$ & Spacing between spatial points & 1  \\ \hline    
    \end{tabular}
    \caption{\new{ABM configuration parameters. We describe each parameter used for ABM configuration and present the values used throughout this study.}}
    \label{tab:ABM_setup}
\end{table}

\old{simulate $R$ identically prepared realizations of the ABM each realization is completed when time reaches $t=1000$. We estimate the total spatiotemporal agent density from each simulation and average over all $R$ simulations to obtain $\mean{\TABM(x,t)} = \{\mean{\TABM(x_i,t_j)}\}_{i=1,\dots,X}^{j=1,\dots,100$, where $x_i=i\Delta x$ and $t_j=(j-1)\Delta t$ for $\Delta x=1$ and $\Delta t=10$. We split $\mean{\TABM(x,t)}$ into its training and testing datasets by setting $\mean{\TABM(x,t)}^{train} = \{\mean{\TABM(x_i,t_j)}\}_{i=1,\dots,X}^{j=1,\dots,\Tftrain}$ and $\mean{\TABM(x,t)}^{test} = \{\mean{\TABM(x_i,t_j)}\}_{i=1,\dots,X}^{j=\Tftrain+1,\dots,\Tftest}$.We set $\Tftrain=75$ and $\Tftest=100$ to place 75\% of data into the training dataset.}}

\section{\old{Data analysis methods} \new{Methods to forecast and predict ABM data}}\label{sec:data_analysis_methods}

\begin{figure}
    \centering
    \includegraphics[width=.97\textwidth]{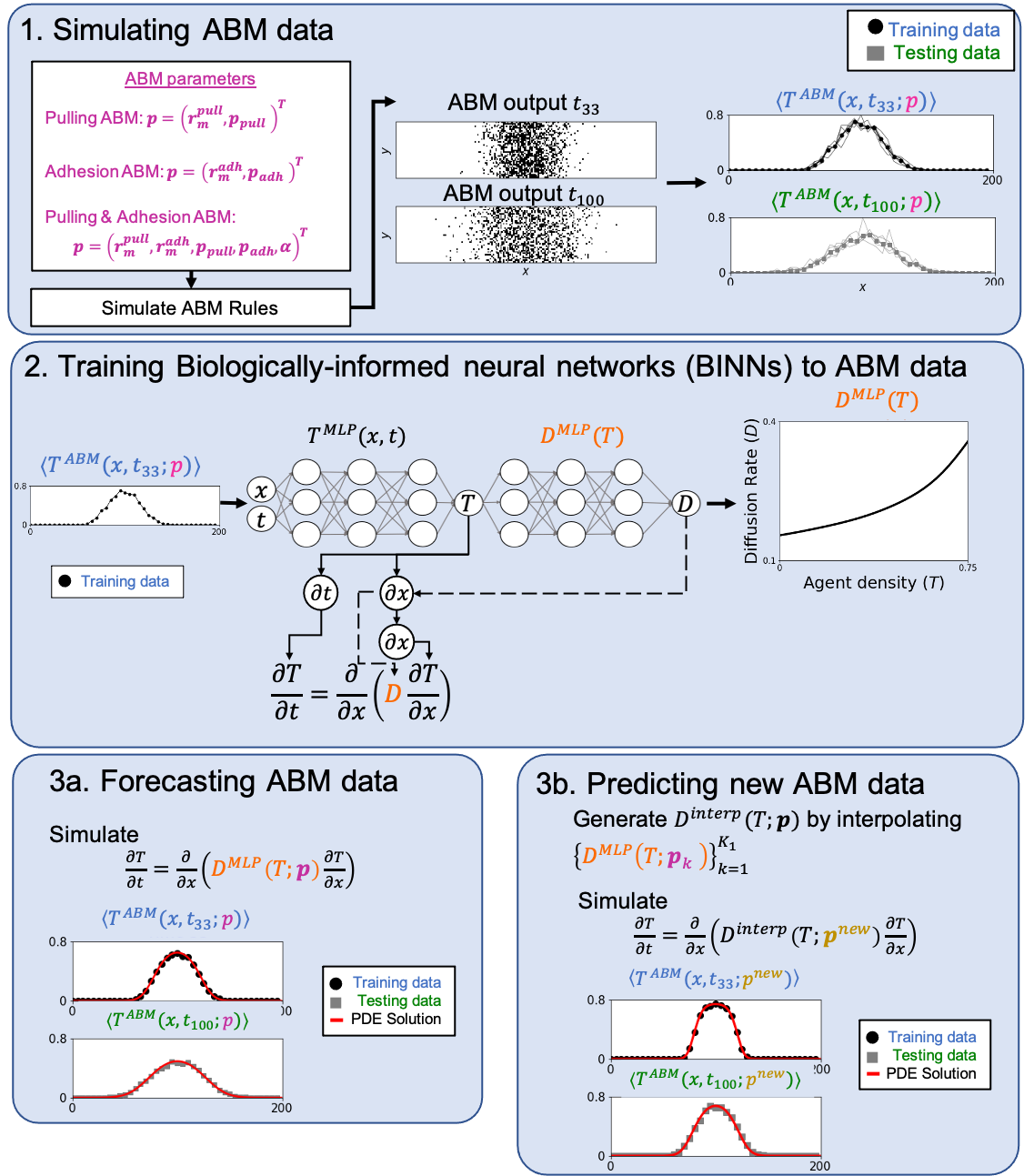}
    \caption{\linespread{1} \small{Forecasting and predicting ABM data with BINNs. 1. \uline{Simulating ABM data}.} For a given parameter, $\Pm$, we simulate the Pulling, Adhesion, or Pulling \& Adhesion ABM. Each model outputs snapshots of agent locations over time; we summarize this data by estimating the average total agent density along the $x$-direction for each snapshot. We perform \old{25} \new{$R$} total ABM simulations \new{(shown as thin lines)} for each $\Pm$ and average the total spatiotemporal agent density to obtain $\mean{\TABM(x,t;\Pm)}$\new{; in this figure, $R=5$}. The first $\Tftrain$ timepoints are placed into a training ABM dataset, and the final $\Tftest$ timepoints are placed into a testing ABM dataset. 2. \uline{Training biologically-informed neural networks (BINNs) to ABM data}. Each BINN model consists of a data-approximating MLP, $\TMLP(x,t)$, and a diffusion-rate-approximating MLP, $\DMLP(T)$. BINN models are trained so that $\TMLP(x,t)\approx\mean{\TABM(x,t;\Pm)}^{train}$ while $\TMLP$ and $\DMLP$ satisfy Equation \eqref{eq:diffusion_framework}. After model training, the inferred $\DMLP(T)$ estimates the agent diffusion rate. 3a. \uline{Forecasting ABM data}. Simulating the diffusion PDE framework with $\DMLP(T)$ allows us to forecast the ABM training and testing data.  3b. \uline{Predicting new ABM data}. We predict the rate of agent diffusion at a new parameter, $\Pmnew$, by interpolating $\DMLP(T;\Pm)$ over several $\Pm$ values to create $\Dinterp(T;\Pm)$. Simulating the diffusion PDE framework with $\Dinterp(T;\Pmnew)$ allows us to predict the new ABM training and testing data. }
    \label{fig:methods_overview}
\end{figure}

\new{In this section, we outline our methodologies for forecasting future ABM data and predicting ABM data at new parameter values. This begins with a description of how we generate ABM data in Section \ref{sec:ABM_data_simulation} followed by an overview of the four methods we use for ABM forecasting in Section \ref{sec:ABM_forecasting_models}. We then describe our approaches for ABM forecasting and prediction in Sections \ref{subsec:Forecasting ABM data} and  \ref{subsec:ABM_prediction}, respectively. We visualize how BINNs can be used for these processes in Figure \ref{fig:methods_overview}.  All methods are implemented using Python (version 3.9.12) with code available on GitHub at \href{https://github.com/johnnardini/Forecasting_predicting_ABMs}{https://github.com/johnnardini/Forecasting\_predicting\_ABMs}.}

\subsection{\new{Simulating ABM data}}\label{sec:ABM_data_simulation}

\new{The process of simulating ABM data is illustrated in Part 1 of Figure \ref{fig:methods_overview}. At the parameter value $\Pm$, we calculate $\mean{\TABM(x,t;\Pm)} = \{\mean{\TABM(x_i,t_j;\Pm)}\}_{i=1,\dots,X}^{j=1,\dots,\new{T_f}}$. For subsequent model training and validation purposes, we split $\mean{\TABM(x,t;\Pm)}$ into training and testing datasets by setting 
\begin{align}
\mean{\TABM(x,t;\Pm)}^{train} &= \left\{\mean{\TABM(x_i,t_j;\Pm)}\right\}_{i=1,\dots,X}^{j=1,\dots,\Tftrain}, \text{ and}\nonumber \\ 
\mean{\TABM(x,t;\Pm)}^{test} &= \left\{\mean{\TABM(x_i,t_j;\Pm)}\right\}_{i=1,\dots,X}^{j=\Tftrain+1,\dots,\Tftrain + \Tftest}. \label{eq:ABM_training_testing_datasets}
\end{align}
Here, $\Tftrain$ and $\Tftest$ denote the number of training and testing timepoints, respectively, and $T_f=\Tftrain + \Tftest$. }

\subsection{\new{Models to forecast ABM data}}\label{sec:ABM_forecasting_models}

\new{We now describe the four models we use to forecast future ABM data. Namely, these models are the mean-field PDE, ANN, BINN, and BINN-guided PDE models.}

\new{The mean-field and BINN-guided PDE models consist of simulating a PDE of the form\footnote{with the exception of the mean-field PDE for the Pulling \& Adhesion ABM, which requires simulating the two-compartment PDE given by Equation \eqref{eq:PullingAdhesionMF} in Section \ref{subsec:ABM_MF_PDE}}:
\begin{align}
    \dfrac{\partial T}{\partial t} &=  \dfrac{\partial}{\partial x} \left( \mathcal{D}(T) \dfrac{\partial T}{\partial x}\right), \label{eq:diffusion_framework_1d}
\end{align}
where $T = T(x,t) = P(x,t) + H(x,t)$ denotes the total agent density over space and time. The form of $\D(T)$ in Equation \eqref{eq:diffusion_framework_1d} changes based on the ABM and the modeling approach being used. For the mean-field PDE, we determine the form of $\D(T)$ by converting discrete ABM rules into their continuous counterparts and invoking the mean-field assumption, which may be invalid at some parameter values. BINNs, on the other hand, are a data-driven approach to infer $\D(T)$ from the data without any such \emph{a priori} assumptions.}

\new{The ANN and BINN models consist of training a prescribed neural network to ABM data and then using the trained neural network to forecast future data.}

\subsubsection{\new{Mean-field PDE Models}}\label{subsec:ABM_MF_PDE}

Here, we present the mean-field PDE models for each case study ABM. More detailed information on how the ABM rules are coarse-grained into these models are provided in electronic supplementary material \ref{app:coarse_graining_ABM_Rules}. Our numerical method to numerically integrate these PDE models is provided in electronic supplementary material \ref{sec:numericalIntegration}.

\textbf{\uline{The Pulling ABM:}} The Pulling ABM includes only pulling agents and consists of Rules A-B from Figure \ref{fig:ABM_rules}. In electronic supplementary material \ref{subsec:Pulling_coarse_graining}, we show that these rules can be coarse grained into the Pulling ABM's mean-field PDE model:
\begin{equation}
    \dfrac{\partial P}{\partial t} =  \nabla \cdot \left( \mathcal{D}^{pull}(P) \nabla P\right), \ \ \  \mathcal{D}^{pull}(P) = \dfrac{\rmp}{4} \left( 1 + 3\ppull P^2 \right) \label{eq:pulling_MF}
\end{equation}
where $P=P(x,y,t)$ denotes the spatiotemporal pulling agent density. \old{In Figure \ref{fig:MF_ABM_compare}(a-f), we find that a simulation of Equation \eqref{eq:pulling_MF} closely matches $P^{(1)}(x,t)$ over time for $\Pm=(\rmp,\ppull)^T=(1.0, 0.5)^T.$}

\textbf{\uline{The Adhesion ABM:}} The Adhesion ABM includes only adhesive agents and consists of Rules C-D from Figure \ref{fig:ABM_rules}. In electronic supplementary material \ref{subsec:Adhesion_coarse_graining}, we show that these rules can be coarse grained into the Adhesion ABM's mean-field PDE model:
\begin{equation}
    \dfrac{\partial H}{\partial t} =  \nabla \cdot \left( \mathcal{D}^{adh}(H) \nabla H\right), \ \ \ \ \mathcal{D}^{adh}(H)=  \dfrac{3\rmh}{4} \left( \padh \left( H-\dfrac{2}{3}\right)^2 + 1 - \dfrac{4\padh}{3}\right) \label{eq:adhesion_MF}
\end{equation}
where $H=H(x,y,t)$ denotes the spatiotemporal adhesive agent density. \old{In Figure \ref{fig:MF_ABM_compare}(g-l), we find that a simulation of Equation \eqref{eq:adhesion_MF} closely matches $H^{(1)}(x,t)$ over time for $\Pm=(\rmh,\padh)^T=(1.0, 0.5)^T.$}

Notice that $\D^{adh}(H)$ from Equation \eqref{eq:adhesion_MF} becomes negative for some density values when $\padh>0.75$. This PDE thus fails to provide an ABM prediction at these parameter values because negative diffusion is ill-posed \cite{anguige_one-dimensional_2009}.

\textbf{\uline{The Pulling \& Adhesion ABM:}} The Pulling \& Adhesion ABM includes both pulling and adhesive agents, and consists of Rules A-F from Figure \ref{fig:ABM_rules}. In electronic supplementary material \ref{subsec:PullingAdhesion_coarse_graining}, we show that these rules can be coarse-grained into the Pulling \& Adhesion ABM's mean-field PDE model: 
\begin{align}
    \dfrac{\partial P}{\partial t} = &\dfrac{\rmp}{4} \nabla \cdot \bigg( (1-T)\nabla P + P\nabla T  \bigg) \nonumber \\
    &+ \padh\dfrac{\rmp}{4} \nabla \cdot \bigg( -3 P(1-T) \nabla H - H (1-T)\nabla P - HP \nabla T  \bigg) \nonumber \\
     &+ \ppull\dfrac{\rmp}{4} \nabla \cdot \bigg( 3P^2 \nabla T  \bigg)  \nonumber \\
    \dfrac{\partial H}{\partial t} =  &\dfrac{\rmh}{4} \nabla \cdot \bigg( (1-T)\nabla H + H\nabla T  \bigg) \nonumber \\
    &+ \padh\dfrac{\rmh}{4} \nabla \cdot \bigg( -4 (1-T)H\nabla H - H^2 \nabla T  \bigg) \nonumber \\
    &+ \ppull\dfrac{\rmp}{4} \nabla \cdot \bigg( -(1-T)H\nabla P + (1-T)P\nabla H + 3HP\nabla T  \bigg). \label{eq:PullingAdhesionMF}
\end{align}
This two-compartment PDE describes the spatiotemporal densities of pulling agents, $P(x,y,t)$, and adhesive agents, $H=H(x,y,t)$. The total agent density is given by $T=T(x,y,t)=H(x,y,t)+P(x,y,t)$.  \old{In Figure \ref{fig:MF_ABM_compare}(m-r), we find that the $P$ and $H$ compartments from a simulation of Equation \eqref{eq:PullingAdhesionMF} closely $P^{(1)}(x,t)$ and $H^{(1)}(x,t)$, respectively, over time for $\Pm=(\rmp, \rmh, \ppull, \padh, \alpha)^T=(1.0, 0.25, 0.33, 0.33, 0.5)^T.$} To the best of our knowledge, it is not possible to convert Rules A-F into a single-compartment PDE model describing $T(x,y,t)$

\subsubsection{\new{The ANN model}}\label{subsec:ANN}

\new{ANNs have recently gained traction as surrogate models for ABMs \cite{larie_use_2021, angione_using_2022}. Here, we consider a simple multilayer perceptron (MLP) model, $\TMLP(x,t)$, to predict the total agent density at the spatiotemporal point $(x,t)$. We provide a brief description of the model architecture and training procedure in this section; more detailed information can be found in electronic supplementary material \ref{app:BINNs}.}

\new{\textbf{\uline{The ANN architecture:}} $\TMLP(x,t)$ has a two-dimensional input, $(x,t)$, and one-dimensional output, $T(x,t).$ This model has three hidden layers,  each with 128 neurons. The hidden layers all have sigmoidal activation functions, and the output layer has a softplus activation function.}

\new{\textbf{\uline{ANN model training:}} The ANN model is trained to minimize}
\new{\begin{equation}
    \mathcal{L}_{ANN} = \mathcal{L}_{WLS} \label{eq:ANN},
\end{equation}
where $\mathcal{L}_{WLS}$ is given by Equation \eqref{eq:L_ols} in electronic supplementary material \ref{app:BINNs}  and computes a weighted mean-squared error (MSE) between $\TMLP(x,t)$ and $\mean{\TABM(x,t)}^{train}$. Here, extra weight is assigned to data from the first timepoint to ensure that $\TMLP$ closely agrees with the ABM's initial data.}

\new{We use the ADAM optimizer with default hyperparameter values to minimize Equation \eqref{eq:ANN}. We perform $10^4$ epochs with an early stopping criterion of $10^3$ epochs.}

\subsubsection{\new{The BINN model}}\label{subsec:Training_BINNs}

\new{We provide a brief overview of our BINN model architecture and training procedure, which closely follow the implementation from the original BINN model study in \cite{lagergren_biologically-informed_2020}. More detailed information can be found in electronic supplementary material \ref{app:BINNs}.}

\new{\textbf{\uline{The BINN architecture:}}} We construct BINN models that consist of two sequential MLP models: $\TMLP(x,t)$ predicts the total agent density at the point $(x,t)$, and $\DMLP(T)$ predicts the agent diffusion rate at the density value $T$ (Part 2 of Figure \ref{fig:methods_overview}). \new{The architecture for $\TMLP(x,t)$ here is identical to the ANN architecture. The architecture for $\DMLP(T)$ also has three hidden layers (each with 128 neurons), and the same hidden and output activation functions. However, this model has a one-dimensional input, $T$, and one-dimensional output, $D(T)$.}

\new{\textbf{\uline{BINN model training:}} The two MLPs comprising the BINN model are trained to concurrently fit the given dataset, $\mean{\TABM(x,t)}^{train}$, and solve the PDE given by 
\begin{equation}
    \dfrac{\partial}{\partial t}\TMLP =  \dfrac{\partial}{\partial x} \left(\DMLP(\TMLP) \dfrac{\partial}{\partial x} \TMLP \right).  \label{eq:diffusion_framework}
\end{equation}
This is achieved by minimizing the following multi-term loss function:
\begin{equation}
    \mathcal{L}_{BINN} = \mathcal{L}_{WLS} + \epsilon\mathcal{L}_{PDE} +\mathcal{L}_{constr}. \label{eq:Loss_total}
\end{equation}
The equation for $\mathcal{L}_{WLS}$ is identical to Equation \eqref{eq:ANN}, $\mathcal{L}_{PDE}$ computes the MSE between the left- and right-hand sides of Equation \eqref{eq:diffusion_framework} to ensure both MLPs satisfy this diffusion framework, and $\mathcal{L}_{constr}$ penalizes the two MLPs for violating user-defined criteria (such as lower and upper bounds on $\DMLP$). The equations for these three terms are provided in Equations \eqref{eq:L_ols}, \eqref{eq:L_pde}, and \eqref{eq:L_constr} from electronic supplementary material \ref{app:BINNs}. The $\epsilon$ parameter is chosen to ensure the $\mathcal{L}_{WLS} \text{ and } \mathcal{L}_{PDE}$ terms are weighted equally.} \old{ We chose to enforce the MLPs satisfy Equation \eqref{eq:diffusion_framework} because the mean-field models for the Pulling ABM and Adhesion ABM from Section \ref{subsec:ABM_MF_PDE} obey this framework with different diffusion rates. }

\new{Following \cite{linka_bayesian_2022}, we minimize Equation \eqref{eq:Loss_total} in a two-step process. In the first process, we minimize Equation \eqref{eq:ANN} over $10^4$ epochs with an early stopping criterion of $10^3$ epochs. In the second process, we minimize Equation \eqref{eq:Loss_total} over $10^6$ epochs with an early stopping criterion of $10^5$ epochs. The ADAM optimizer is used during both steps with its default hyperparameter values.}

\subsubsection{\new{The BINN-guided PDE model}}\label{subsec:BG_PDE}

\new{BINN models are trained to satisfy Equation \eqref{eq:diffusion_framework}. The \emph{BINN-guided PDE model} computes this learned equation by simulating Equation \eqref{eq:diffusion_framework_1d} with $\mathcal{D}(T) = \DMLP(T)$. Our numerical method to numerically integrate this PDE is provided in electronic supplementary material \ref{sec:numericalIntegration}.}

\subsection{Forecasting \new{future} ABM data} \label{subsec:Forecasting ABM data}

\new{We use the four models introduced in Section \ref{sec:ABM_forecasting_models} to forecast future ABM data (Part 3a of Figure \ref{fig:methods_overview}). In \emph{forecasting}, we assess the ability of a model to compute future ABM data at a fixed parameter value from previous ABM data. This could correspond to inferring the future behavior of a computationally-intensive ABM simulation or an expensive experimental procedure.}

\new{We perform ABM forecasting by training each model to the training ABM dataset and then computing the model prediction over all space- and timepoints. The mean-field PDE model does not require any model training because we can directly compute it from the ABM parameter values. We then partition each model's prediction into training and testing datasets to match the ABM training and testing datasets from Equation \eqref{eq:ABM_training_testing_datasets}. We report the training MSE from each model prediction as:
$$ \new{ \dfrac{1}{X\Tftrain} \sum_{i=1}^{X} \sum_{j = 1}^{\Tftrain} \left( T^{model}(x_i,t_j) - \mean{\TABM(x_i,t_j)} \right)^2},$$ and the testing MSE as: $$ \new{\dfrac{1}{X\Tftest} \sum_{i=1}^{X} 
\sum_{j = \Tftrain+1}^{T_f} \left( T^{model}(x_i,t_j) - \mean{\TABM(x_i,t_j)} \right)^2}.$$}

\subsection{Predicting new ABM data \new{using BINN-guided PDE models}}\label{subsec:ABM_prediction}

\new{We combine BINN modeling, multivariate interpolation, and numerical integration of PDEs to predict new ABM data (Part 3b of Figure \ref{fig:methods_overview}). In \emph{predicting}, we assess the ability of our proposed approach to compute ABM data at a parameter value that has not been seen previously. This could correspond to exploring an ABMs' parameter space, or predicting the output of an experimental procedure for different experimental conditions, such as drug concentration or the initial number of agents.}

We perform multivariate interpolation using BINNs' computed diffusion rates to predict density-dependent diffusion rates for new ABM data. We define a prior parameter collection and a new parameter collection as 
$$ \mathcal{P}^{prior} = \{\Pm_k\}_{k=1}^{K_1} \text{ and } \mathcal{P}^{new}=\{\Pmnew_k\}_{k=1}^{K_2}.$$ Our workflow for predicting ABM data from $\mathcal{P}^{new}$ proceeds as follows:
\begin{enumerate}
    \item Generate the prior and new ABM data collections by simulating the ABM at all parameters from the prior and new parameter collections:
$$\mathcal{T}^{prior} = \bigg\{\mean{\TABM(x,t;\Pm_k)}\bigg\}_{k=1}^{K_1} \text{ and } \mathcal{T}^{new} = \bigg\{\mean{\TABM(x,t;\Pmnew_k)}\bigg\}_{k=1}^{K_2}.$$ 
\item Train a BINN model to each $k^{th}$ training ABM dataset from $\mathcal{T}^{prior}$ and extract $\DMLP(T;\Pm_k)$ from the trained BINN model.
\item Perform multivariate interpolation on $\{\DMLP(T;\Pm_k)\}_{k=1}^{K_1}$ to create an interpolant, $\mathcal{D}^{interp}(T;\Pm)$, that matches the concatenated vector $[T,\Pm_k]$ to the diffusion rate $\mathcal{D}^{MLP}(T;\Pm_k)$ for $k=1,\dots,K_1$. 
\item Predict the new ABM dataset, $\mean{\TABM(x,t;\Pmnew_k)}$, by simulating Equation \eqref{eq:diffusion_framework} with $\mathcal{D}=\mathcal{D}^{interp}(T;\Pmnew_k)$ to create $T^{interp}(x,t;\Pmnew_k)$. Partition $T^{interp}(x,t;\Pmnew_k)$ into its training and testing datasets to match the ABM data's training and testing datasets.
\item Compute the training and testing MSEs between $T^{interp}(x,t;\Pmnew_k)$ and $\mean{\TABM(x,t;\Pmnew_k)}$ to summarize the predictive performance of $T^{interp}(x,t;\Pmnew_k)$ for $k=1,\dots,K_2$.
\end{enumerate}
We implement multi-dimensional radial basis function interpolation using Sci-kit Learn's (version 0.24.2) \textbf{RBFInterpolator} command  to create $\Dinterp(T;\Pm)$.

\section{Results}\label{sec:results}

\subsection{\new{Mean-field and BINN-guided PDEs accurately forecast baseline ABM simulations}} \label{sec:results_baseline_forecasting}

\begin{figure}
    \centering
    \includegraphics[width=\textwidth]{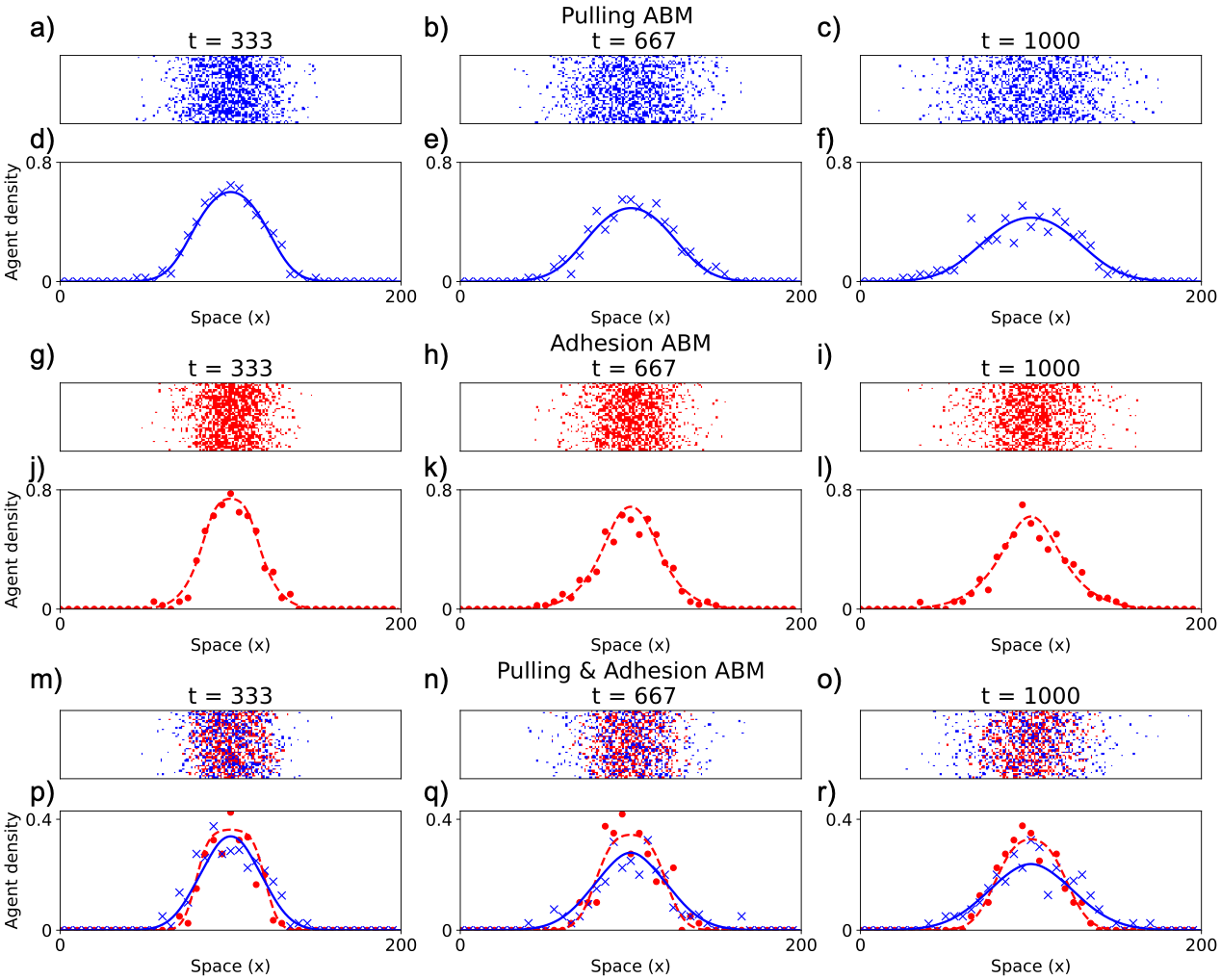}
    \caption{Baseline ABM simulation snapshots and the mean-field PDE models for the Pulling, Adhesion, and Pulling \& Adhesion ABMs. Blue pixels denote pulling agents and red pixels denote adhesive agents. All ABMs were simulated on rectangular  200$\times$40 lattices. (a-c) Snapshots of the Pulling ABM for $\rmp=1.0, \ppull=0.5$. (d-f) The output spatiotemporal pulling agent density (blue `x' marks) is plotted against the solution of the mean-field PDE (solid blue line) given by Equation \eqref{eq:pulling_MF}. (g-i) Snapshots of the Adhesion ABM for $\rmh=1.0, \padh=0.5$. (j-l) The output spatiotemporal adhesive agent density (red dots) is plotted against the solution of the mean-field PDE (dashed red line) given by Equation \eqref{eq:adhesion_MF}. (m-o) Snapshots of the Pulling \& Adhesion ABM for $\rmp=1.0, \rmh=0.25, \ppull=0.33, \padh = 0.33, \alpha = 0.5$. (p-r) The output spatiotemporal pulling and adhesive agent densities are plotted against the solution of the mean-field PDE given by Equation \eqref{eq:PullingAdhesionMF}.}
    \label{fig:MF_ABM_compare}
\end{figure}

\new{We simulated the three case study ABMs using the configuration values provided in Table \ref{tab:ABM_setup}. These values were chosen to match previous studies \cite{chappelle_pulling_2019, simpson_reliable_2022}. For ABMs of collective migration, one often chooses a large spatiotemporal domain to ensure ample ABM behavior is observed (e.g., the population spreads) while ensuring the boundary does not affect this behavior. In Table \ref{tab:model_predictions}, we provide baseline model parameter values for each case study ABM; these values were arbitrarily chosen to demonstrate typical ABM behavior \new{characterized by moderate population spread}. The ABM outputs are depicted against each ABM's mean-field PDE in Figure \ref{fig:MF_ABM_compare}. The mean-field PDE models accurately describe the baseline simulations for all three ABMs.}

\begin{figure}
    \centering
    \includegraphics[width=.9\textwidth]{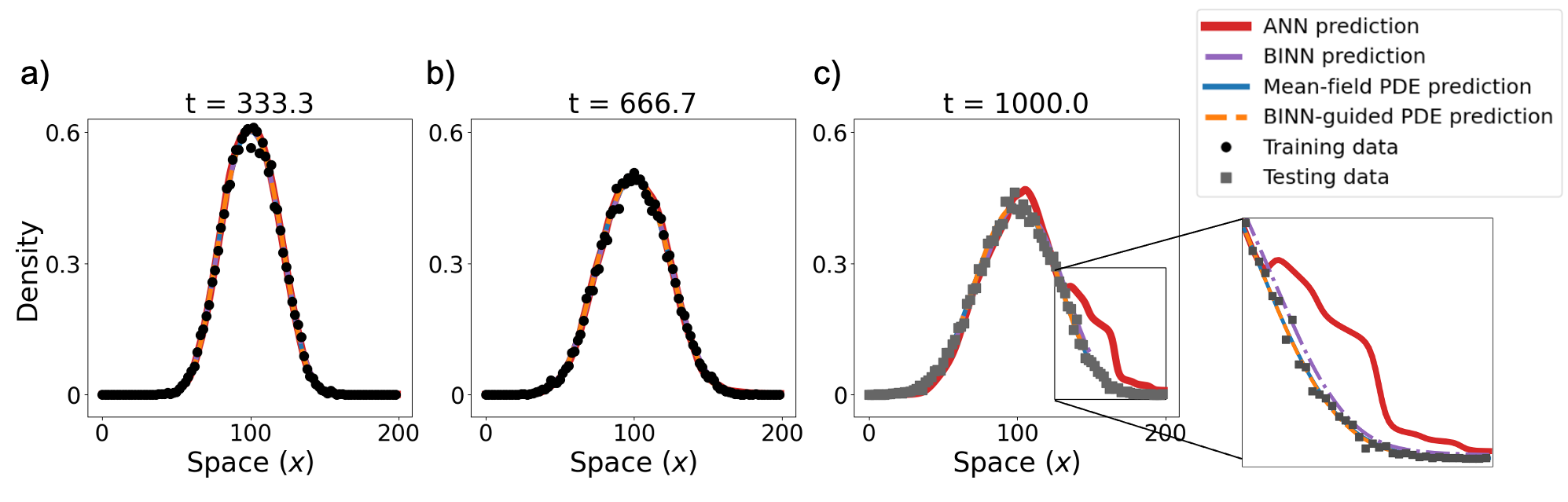}
    \caption{Forecasting Pulling ABM data with neural networks and PDEs. ANN and BINN models were trained to fit $\mean{\TABM(x,t)}^{train}$ from the Pulling ABM with $\Pm = (\rmp,\ppull)^T = (1.0, 0.5)^T$. These two neural networks and the mean-field and BINN-guided PDE simulations were then used to forecast (a-b) $\mean{\TABM(x,t)}^{train}$ and (c) $\mean{\TABM(x,t)}^{test}$. \old{The mean-field PDE simulation is not plotted here because it is visually indistinguishable from the BG PDE simulation.}}
    \label{fig:ANN_BINN_PDE_prediction}
\end{figure}

We investigate the performance of the mean-field PDE, ANN, BINN, and BINN-guided PDE models in forecasting Pulling ABM data \new{from the baseline parameter values provided in Table \ref{tab:model_predictions}}. Visual inspection suggests that all four models match the ABM training data well (Figure \ref{fig:ANN_BINN_PDE_prediction}(a-b)\old{; the mean-field PDE  is not plotted in this figure because it is visually indistinguishable from the BINN-guided PDE}). The computed training MSE values reveal that the mean-field and BINN-guided PDEs outperform the neural networks in describing this data (Table \ref{tab:model_predictions}). The BINN, BINN-guided PDE, and mean-field PDE all accurately forecast the testing data (Figure \ref{fig:ANN_BINN_PDE_prediction}(c)), but the two PDE models achieve smaller testing MSE values than the BINN model (Table \ref{tab:model_predictions}). The ANN's prediction for the testing data has a protrusion that overpredicts all data for $x>125$ (Figure \ref{fig:ANN_BINN_PDE_prediction}(c) inset), which causes this model's computed testing MSE value to be almost an order of magnitude higher than all others. \old{We simulated the Pulling ABM with $\Pm = (\rmp,\ppull)^T = (1.0, 0.5)^T$ to generate the ABM data. The ANN was trained to minimize the loss function $\mathcal{L}_{WLS}$ from Supplementary Equation \eqref{eq:L_ols}, whereas the BINN was trained to minimize $\mathcal{L}_{BINN}$ from Supplementary Equation \eqref{eq:Loss_total}. Both PDE models simulate Equation \eqref{eq:diffusion_framework_1d}: for the BINN-guided PDE, $\D = \DMLP$ from the trained BINN model; for the mean-field PDE, $\D=\D^{pull}$ from Equation \eqref{eq:pulling_MF}.} \new{We obtain similar results when using the four models to predict data from the Adhesion ABM and Pulling \& Adhesion ABM at their baseline parameter values (Table \ref{tab:model_predictions} and Supplementary Figure \ref{fig:forecasting_baseline_sims}).}

\begin{table}[]
    \centering
    \begin{tabular}{|l|c|c|}
    \hline
         \textbf{Forecasting model} & \textbf{Training MSE} & \textbf{Testing MSE}  \\ \hline
         \multicolumn{3}{|c|}{The Pulling ABM}\\
         \multicolumn{3}{|c|}{with baseline parameters $ \Pm = (\rmp, \ppull)^T = (1.0, 0.5)^T$} \\\hline
         ANN & $1.17\times10^{-4}$ & $9.36\times10^{-4}$ \\\hline
	BINN & $9.32\times10^{-5}$ & $1.47\times10^{-4}$ \\\hline
         Mean-field PDE & $7.45\times10^{-5}$ & $1.00\times10^{-4}$ \\\hline
         BINN-guided PDE & $7.64\times10^{-5}$ & $1.02\times10^{-4}$ \\\hline
         
         \multicolumn{3}{|c|}{The Adhesion ABM}\\ 
         \multicolumn{3}{|c|}{with baseline parameters $\Pm = (\rmh, \padh)^T = (1.0, 0.5)^T$} \\\hline
         ANN & $1.55\times10^{-4}$ & $1.84\times10^{-3}$ \\\hline
         BINN & $8.54\times10^{-5}$ & $1.50\times10^{-4}$ \\\hline
         Mean-field PDE & $7.18\times10^{-5}$ & $9.21\times10^{-5}$ \\\hline
         BINN-guided PDE & $7.43\times10^{-5}$ & $1.02\times10^{-4}$ \\\hline
                  
         \multicolumn{3}{|c|}{The Pulling \& Adhesion ABM}\\
         \multicolumn{3}{|c|}{with baseline parameters}\\ 
         \multicolumn{3}{|c|}{$\Pm = (\rmp, \rmh, \ppull, \padh, \alpha)^T = (1.0, 0.25, 0.33, 0.33, 0.5)^T$}\\\hline
         ANN & $1.25\times10^{-4}$ & $2.67\times10^{-3}$ \\\hline
         BINN & $9.65\times10^{-5}$ & $9.96\times10^{-5}$ \\\hline
         Mean-field PDE & $7.50\times10^{-5}$ & $8.55\times10^{-5}$ \\\hline
         BINN-guided PDE & $6.55\times10^{-5}$ & $9.11\times10^{-5}$ \\\hline
         
    \end{tabular}
    \caption{Computed training and testing MSE values. Computed MSE values when forecasting $\mean{\TABM(x,t)}^{train}$ and $\mean{\TABM(x,t)}^{test}$ from the three ABMs at their baseline parameter values. We used an ANN, BINN, mean-field PDE, and BINN-guided PDE to forecast each baseline ABM dataset.}
    \label{tab:model_predictions}
\end{table}

\subsection{Forecasting ABM data \new{for many parameter values} with BINN-guided and mean-field PDE simulations}\label{subsec:results_ABMForecasting}

We now investigate the performance of BINN-guided and mean-field PDE simulations in forecasting ABM datasets \new{over a wide range of parameter values} for all three case study ABMs. \new{We only consider the two PDE models (and exclude the neural network models) in this section due to their strong forecasting performance in Section \ref{sec:results_baseline_forecasting}.}

\subsubsection{The BINN-guided and mean-field PDEs both accurately forecast Pulling ABM data}\label{subsubsec:Forecast_pulling_ABM}

\begin{figure}
    \centering
    \includegraphics[width=\textwidth]{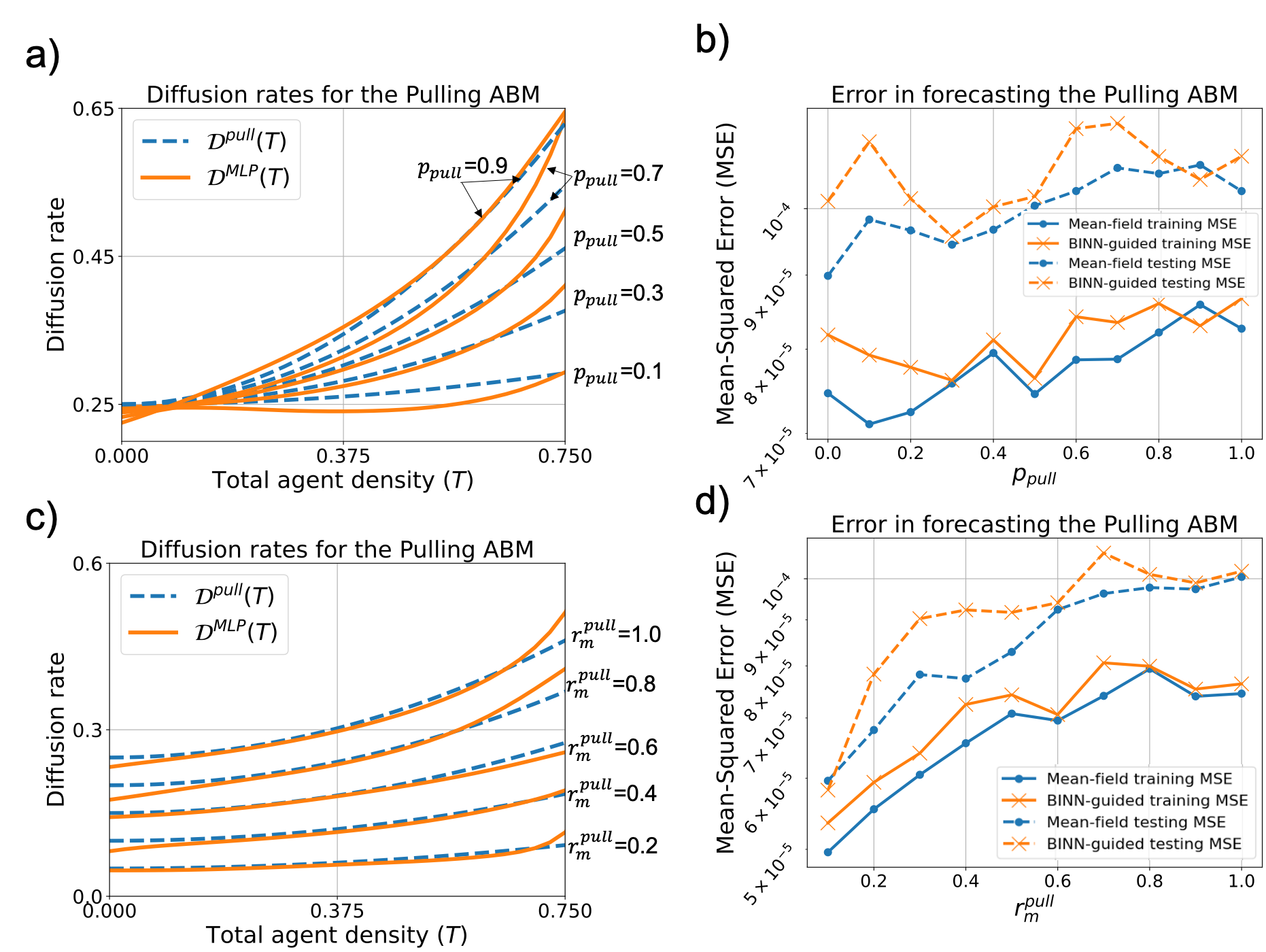}
    \caption{Forecasting Pulling ABM data with the mean-field (MF) and BINN-guided PDEs. (a) Plots of the mean-field diffusion rate, $\D^{pull}(T)$, from Equation \eqref{eq:pulling_MF} and the \new{BINN-guided diffusion rate}, $\DMLP(T)$, for $\ppull=0.1, 0.3, \dots, 0.9$ (results not shown for $\ppull=0.0, 0.2, \dots, 1.0$ for visual ease) while fixing $\rmp$ \new{at its baseline value of 1.0}. \new{The horizontal axis ends at 0.75 instead of 1.0 because the ABM simulations begin with a density of 0.75 and will rarely exceed this initial value. The BINN cannot reliably predict the diffusion rate for densities outside the values observed in the data.} (b) Plots of the mean-field and BINN-guided PDEs' computed training and testing MSE values while varying $\ppull$ and fixing $\rmp=1.0$. (c) Plots of $\D^{pull}(T)$ and $\DMLP(T)$ for $\rmp=0.2, 0.4, \dots, 1.0$ while fixing $\ppull$ \new{at its baseline value of 0.5}. (d) Plots of the mean-field and BINN-guided PDEs' computed training and testing MSE values while varying $\rmp$ and fixing $\ppull=0.5$.}
    \label{fig:pulling_vary_ppull}
\end{figure}

The parameters for the Pulling ABM are $\Pm=(\rmp,\ppull)^T$. To evaluate the BINN-guided and mean-field PDE models' performances in forecasting Pulling ABM data over a range of agent pulling parameter values, we computed eleven ABM datasets by varying $\ppull=0.0, 0.1, 0.2, \dots, 1.0$ while fixing $\rmp$ \new{at its baseline value of 1.0}. The inferred rates of agent diffusion from both models propose that agents diffuse slower for low densities and faster for high densities (Figure \ref{fig:pulling_vary_ppull}(a)). \new{While the mean-field diffusion rate at $\ppull=0$ is constant, BINNs do not use this \emph{a priori} information. Instead, their flexible nature leads to them learning a different diffusion rate from the data.}  The two PDE models achieve comparable training and testing MSE values for all values of $\ppull$, though the mean-field PDE usually attains slightly smaller values (Figure \ref{fig:pulling_vary_ppull}(b)). Snapshots of both simulated PDE models against data shows that their ABM predictions are visually indistinguishable (Supplementary Figure \ref{fig:pulling_PDE_simulations}(a-c)).  

To evaluate both PDE models' performances over a range of pulling agent migration values, we computed 10 Pulling ABM datasets with $\rmp=0.1, 0.2, \dots, 1.0$ while fixing $\ppull$ \new{at its baseline value of 0.5}. We find close agreement between both models' inferred diffusion rates for all values (Figure \ref{fig:pulling_vary_ppull}(c)). Both models achieve similar computed training and testing MSE values (Figure \ref{fig:pulling_vary_ppull}(d)). Snapshots of both simulated PDE models against data reveals that their ABM predictions are visually indistinguishable (Supplementary Figure \ref{fig:pulling_PDE_simulations}(d-f)). 

\subsubsection{BINN-guided PDEs accurately forecast Adhesion ABM data when the mean-field PDE is ill-posed}\label{subsec:adhesion_ABM_forecasting}

\begin{figure}
    \centering
    \includegraphics[width=.9\textwidth]{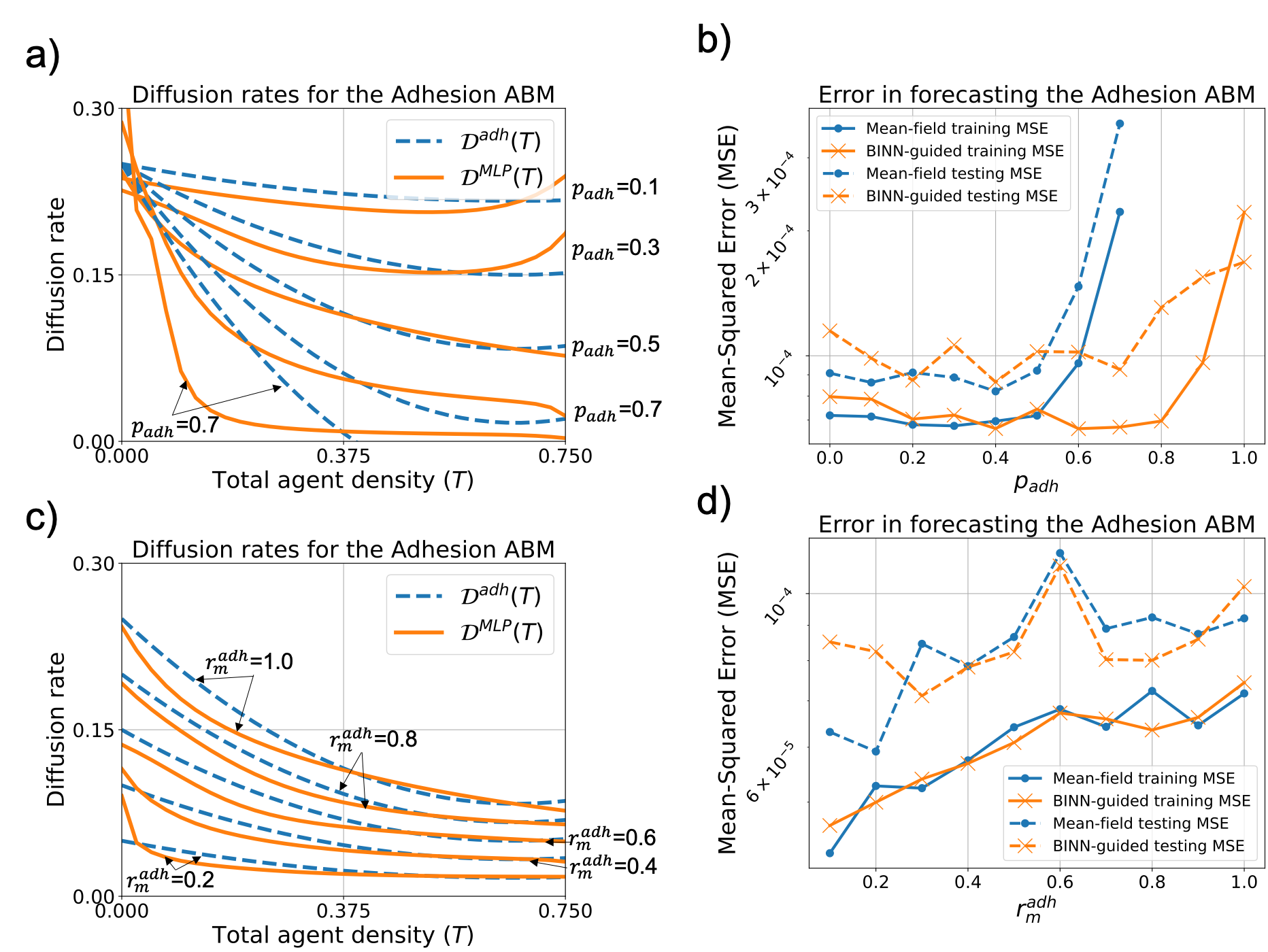}
    \caption{Forecasting Adhesion ABM data with the mean-field (MF) and BINN-guided PDEs. (a) Plots of the mean-field diffusion rate, $\D^{adh}(T)$, from Equation \eqref{eq:adhesion_MF} and the \new{BINN-guided diffusion rate}, $\DMLP(T)$, for $\padh=0.1, 0.3, \dots, 0.9$ (results not shown for $\padh=0.0, 0.2, \dots, 1.0$ for visual ease) while fixing $\rmh$ \new{at its baseline value of 1.0}. (b) Plots of the mean-field and BINN-guided PDEs' computed training and testing MSE values while varying $\padh$ and fixing $\rmh=1.0$. (c) Plots of $\D^{adh}(T)$ and $\DMLP(T)$ for $\rmh=0.2, 0.4, \dots, 1.0$ while fixing $\padh$ \new{at its baseline value of 0.5}. (d) Plots of the mean-field and BINN-guided PDEs' computed training and testing MSE values while varying $\rmh$ and fixing $\padh=0.5$.}
    \label{fig:adhesion_vary_padh}
\end{figure}

The parameters for the pulling ABM are $\Pm=(\rmh,\padh)^T$. To evaluate the BINN-guided and mean-field PDE models' performances over a range of agent adhesion parameter values, we computed eleven ABM datasets by varying $\padh=0.0, 0.1, 0.2, \dots, 1.0$ while fixing $\rmh$ \new{at its baseline value of 1.0}. The inferred rates of agent diffusion from both models decrease with agent density for most values of $\padh$ (Figure \ref{fig:adhesion_vary_padh}(a)). When $\padh=0$, the BINN-guided diffusion rate is slightly increasing  and the mean-field model's diffusion rate is constant. The BINN-guided diffusion rates decline faster with agent density than the corresponding mean-field diffusion rates for low density values. We computed the training and testing MSEs for both models for all values of $\padh$ (Figure \ref{fig:adhesion_vary_padh}(b)) and partition the results as follows :
 \begin{itemize}
     \item \textbf{When $\boldsymbol{\padh<0.5}$}: both models achieve similar training MSE values near $7\times10^{-5}$ and  testing MSE values around $10^{-4}$.
     \item \textbf{When $\boldsymbol{0.5\le \padh \le 0.75}$:} the mean-field PDE models' training and testing MSE values increase with $\padh$, with a maximum computed value above $3\times10^{-4}$. The BINN-guided PDE model's training and testing MSE values remain near $7\times10^{-5}$ and $10^{-4}$, respectively.
     \item \textbf{When $\boldsymbol{\padh>0.75}$:} the mean-field PDE model is ill-posed and cannot forecast this ABM data. The BINN-guided PDE model's computed training and testing MSE values increase with $\padh$ and have a maximum computed value of $2\times10^{-4}$.
 \end{itemize}
Close inspection of snapshots from both PDE model simulations against ABM data from $\padh=0.7$ reveals that the mean-field PDE model slightly overpredicts the data at high densities above 0.5 and low densities below 0.1, whereas the BINN-guided PDE closely matches the data (Supplementary Figure \ref{fig:adhesion_PDE_simulations}(a-c)).

To evaluate both PDE models' performances over a range of adhesive agent migration values, we computed ten ABM datasets with $\rmh=0.1, 0.2, \dots, 1.0$ while fixing $\padh$ \new{at its baseline value of 0.5}. Both PDEs achieve similar computed training and testing MSE values for most values of $\rmh$ (Figure \ref{fig:adhesion_vary_padh}(d)). When $\rmh=0.1$, however, the BINN-guided PDE's testing MSE value is close to $10^{-4}$, whereas the mean-field PDE attains a lower testing MSE value near $6\times10^{-5}$. Despite these differences, the two model simulations appear similar at these parameter values (Supplementary Figure \ref{fig:adhesion_PDE_simulations}(d-f)).

\subsubsection{BINN-guided PDEs accurately forecast Pulling \& Adhesion ABM data with a one-compartment model}

The parameters for the Pulling \& Adhesion ABM are $\Pm=(\rmp,\rmh,\ppull,\padh,\alpha)^T$. We evaluate the performance of the BINN-guided and mean-field DE models in forecasting data from the Pulling \& Adhesion ABM.  We created 48 ABM datasets by fixing the baseline parameter values at $\Pm_{base} = (1.0, 0.25, 0.33, 0.33, 0.5)^T$ and then varying each parameter individually. We vary $\rmp=0.5, 0.6,\dots,1.5$; $\rmh=0.0, 0.1,\dots,1.0$; $\ppull=0.1, 0.2, \dots , 0.6, 0.67$; $\padh=0.1, 0.2, \dots , 0.6, 0.67$; and $\alpha=0.0, 0.1,\dots,1.0$. These parameter values were chosen to always satisfy $\ppull+\padh \le 1.$

\begin{figure}
    \centering
    \includegraphics[width=.95\textwidth]{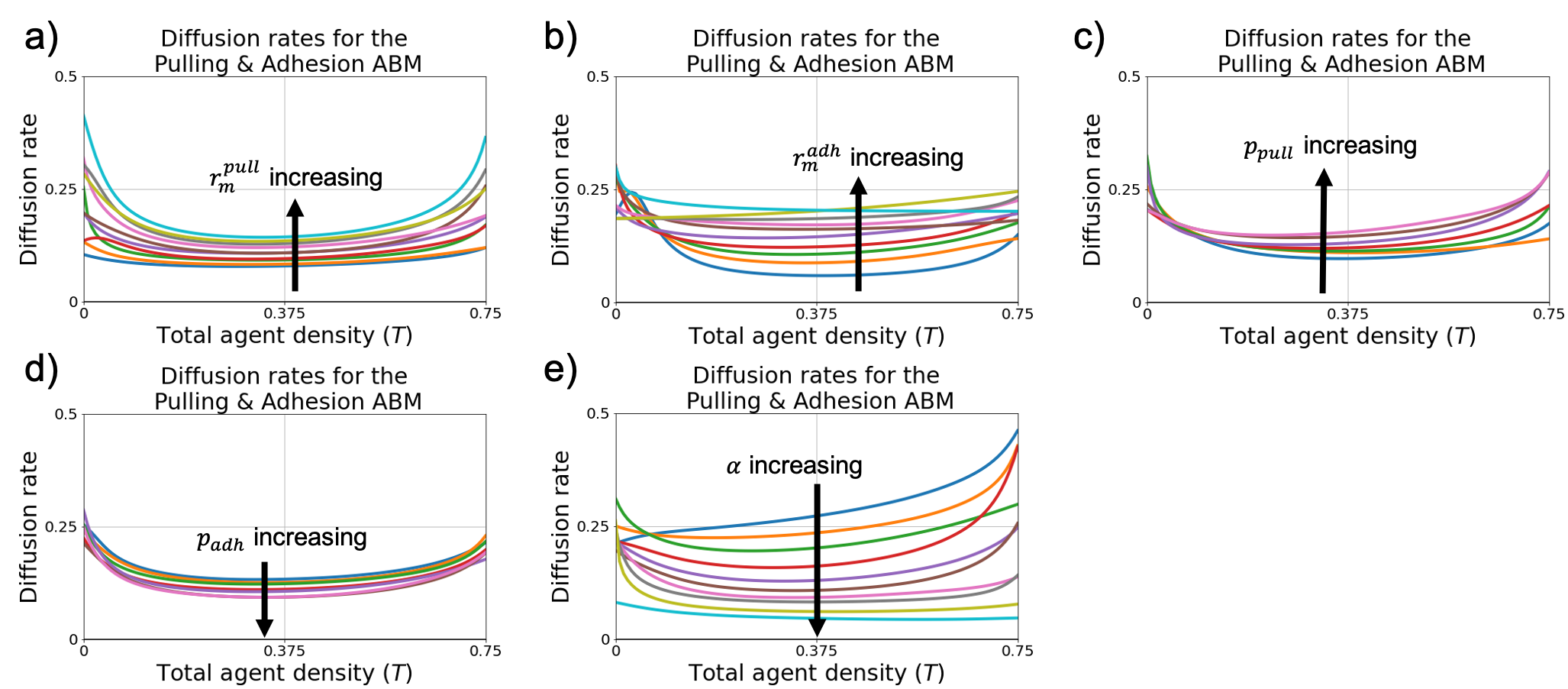}
    \caption{The \new{BINN-guided diffusion rates} for the Pulling \& Adhesion ABM data. Plots of the \new{BINN-guided diffusion rate}, $\DMLP(T)$, when varying (a) $\rmp$, (b) $\rmh$, (c) $\ppull$, (d) $\padh$, and (e) $\alpha$.}
    \label{fig:heterogeneous_diffusion_rates}
\end{figure}

The BINN models' inferred diffusion rates, $\DMLP(T;\Pm)$, are often U-shaped with larger diffusion values at low and high agent densities and smaller values at intermediate densities (Figure \ref{fig:heterogeneous_diffusion_rates}). This U-shape tends to increase for larger values of $\rmp, \rmh, \text{ and }\ppull$ and decrease for larger values of $\padh$ and $\alpha$. The inferred diffusion rates appear most sensitive to changes in the $\alpha$ parameter: at $\alpha=0.0$, $\DMLP(T;\Pm)$ strictly increases with agent density and attains an average value of 0.289; at $\alpha=1.0$, $\DMLP(T;\Pm)$ is strictly decreasing and has an average value of 0.051. The inferred diffusion rate is also sensitive to the $\rmh$ and $\rmp$ parameters: varying $\rmh$ primarily alters the BINN diffusion rate at intermediate agent density values, whereas varying $\rmp$ changes the BINN diffusion rate at low and high agent densitiy values.

\begin{figure}
    \centering
    \includegraphics[width=.95\textwidth]{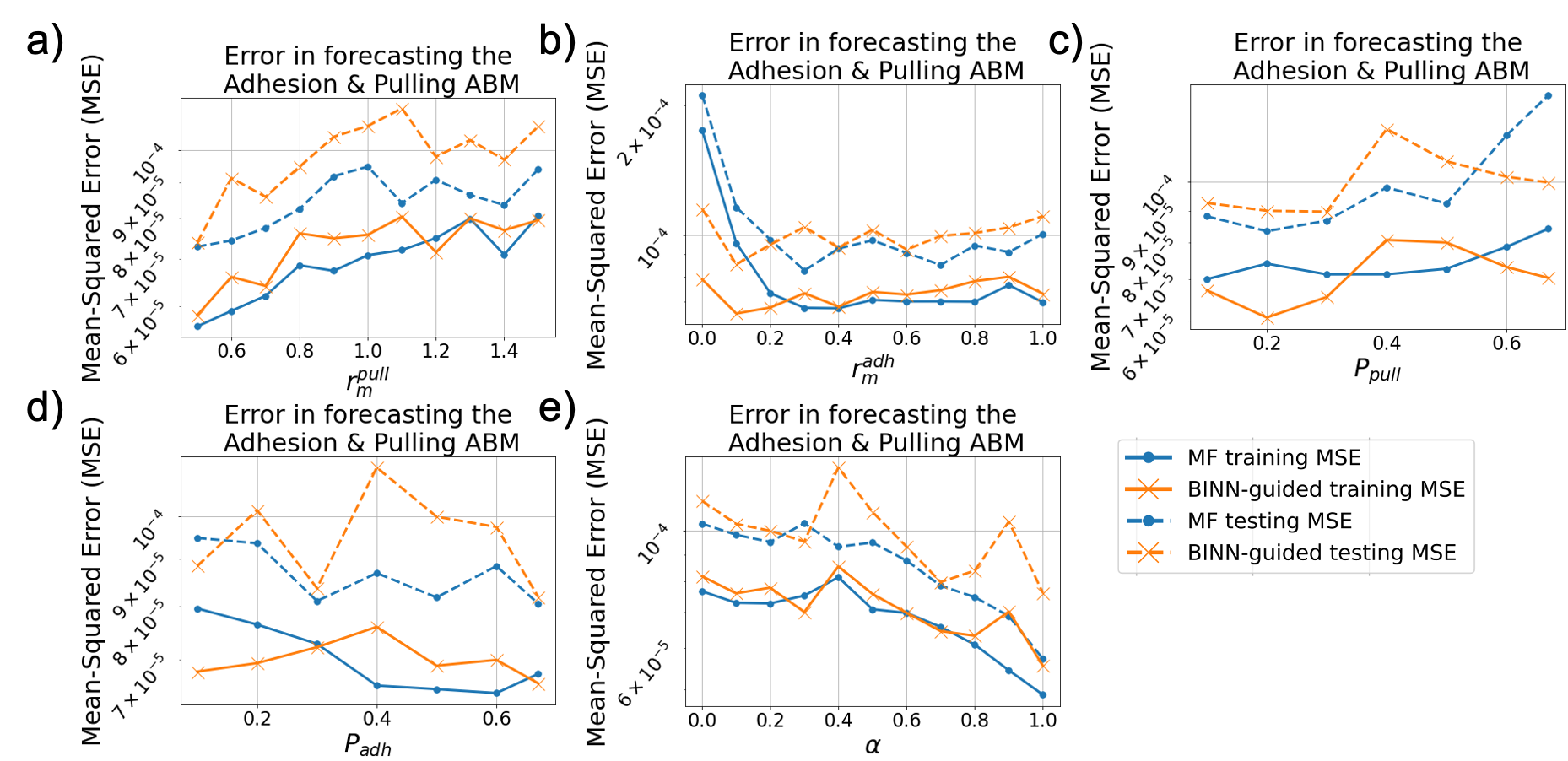}
    \caption{Forecasting Pulling \& Adhesion ABM data with the mean-field and BINN-guided PDEs. Plots of the mean-field and BINN-guided PDEs' computed training and testing values while varying (a) $\rmp$, (b) $\rmh$, (c) $\ppull$, (d) $\padh$, and (e) $\alpha$.}
    \label{fig:heterogeneous_forecasting_MSEs}
\end{figure}

The BINN-guided PDE computes a single compartment to forecast the total agent density, $T(x,t)$,  whereas the mean-field PDE computes two compartments forecasting the Pulling and Adhesive agent densities, $P(x,t)$ and $H(x,t)$, respectively. We forecast the total agent density with the mean-field PDE by setting $T(x,t)=P(x,t)+H(x,t)$. The two PDE models achieve similar training MSE values for most parameter values that we considered (Figure \ref{fig:heterogeneous_forecasting_MSEs}). The mean-field model's testing MSE values are often smaller than the BINN-guided testing MSE values, though the BINN-guided PDE also achieves small testing MSE values. For example, both PDE simulations accurately predict ABM data when $\padh$ is set to $0.4$, but visualizing both PDE simulations shows that the mean-field PDE better matches the elbow of the data than the BINN-guided PDE (Supplementary Figure \ref{fig:Pullingadhesion_PDE_simulations}(a-c)). The BINN-guided PDE outperforms the mean-field PDE in forecasting data for small values of $\rmh$: plotting both PDE simulations against data from $\rmh=0.1$ shows that the mean-field PDE underpredicts the largest agent density values, while the BINN-guided PDE accurately matches this data (Supplementary Figure \ref{fig:Pullingadhesion_PDE_simulations}(d-f)).

\subsection{Predicting ABM data at new parameter values}

We now examine how performing multivariate interpolation on several BINN-guided diffusion rates, $\DMLP(T;\Pm)$, can aid the prediction of previously-unseen ABM data at new parameter values (see Section \ref{subsec:ABM_prediction} for implementation details). 

We predict new data from the Adhesion and Pulling \& Adhesion ABMs in this section. We do not include the Pulling ABM in this work because the mean-field PDE model accurately forecasted ABM data for all parameter values that we considered in Section \ref{subsubsec:Forecast_pulling_ABM}.

\subsubsection{Predicting Adhesion ABM data}
 
The parameters for the Adhesion ABM are $\Pm = (\rmh,\padh)^T$. We perform ABM data prediction for $\padh\ge0.5$ in this section because we found that the mean-field PDE model accurately forecasted ABM data for $\padh \le0 .5$ in Section \ref{subsec:adhesion_ABM_forecasting}. 
 
\begin{figure}
    \centering
    \includegraphics[width=.95\textwidth]{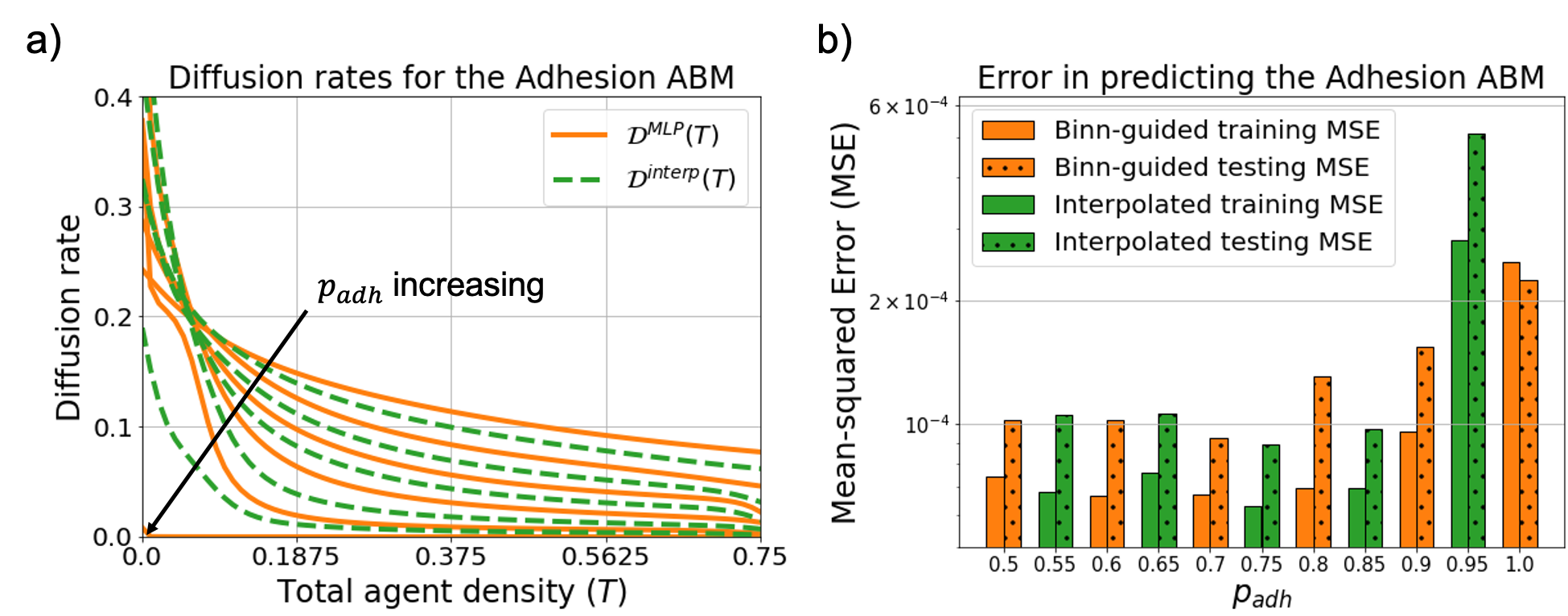}
    \caption{Predicting Adhesion ABM data with BINN-guided PDEs and multivariate interpolation for new $\padh$ values. The parameters for the Adhesion ABM are given by $\Pm=(\rmh,\padh)^T.$ Here, we vary $\padh$ while fixing $\rmh$ \new{at its baseline value of 1.0}. The prior data collection consists of $\padh=0.5, 0.6, \dots , 1.0$ and the new data collection consists of $\padh=0.55, 0.65, \dots , 0.95$ (a) Plots of the learned $\DMLP(T;\Pm)$ diffusion rates for the prior data collection. We performed multivariate interpolation on these rates to obtain $\Dinterp(T;\Pm)$, which we plot for the new data collection. (b) Plots of the BINN-guided PDEs' computed training and testing values on the prior data collection, and the interpolated PDE's training and testing values on the new data collection.}
    \label{fig:adhesion_interpolation_padh}
\end{figure}

We first predict ABM data when varying $\padh$ and fixing $\rmh$. The prior data collection consists of \new{$K_1=6$} ABM datasets generated by varying $\padh=0.5, 0.6, 0.7, \dots, 1.0$ while fixing $\rmh$ \new{at its baseline value of 1.0}; the new data collection consists of \new{$K_2=5$} ABM datasets generated by varying $\padh=0.55, 0.65, 0.75, 0.85, \text{ and } 0.95$ while fixing $\rmh$ \new{at its baseline value of 1.0}. We performed multivariate interpolation over the six inferred $\DMLP(T;\Pm)$ terms from the prior data collection to generate $\D^{interp}(T;\Pm)$. We use this interpolant to predict the diffusion rates for all parameters from the new data collection (Figure \ref{fig:adhesion_interpolation_padh}(a)). All interpolated diffusion rates decrease with agent density and tend to fall with larger $\padh$ values. Most of the computed training and testing MSE values on the new data collection are comparable to their counterparts from the prior data collection (Figure \ref{fig:adhesion_interpolation_padh}(b)). The lone exception occurs at $\padh=0.95$, where the testing MSE exceeds $5\times10^{-4}$ while the testing MSEs at $\padh=0.9 \text{ and } 1.0$ do not exceed $2.5\times10^{-4}$. Visual inspection of the simulated PDE prediction against ABM data at $\padh=0.95$ reveals that it matches the data well but slightly mispredicts the data's heel at later time points (Supplementary Figure \ref{fig:Adhesion_PDE_predictions_Pm_fixed}(a-c)).

\begin{figure}
    \centering
    \includegraphics[width=.95\textwidth]{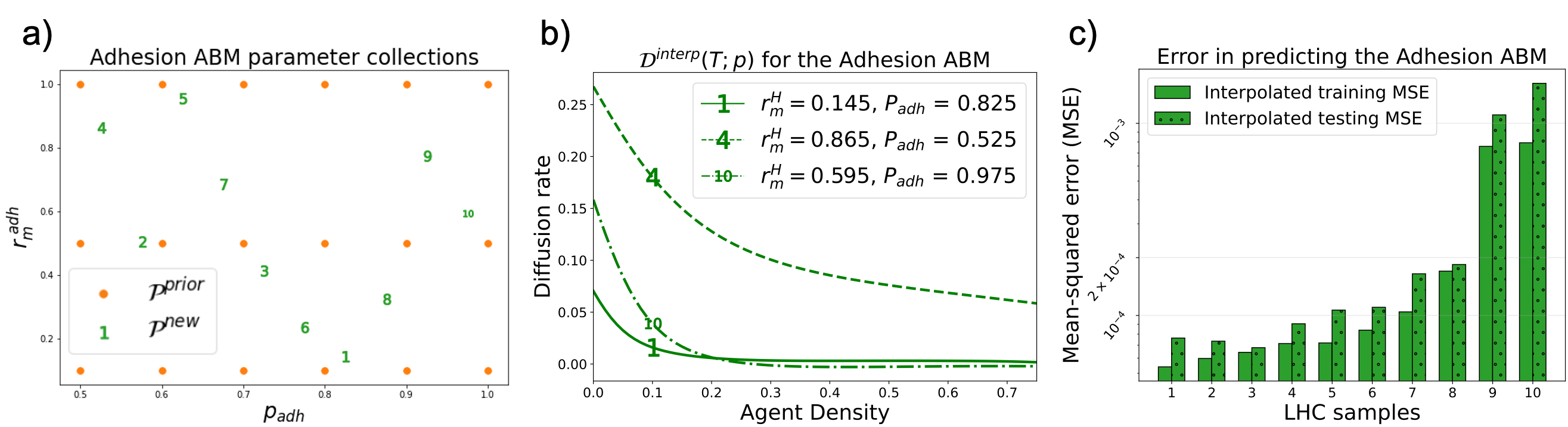}
    \caption{Predicting Adhesion ABM data with BINN-guided PDEs and multivariate interpolation for new $\rmh$ and $\padh$ values. The parameters for the Adhesion ABM are given by $\Pm=(\rmh,\padh)^T.$ Here, we vary both parameters. (a) The prior data collection consists of $\rmh=0.1, 0.5, 1.0$ and $\padh=0.5, 0.6, \dots , 1.0$ and the new data collection consists of a Latin hypercube (LHC) sampling of $\Pm\in[0.1, 1.0]\times[0.5,1.0]$ with \new{$K_2=10$} samples. (b) We performed multivariate interpolation on the $\DMLP(T;\Pm)$ rates on the prior data collection to obtain $\Dinterp(T;\Pm)$. We plot three illustrative $\Dinterp(T;\Pm)$ values from the new data collection. (c) Plots of the interpolated PDE's training and testing values on the new data collection.}
    \label{fig:adhesion_interpolation_rmh_padh}
\end{figure}

We next predict ABM data when varying both $\rmh$ and $\padh$. The prior data collection consists of \new{$K_1=18$} ABM datasets generated by varying $\rmh=0.1, 0.5, 1.0$ and $\padh=0.5, 0.6, \dots, 1.0$; the new data collection consists of \new{$K_2=10$} ABM datasets generated from a latin hypercube sampling of $(\rmh, \padh) \in [0.1,1.0] \times [0.5,1.0]$ (Figure \ref{fig:adhesion_interpolation_rmh_padh}(a) and Supplementary Table \ref{tab:Adhesion_LHC_sample}).  We performed multivariate interpolation over each $\DMLP(T;\Pm)$ from the prior data collection to generate $\D^{interp}(T;\Pm)$. The predicted diffusion rates for the new data collection decrease with agent density, rise for larger $\rmh$ values, and decrease faster for larger $\padh$ values (Figure \ref{fig:adhesion_interpolation_rmh_padh}(b)). We order the parameters from the new data collection by increasing training MSE values (Figure \ref{fig:adhesion_interpolation_rmh_padh}(c)). The four lowest training and testing MSE values are all below $1\times10^{-4}$, the eight lowest are all below $2\times10^{-4}$, and the highest testing MSE value reaches $1.6\times10^{-3}$. Visual inspection of the interpolated PDE prediction with the highest testing MSE value reveals that this simulation mispredicts the data's heel but otherwise matches the ABM data well (Supplementary Figure \ref{fig:Adhesion_PDE_predictions}(a-c)). Visual inspection of the interpolated PDE prediction with the third-highest MSE value shows that this simulation accurately matches the ABM data (Supplementary Figure \ref{fig:Adhesion_PDE_predictions}(d-f)).

\subsubsection{Predicting Adhesion \& Pulling ABM data}

The parameters for the Pulling \& Adhesion ABM are $\Pm = (\rmp, \rmh, \ppull, \padh, \alpha)^T$. We perform ABM data prediction over a large range of parameter values to determine if the one-compartment BINN-guided PDE simulations can predict this ABM's data, which results from two interacting subpopulations.

\begin{figure}
    \centering
    \includegraphics[width=.9\textwidth]{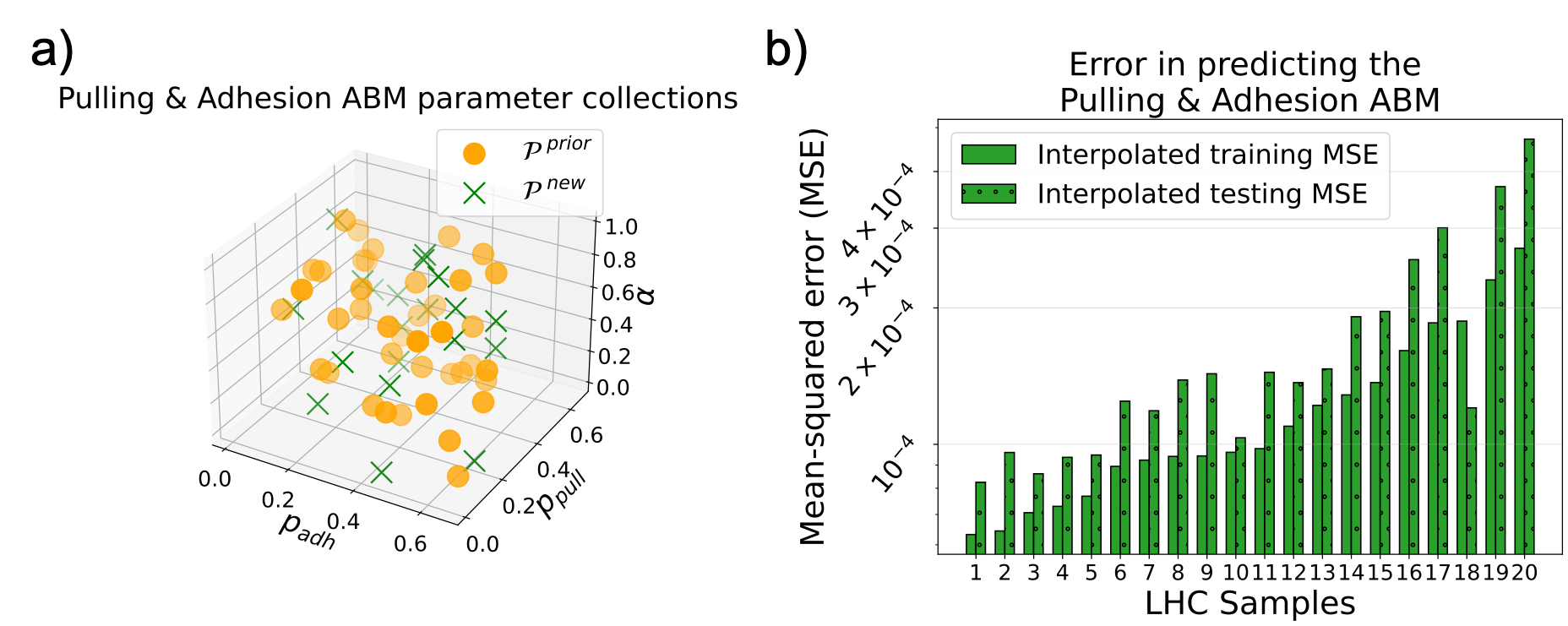}
    \caption{Predicting Pulling \& Adhesion ABM data for new $\ppull, \padh$, and $\alpha$ values. The parameters for the Adhesion ABM are given by $\Pm = (\rmh, \rmp, \padh, \ppull, \alpha)^T.$ Here, we vary $\ppull,\padh,$ and $\alpha$ while fixing $\rmp$ and $\rmh$ \new{at their baseline values of 1.0 and 0.25, respectively}. (a) The prior data consists of a Latin hypercube (LHC) sampling of $(\ppull,\padh,\alpha)\in[0, 0.67]\times[0, 0.67]\times[0,1]$ with \new{$K_1=40$} samples and the new data consists of a LHC sampling of the same domain with \new{$K_2=20$} samples. (b) Plots of the interpolated PDE's training and testing values on the new data, arranged by increasing training MSE values.}
    \label{fig:PullingAdhesionPrediction}
\end{figure}

We perform multivariate interpolation over the $\ppull, \padh, $ and $\alpha$ parameters while fixing $\rmp$ and $\rmh$ \new{at their baseline values of 1.0 and 0.25, respectively}. The prior and new data collections consist of \new{$K_1=40$} and \new{$K_2=20$} ABM parameter combinations, respectively, that were generated from Latin hypercube samplings of $(\ppull, \padh, \alpha) \in [0, 0.67]\times[0, 0.67]\times[0,1]$ (Figure \ref{fig:PullingAdhesionPrediction}(a) and Supplementary Tables \ref{tab:PullingAdhesionLHCprior} and \ref{tab:PullingAdhesionLHCnew}). We chose samplings where $\ppull + \padh \le 1.0$ for all samples. The computed training and testing MSE values for the new parameter collection suggest all simulated PDE predictions accurately match the ABM data at those parameters (Figure \ref{fig:PullingAdhesionPrediction}(b)). Of the \new{$K_2=20$} computed testing MSE values in the new data collection, four are below $1\times 10^{-4}$, 16 are below $2\times 10^{-4}$, and all are below $5\times10^{-4}$. The highest and third highest testing MSE value results from $(\ppull, \padh, \alpha) = (0.218, 0.553, 0.675)$ and $(0.251, 0.486, 0.975)$, respectively. Visually inspecting the interpolated PDE predictions from these parameter values against ABM data reveals that both match the data well, though the worst prediction overpredicts the largest ABM density values (Supplementary Figure \ref{fig:PullingAdhesion_PDE_predictions}). 

\subsection{Comparing the computational expense of each modeling approach}

We finish with a discussion on the computational expense of all approaches discussed in this work (Table \ref{tab:wall_time_estimation} and Supplementary Figure \ref{fig:wall_time_estimation}). We recorded the computed wall times to simulate each ABM, train each BINN model, and simulate each PDE from Section \ref{subsec:results_ABMForecasting}. Averaging across all ABMs suggests that the average ABM dataset took 40.0 minutes to generate with a standard deviation of 15.6 minutes. The average mean-field PDE model simulations for the Pulling ABM and the Adhesion ABM took 0.6 and 0.5 seconds to complete, respectively, which are about 4,000 and 4,500 times faster than the average ABM simulation time. The average mean-field PDE model simulation time for the Pulling \& Adhesion ABM was 4.7 seconds, which is 542 times faster than the average ABM simulation time. Training a BINN model is the most time-consuming task with an average time of 11.2 hours across all ABMs with a standard deviation of 4.32 hours. The average BINN-guided PDE simulation takes 82.9 seconds with a standard deviation of 77.12 seconds, which is approximately 28 times faster than simulating the ABM. 

\begin{table}[]
    \centering
    \begin{tabular}{|c|c|c|c|c|}
    \hline
     \textbf{ABM Name} & \textbf{ABM simulation} & \textbf{MF PDE simulation}& \textbf{BINN Training}  & \textbf{BG PDE simulation} \\ \hline
    Adhesion & 37.5 (15.4) minutes & 0.5 (0.15) seconds & 10.6 (4.44) hours  & 16.9 (23.65) seconds \\ \hline
    Pulling & 39.9 (15.8) minutes & 0.6 (0.20) seconds & 10.0 (3.99) hours & 164.8 (156.9) seconds \\ \hline
    Pulling \& Adhesion & 42.5 (15.52) minutes & 4.7 (1.20) seconds & 13.1 (4.54) hours  & 66.9 (50.81) seconds \\ \hline
    Average & 40.0 minutes & 1.9 seconds & 11.2 hours & 82.9 seconds \\ \hline
    \end{tabular}
    \caption{Computational expenses of each modeling approach. The mean wall time computations (standard deviation in parentheses) for ABM simulations, BINN training, mean-field (MF) PDE simulations, and BINN-guided (BG) PDE simulations for all three ABMs. \new{The last row depicts the average mean computation time across all three ABMs.}}
    \label{tab:wall_time_estimation}
\end{table}

\section{Discussion and Future Work} \label{sec:discussion}

In this work, we introduced how BINNs can be used to learn BINN-guided PDE models from simulated ABM data. BINN-guided PDE model simulations provide a new approach for forecasting and predicting ABM data. This methodology works by training a BINN model to match simulated ABM data while also obeying a pre-specified PDE model framework. After model training, future ABM data can be forecasted by simulating the BINN-guided PDE. Predicting ABM data at new parameters can be performed by simulating the pre-specified PDE framework with an interpolated modeling term. This model term is computed by interpolating over several learned BINN model terms and the parameter values that led to these terms.

It is challenging to predict how model parameters affect ABMs' output behavior due to their heavy computational nature. Mathematical modelers often address this limitation by coarse-graining ABM rules into computationally-efficient mean-field DE models. Unfortunately, these DE models may give misleading ABM predictions; furthermore, they can be ill-posed for certain parameter values \cite{anguige_one-dimensional_2009,baker_correcting_2010}. Here, we demonstrated that BINN-guided PDE models accurately forecast future ABM data and predict ABM data from new parameter values. One benefit of this BINN-guided approach for ABM prediction is that BINNs can, in theory, be trained to simulated data from complex ABMs because BINN models are agnostic to the ABM rules. This is in contrast to the coarse-graining approach, which is limited to ABMs with simple rules to ensure a final PDE model can be recovered.

A limitation of the BINN-guided approach for ABM forecasting and prediction is the computational expense of BINN model training. The average BINN training procedure in this study took 11.2 hours, which is 17 times longer than the average ABM data generation time of 40 minutes. Once a BINN model has been trained, however, the average BINN-guided PDE simulation took 83 seconds, which is 28 times faster than the average time to generate an ABM dataset. One possible source of these long BINN training times is our chosen BINN model architecture, which consists of over over 50,000 parameters to train. Kaplarevi-Malii et al.\ \cite{kaplarevic-malisic_identifying_2023} proposed a genetic algorithm to identify the optimal model archictecture for PINN models. In future work, we plan to implement this algorithm to identify simpler BINN model architectures that can be efficiently trained to learn predictive PDE models for ABMs. 

\old{We did not apply this methodology to experimental data in this work, however, it would} \new{This work was purely computational, as we applied all prediction methodologies to simulated ABM data. It will} be interesting \new{in the future} to \old{use BINN-guided PDE model simulations to analyze} \new{validate the BINN-guided methodology on} experimental data. \old{Stochastic ABMs are frequently used to simulate barrier and scratch assay experiments \cite{vo_quantifying_2015}.} Performing data-driven modeling techniques, such as parameter estimation, is challenging for ABMs due to their long simulation times\new{. Our results suggest that BINN-guided PDE models may advance parameter estimation for ABMs by providing an accurate and efficient ABM surrogate model}. \old{We can imagine reducing the computational cost of these processes with the aid of BINNs by training BINN models to a small number of ABM simulations and then comparing BINN-guided PDE simulations to experimental data in place of ABM simulations.} \new{For example, a typical approximate Bayesian computation (ABC) for parameter estimation requires performing 10,000 ABM simulations \cite{nguyen_quantifying_2024}, which would require more than 6,600 computational hours. If we instead simulate the ABM at 10 parameter combinations, train BINN models to these data, and then use 10,000 interpolated BINN-guided PDE model simulations for ABC, then this total process would take 349 hours, a 19-fold reduction in time. This  process will become even more efficient with new methodologies to expedite BINN model training.}

\old{Future work will include developing new BINN model architectures to enhance our understanding of heterogeneous ABMs. In this study, we simulated an ABM with two agent types to model a heterogeneous population and demonstrated that BINNs can learn interpretable one-compartment PDE models to forecast and predict simulated data from this ABM. In a future study, we plan to update the BINN modeling architecture to predict the density of each agent type using a multi-compartment PDE framework with separate model terms for each agent type. This will allow us to interpret how the input model parameters affect the modeling terms for each agent type. We could quantify the uncertainty of each modeling term by incorporating the Bayesian inference techniques from \cite{linka_bayesian_2022} into our analysis. There are thus many ways to extend the methods proposed in this work, which will allow us to study more complex ABMs and improve our understanding of the learned DE models.}

\textbf{Case study: collective migration.} We studied three case study ABMs that are applicable to cell biological experiments, such as barrier and scratch assays. Each ABM consists of rules governing how key cellular interactions (namely, pulling and adhesion) impact the collective migration of cell populations during these experiments \cite{nardini_modeling_2016, thompson_modelling_2012}. \old{We demonstrated the abilities of mean-field and BINN-guided PDE models in forecasting and predicting simulated data from these three case study ABMs. Visual inspection of the diffusion terms from both PDE models enables the interpretation of how cell-level interactions scale into population-level behavior for each ABM.}  Table \ref{tab:case_study_overview} summarizes the predictive and interpretative capabilities of the mean-field and BINN-guided PDE models for the three case study ABMs. For the Pulling ABM, \new{both models use interpretable one-compartment PDEs that accurately predict ABM behavior for all parameter values.} \old{the trained BINN models learn diffusion rates that are similar to the mean-field PDE diffusion rates for all parameter values (Figures \ref{fig:pulling_vary_ppull}(a) and (c)); both models perform similarly in forecasting future ABM data (Figures \ref{fig:pulling_vary_ppull}(b) and (d)). Both models are one-compartment PDEs that can easily be interpreted via visualization of their diffusion rates. The diffusion rates from both approaches propose that population diffusion increases with agent density, the rate of agent migration, and the probability of agent pulling.} For the Adhesion ABM, \new{the mean-field PDE predictions become less accurate for $\padh\in[0.5,0.75]$ and are ill-posed for $\padh>0.75$, whereas the BINN-guided PDEs make accurate predictions for $\padh\le0.9$.} \old{both PDE models are able to accurately forecast ABM data for adhesion probabilities below 0.5 (Figure \ref{fig:adhesion_vary_padh}(b) and (d)). The mean-field PDE's predictions become worse when this probability exceeds 0.5, however, and this model is ill-posed when this probability exceeds 0.75. BINN-guided PDEs, on the other hand,  are well-posed for all adhesion probabilities, and their ABM forecasts and predictions are accurate when this probability is below 0.9 (Figures \ref{fig:adhesion_vary_padh}(b) and (d), \ref{fig:adhesion_interpolation_padh}(b), and \ref{fig:adhesion_interpolation_rmh_padh}(c)). The diffusion rates from both PDEs agree that population diffusion decreases when either agent density or the probability of agent adhesion increases (Figure \ref{fig:adhesion_vary_padh}(a) and (c)). The BINN-guided PDE proposes that population diffusion decreases with agent density faster than the mean-field PDE suggests. Both models suggest that population diffusion increases with the rate of adhesive agent migration.} For the Pulling \& Adhesion ABM, both PDE models accurately forecast the total ABM data for most parameter values considered. \old{ (Figure \ref{fig:heterogeneous_forecasting_MSEs}).} The mean-field PDE model \new{is not interpretable, as it contains two compartments that consist of many terms. The BINN-guided PDE, on the other hand, achieves similar accuracy to the mean-field PDE with an interpretable one-compartment PDE model.} \old{for this ABM contains two compartments, one for for pulling agents and another for adhesive agents. The BINN-guided PDE, however, achieves similar accuracy to the mean-field PDE using an interpretable one-compartment PDE model whose diffusion rate describes how the total population diffuses (Figure \ref{fig:heterogeneous_diffusion_rates}). We found that many (but not all) learned diffusion rates are U-shaped, meaning that agent diffusion decreases with agent density at low density values and increases at high density values (Figure \ref{fig:heterogeneous_diffusion_rates}). Visualizing the learned diffusion rates for many parameter values reveals that this U-shape increases when we increase the rates of pulling agent migration, adhesive agent migration, and the probability of agent pulling; the U-shape decreases when we increase the probability of adhesion or the proportion of adhesive agents in the simulation.  As opposed to the BINN-guided PDE model.}

\begin{table}
    \centering
    \begin{tabular}{|l|c|c|}
        \hline
         & \textbf{ABM prediction} & \textbf{Interpretability} \\\hline
        \multirow{2}{*}{\textbf{Pulling ABM}} &  MF PDE accurate for all parameters  & MF PDE is interpretable \\
           & BG PDE accurate for all parameters &  BG PDE is interpretable\\\hline
        \multirow{2}{*}{\textbf{Adhesion ABM}} &  MF PDE accurate for $\padh\le0.5$  & MF PDE is interpretable \\
           & BG PDE accurate for $\padh\le0.9$ &  BG PDE is interpretable\\\hline
        \multirow{2}{*}{\textbf{Pulling \& Adhesion ABM}} &  MF PDE accurate for all parameters  & MF PDE not interpretable \\
           & BG PDE accurate for all parameters &  BG PDE is interpretable\\\hline
    \end{tabular}
    \caption{Highlighting the ability of mean-field (MF) and BINN-guided (BG) PDEs to accurately forecast simulated ABM data with interpretable PDE models.}
    \label{tab:case_study_overview}
\end{table}

\new{We compared the performance of the mean-field and BINN-guided PDE models throughout this work. We emphasize, however, that these two approaches are complementary, and our thorough investigation highlights the strengths and limitations of each model. The mean-field PDE is fast to simulate but can provide inaccurate, ill-posed, and/or uninterpretable ABM predictions. The BINN-guided PDE accurately predicts ABM behavior with an interpretable PDE, but current BINN model training times are lengthy. We encourage modelers to refer to these guidelines when deciding which approach to use for their future applications.}

\old{In future work, we plan to study more complex ABMs of collective migration using the BINNs methodology. \old{Chappelle and Yates \cite{chappelle_pulling_2019} considered rules where multiple agents interact during cell pulling events. They found that the predictive accuracy of mean-field PDE models declines as more agents become involved in pulling processes. In our study, we focused on simple cell-cell interaction rules between two agents: in future work, we will investigate how BINN-guided PDEs change with the number of agents involved in each interaction. Gallaher et al. \cite{gallaher_spatial_2018} used a complex two-dimensional spatial ABM to study how spatial heterogeneity and evolutionary dynamics impact a tumor's response to adaptive cancer therapies. Each agent in their model has its own internal growth rate and response to drug exposure. It would be interesting to determine if the BINNs approach from our study can be used to describe the complex spread, growth, and sensitivity of this model's simulated tumors in response to drug treatment.} Agent proliferation is an important component of collective migration that we did not consider in our study. Many previous studies have shown that coarse-grained DE models fail to accurately describe ABM simulations when agents proliferate quickly \cite{baker_correcting_2010}. It would be straightforward to extend our methods to predict ABM data from models where agents migrate and proliferate by adding a population growth term into the PDE framework during BINN training. }

\vspace{.25in}

\noindent\textbf{Acknowledgements:} The author thanks R. Baker, J. Gevertz, and S. Nardini for helpful discussion and commentary.

\noindent\textbf{Data Availability statement:} All code and simulated data for this work is publicly available at \\
\href{https://github.com/johnnardini/Forecasting_predicting_ABMs}{https://github.com/johnnardini/Forecasting\_predicting\_ABMs}.

\noindent\textbf{Conflict of interest:} The author declares no conflict of interest

\noindent\textbf{Funding Statement:} The author acknowledges use of the ELSA high performance computing cluster at The College of New Jersey for conducting the research reported in this paper. This cluster is funded in part by the National Science Foundation under grant numbers OAC-1826915 and OAC-2320244.

\bibliographystyle{unsrt}
\bibliography{references.bib}  

\begin{thebibliography}{10}

\bibitem{anguige_one-dimensional_2009}
Kathleen Anguige and Christian Schmeiser.
\newblock A one-dimensional model of cell diffusion and aggregation,
  incorporating volume filling and cell-to-cell adhesion.
\newblock {\em Journal of Mathematical Biology}, 58(3):395, March 2009.

\bibitem{brauer_mathematical_2019}
Fred Brauer, Carlos Castillo-Chavez, and Zhilan Feng.
\newblock {\em Mathematical models in epidemiology}, volume~69 of {\em Texts in
  {Applied} {Mathematics}}.
\newblock Springer, New York, NY, 2019.

\bibitem{gibbs_coexistence_2022}
Theo Gibbs, Simon~A. Levin, and Jonathan~M. Levine.
\newblock Coexistence in diverse communities with higher-order interactions.
\newblock {\em Proceedings of the National Academy of Sciences},
  119(43):e2205063119, October 2022.
\newblock Publisher: Proceedings of the National Academy of Sciences.

\bibitem{huppert_mathematical_2013}
A.~Huppert and G.~Katriel.
\newblock Mathematical modelling and prediction in infectious disease
  epidemiology.
\newblock {\em Clinical Microbiology and Infection}, 19(11):999--1005, November
  2013.

\bibitem{nardini_modeling_2016}
John~T. Nardini, Douglas~A. Chapnick, Xuedong Liu, and David~M. Bortz.
\newblock Modeling keratinocyte wound healing: cell-cell adhesions promote
  sustained migration.
\newblock {\em Journal of Theoretical Biology}, 400:103--117, July 2016.

\bibitem{raja_noureen_swapping_2023}
Shahzeb Raja~Noureen, Jennifer~P. Owen, Richard~L. Mort, and Christian~A.
  Yates.
\newblock Swapping in lattice-based cell migration models.
\newblock {\em Physical Review E}, 107(4):044402, April 2023.

\bibitem{xiao_modeling_2001}
Yanni Xiao and Lansun Chen.
\newblock Modeling and analysis of a predator–prey model with disease in the
  prey.
\newblock {\em Mathematical Biosciences}, 171(1):59--82, May 2001.

\bibitem{baker_correcting_2010}
Ruth~E. Baker and Matthew~J. Simpson.
\newblock Correcting mean-field approximations for birth-death-movement
  processes.
\newblock {\em Physical Review E}, 82(4):041905, October 2010.

\bibitem{grimm_pattern-oriented_2005}
Volker Grimm, Eloy Revilla, Uta Berger, Florian Jeltsch, Wolf~M. Mooij,
  Steven~F. Railsback, Hans-Hermann Thulke, Jacob Weiner, Thorsten Wiegand, and
  Donald~L. DeAngelis.
\newblock Pattern-oriented modeling of agent-based complex systems: lessons
  from ecology.
\newblock {\em Science}, 310(5750):987--991, November 2005.
\newblock Publisher: American Association for the Advancement of Science.

\bibitem{marshall_formalizing_2015}
Brandon D.~L. Marshall and Sandro Galea.
\newblock Formalizing the role of agent-based modeling in causal inference and
  epidemiology.
\newblock {\em American Journal of Epidemiology}, 181(2):92--99, January 2015.

\bibitem{tracy_agent-based_2018}
Melissa Tracy, Magdalena Cerdá, and Katherine~M. Keyes.
\newblock Agent-based modeling in public health: current applications and
  future directions.
\newblock {\em Annual Review of Public Health}, 39(1):77--94, 2018.
\newblock \_eprint: https://doi.org/10.1146/annurev-publhealth-040617-014317.

\bibitem{chappelle_pulling_2019}
George Chappelle and Christian~A. Yates.
\newblock Pulling in models of cell migration.
\newblock {\em Physical Review E}, 99(6):062413, June 2019.

\bibitem{simpson_reliable_2022}
Matthew~J. Simpson, Ruth~E. Baker, Pascal~R. Buenzli, Ruanui Nicholson, and
  Oliver~J. Maclaren.
\newblock Reliable and efficient parameter estimation using approximate
  continuum limit descriptions of stochastic models.
\newblock {\em Journal of Theoretical Biology}, 549:111201, September 2022.

\bibitem{nardini_learning_2021}
John~T. Nardini, Ruth~E. Baker, Matthew~J. Simpson, and Kevin~B. Flores.
\newblock Learning differential equation models from stochastic agent-based
  model simulations.
\newblock {\em Journal of The Royal Society Interface}, 18(176):rsif.2020.0987,
  20200987, March 2021.

\bibitem{kieu_dealing_2020}
Le-Minh Kieu, Nicolas Malleson, and Alison Heppenstall.
\newblock Dealing with uncertainty in agent-based models for short-term
  predictions.
\newblock {\em Royal Society Open Science}, 7(1):191074, January 2020.

\bibitem{larie_use_2021}
Dale Larie, Gary An, and R.~Chase Cockrell.
\newblock The use of artificial neural networks to forecast the behavior of
  agent-based models of pathophysiology: an example utilizing an agent-based
  model of sepsis.
\newblock {\em Frontiers in Physiology}, 12:716434, October 2021.

\bibitem{thompson_modelling_2012}
Robin~N. Thompson, Christian~A. Yates, and Ruth~E. Baker.
\newblock Modelling cell migration and adhesion during development.
\newblock {\em Bulletin of Mathematical Biology}, 74(12):2793--2809, December
  2012.

\bibitem{vandenheuvel_pushing_2024}
Daniel~J. VandenHeuvel, Pascal~R. Buenzli, and Matthew~J. Simpson.
\newblock Pushing coarse-grained models beyond the continuum limit using
  equation learning.
\newblock {\em Proceedings of the Royal Society A: Mathematical, Physical and
  Engineering Sciences}, 480(2281):20230619, January 2024.

\bibitem{brunton_discovering_2016}
Steven~L. Brunton, Joshua~L. Proctor, and J.~Nathan Kutz.
\newblock Discovering governing equations from data by sparse identification of
  nonlinear dynamical systems.
\newblock {\em Proceedings of the National Academy of Sciences},
  113(15):3932--3937, April 2016.

\bibitem{kaiser_sparse_2018}
E.~Kaiser, J.~Nathan Kutz, and Steven~L. Brunton.
\newblock Sparse identification of nonlinear dynamics for model predictive
  control in the low-data limit.
\newblock {\em Proceedings of the Royal Society A: Mathematical, Physical and
  Engineering Sciences}, 474(2219):20180335, November 2018.

\bibitem{rudy_data-driven_2019}
Samuel Rudy, Alessandro Alla, Steven~L. Brunton, and J.~Nathan Kutz.
\newblock Data-driven identification of parametric partial differential
  equations.
\newblock {\em SIAM Journal on Applied Dynamical Systems}, 18(2):643--660,
  January 2019.
\newblock Publisher: Society for Industrial and Applied Mathematics.

\bibitem{champion_data-driven_2019}
Kathleen Champion, Bethany Lusch, J.~Nathan Kutz, and Steven~L. Brunton.
\newblock Data-driven discovery of coordinates and governing equations.
\newblock {\em Proceedings of the National Academy of Sciences},
  116(45):22445--22451, November 2019.
\newblock Publisher: National Academy of Sciences Section: Physical Sciences.

\bibitem{mangan_inferring_2016}
Niall~M. Mangan, Steven~L. Brunton, Joshua~L. Proctor, and J.~Nathan Kutz.
\newblock Inferring biological networks by sparse identification of nonlinear
  dynamics.
\newblock {\em IEEE Transactions on Molecular, Biological and Multi-Scale
  Communications}, 2(1):52--63, June 2016.
\newblock Conference Name: IEEE Transactions on Molecular, Biological and
  Multi-Scale Communications.

\bibitem{mangan_model_2017}
Niall~M. Mangan, J.~Nathan Kutz, Steven~L. Brunton, and Joshua~L. Proctor.
\newblock Model selection for dynamical systems via sparse regression and
  information criteria.
\newblock {\em Proceedings of the Royal Society A: Mathematical, Physical and
  Engineering Sciences}, 473(2204):20170009, August 2017.

\bibitem{messenger_weak_2021-1}
Daniel~A. Messenger and David~M. Bortz.
\newblock Weak {SINDy}: galerkin-based data-driven model selection.
\newblock {\em Multiscale Modeling \& Simulation}, 19(3):1474--1497, January
  2021.

\bibitem{messenger_weak_2021}
Daniel~A. Messenger and David~M. Bortz.
\newblock Weak {SINDy} for partial differential equations.
\newblock {\em Journal of Computational Physics}, 443:110525, October 2021.

\bibitem{lagergren_learning_2020}
John~H. Lagergren, John~T. Nardini, G.~Michael~Lavigne, Erica~M. Rutter, and
  Kevin~B. Flores.
\newblock Learning partial differential equations for biological transport
  models from noisy spatio-temporal data.
\newblock {\em Proceedings of the Royal Society A: Mathematical, Physical and
  Engineering Sciences}, 476(2234):20190800, February 2020.

\bibitem{lagergren_biologically-informed_2020}
John~H. Lagergren, John~T. Nardini, Ruth~E. Baker, Matthew~J. Simpson, and
  Kevin~B. Flores.
\newblock Biologically-informed neural networks guide mechanistic modeling from
  sparse experimental data.
\newblock {\em PLOS Computational Biology}, 16(12):e1008462, December 2020.
\newblock Publisher: Public Library of Science.

\bibitem{nardini_learning_2020}
John~T. Nardini, John~H. Lagergren, Andrea Hawkins-Daarud, Lee Curtin, Bethan
  Morris, Erica~M. Rutter, Kristin~R. Swanson, and Kevin~B. Flores.
\newblock Learning equations from biological data with limited time samples.
\newblock {\em Bulletin of Mathematical Biology}, 82(9):119, September 2020.

\bibitem{rudy_data-driven_2017}
Samuel~H. Rudy, Steven~L. Brunton, Joshua~L. Proctor, and J.~Nathan Kutz.
\newblock Data-driven discovery of partial differential equations.
\newblock {\em Science Advances}, 3(4):e1602614, April 2017.
\newblock Publisher: American Association for the Advancement of Science
  Section: Research Article.

\bibitem{messenger_learning_2022-1}
Daniel~A. Messenger, Graycen~E. Wheeler, Xuedong Liu, and David~M. Bortz.
\newblock Learning anisotropic interaction rules from individual trajectories
  in a heterogeneous cellular population.
\newblock {\em Journal of The Royal Society Interface}, 19(195):20220412,
  October 2022.

\bibitem{messenger_learning_2022}
Daniel~A. Messenger and David~M. Bortz.
\newblock Learning mean-field equations from particle data using {WSINDy}.
\newblock {\em Physica D: Nonlinear Phenomena}, 439:133406, November 2022.

\bibitem{supekar_learning_2023}
Rohit Supekar, Boya Song, Alasdair Hastewell, Gary P.~T. Choi, Alexander
  Mietke, and Jörn Dunkel.
\newblock Learning hydrodynamic equations for active matter from particle
  simulations and experiments.
\newblock {\em Proceedings of the National Academy of Sciences},
  120(7):e2206994120, February 2023.

\bibitem{cai_physics-informed_2021}
Shengze Cai, Zhicheng Wang, Sifan Wang, Paris Perdikaris, and George~Em
  Karniadakis.
\newblock Physics-informed neural networks for heat transfer {Problems}.
\newblock {\em Journal of Heat Transfer}, 143(6):060801, June 2021.

\bibitem{kaplarevic-malisic_identifying_2023}
Ana Kaplarevi{\'c}-Malisi{\'c}, Branka Andrijevi{\'c}, Filip Bojovi{\'c}, Srdan
  Nikoli{\'c}, Lazar Krsti{\'c}, Boban Stojanovi{\'c}, and Milo{\v s}
  Ivanovi{\'c}.
\newblock Identifying optimal architectures of physics-informed neural networks
  by evolutionary strategy.
\newblock {\em Applied Soft Computing}, page 110646, July 2023.

\bibitem{linka_bayesian_2022}
Kevin Linka, Amelie Sch{\"a}fer, Xuhui Meng, Zongren Zou, George~Em
  Karniadakis, and Ellen Kuhl.
\newblock Bayesian physics informed neural networks for real-world nonlinear
  dynamical systems.
\newblock {\em Computer Methods in Applied Mechanics and Engineering},
  402:115346, December 2022.

\bibitem{raissi_physics-informed_2019}
M.~Raissi, P.~Perdikaris, and G.E. Karniadakis.
\newblock Physics-informed neural networks: a deep learning framework for
  solving forward and inverse problems involving nonlinear partial differential
  equations.
\newblock {\em Journal of Computational Physics}, 378:686--707, February 2019.

\bibitem{shin_convergence_2020}
Yeonjong Shin, Jerome Darbon, and George~Em Karniadakis.
\newblock On the convergence of physics informed neural networks for linear
  second-order elliptic and parabolic type {PDEs}.
\newblock {\em Communications in Computational Physics}, 28(5):2042--2074, June
  2020.
\newblock arXiv: 2004.01806.

\bibitem{johnston_mean-field_2012}
Stuart~T. Johnston, Matthew~J. Simpson, and Ruth~E. Baker.
\newblock Mean-field descriptions of collective migration with strong adhesion.
\newblock {\em Physical Review E}, 85(5):051922, May 2012.

\bibitem{decaestecker_can_2007}
Christine Decaestecker, Olivier Debeir, Philippe Van~Ham, and Robert Kiss.
\newblock Can anti‐migratory drugs be screened in vitro? {A} review of {2D}
  and {3D} assays for the quantitative analysis of cell migration.
\newblock {\em Medicinal Research Reviews}, 27(2):149--176, March 2007.

\bibitem{das_ring_2015}
Asha~M. Das, Alexander M.~M. Eggermont, and Timo L.~M. ten Hagen.
\newblock A ring barrier–based migration assay to assess cell migration in
  vitro.
\newblock {\em Nature Protocols}, 10(6):904--915, June 2015.
\newblock Number: 6 Publisher: Nature Publishing Group.

\bibitem{kashef_quantitative_2015}
Jubin Kashef and Clemens~M. Franz.
\newblock Quantitative methods for analyzing cell–cell adhesion in
  development.
\newblock {\em Developmental Biology}, 401(1):165--174, May 2015.

\bibitem{venhuizen_making_2017}
Jan-Hendrik Venhuizen and Mirjam~M. Zegers.
\newblock Making {heads} or {tails} of {it}: {cell}–{cell} {adhesion} in
  {cellular} and {supracellular} {polarity} in {collective} {migration}.
\newblock {\em Cold Spring Harbor Perspectives in Biology}, 9(11):a027854,
  November 2017.

\bibitem{vishwakarma_mechanobiology_2020}
Medhavi Vishwakarma, Joachim~P. Spatz, and Tamal Das.
\newblock Mechanobiology of leader-follower dynamics in epithelial cell
  migration.
\newblock {\em Current Opinion in Cell Biology}, 66:97--103, October 2020.

\bibitem{janiszewska_cell_2020}
Michalina Janiszewska, Marina~C. Primi, and Tina Izard.
\newblock Cell adhesion in cancer: beyond the migration of single cells.
\newblock {\em Journal of Biological Chemistry}, 295(8):2495--2505, February
  2020.

\bibitem{rothenberg_rap1_2023}
Katheryn~E. Rothenberg, Yujun Chen, Jocelyn~A. McDonald, and Rodrigo
  Fernandez-Gonzalez.
\newblock Rap1 coordinates cell-cell adhesion and cytoskeletal reorganization
  to drive collective cell migration in vivo.
\newblock {\em Current Biology}, 33(13):2587--2601.e5, July 2023.

\bibitem{angione_using_2022}
Claudio Angione, Eric Silverman, and Elisabeth Yaneske.
\newblock Using machine learning as a surrogate model for agent-based
  simulations.
\newblock {\em PLOS ONE}, 17(2):e0263150, February 2022.

\bibitem{nguyen_quantifying_2024}
Kyle~C. Nguyen, Carter~D. Jameson, Scott~A. Baldwin, John~T. Nardini, Ralph~C.
  Smith, Jason~M. Haugh, and Kevin~B. Flores.
\newblock Quantifying collective motion patterns in mesenchymal cell
  populations using topological data analysis and agent-based modeling.
\newblock {\em Mathematical Biosciences}, 370:109158, April 2024.

\bibitem{kurganov_new_2000}
Alexander Kurganov and Eitan Tadmor.
\newblock New high-resolution central schemes for nonlinear conservation laws
  and convection–diffusion equations.
\newblock {\em Journal of Computational Physics}, 160(1):241--282, May 2000.

\bibitem{petzold_automatic_1983}
Linda Petzold.
\newblock Automatic selection of methods for solving stiff and nonstiff systems
  of ordinary differential equations.
\newblock {\em SIAM Journal on Scientific and Statistical Computing},
  4(1):136--148, March 1983.

\end{thebibliography}

\clearpage
\appendix

\tableofcontents

\section{ABM Rules} \label{app:ABM_Rules}

\subsection{The Pulling Model}\label{subsubsec:Pulling_model}

The Pulling model consists of pulling agents that migrate with rate\footnote{Meaning that pulling agents attempt to migrate over an infinitesimal time interval of length $dt$ with probability $\rmp dt$.} $\rmp$ and perform rules A and B from Figure \ref{fig:ABM_rules}. Suppose a pulling agent at lattice site $(i,j)$ chooses to move rightwards into site $(i+1,j)$. If the lattice site $(i-1,j)$ is unoccupied, then the agent performs Rule A and moves into site $(i+1,j)$. If the lattice site $(i-1,j)$ is occupied, then the agent attempts Rule B on agent pulling. This event succeeds with probability $\ppull$, and the agent moves to site $(i+1,j)$ and pulls its neighbor into lattice site $(i,j)$. This event fails with probability $1-\ppull$, in which the agent moves into site $(i+1,j)$ but the neighbor remains at lattice site $(i-1,j)$. These rules can be described by the following trimolecular reaction rates:

\begin{align}
    0_{i-1,j} + P_{i,j} + 0_{i+1,j}& \xrightarrow{\rmp / 4} &0_{i-1,j} + 0_{i,j} + P_{i+1,j}, \tag{Rule\ A} \label{eq:ruleA}\\
    P_{i-1,j} + P_{i,j} + 0_{i+1,j}& \xrightarrow{ \ppull\rmp / 4  } &0_{i-1,j} + P_{i,j} + P_{i+1,j}, \tag{Rule\ B.1}\label{eq:ruleB1}\\ 
    P_{i-1,j} + P_{i,j} + 0_{i+1, j}& \xrightarrow{ (1 -\ppull)\rmp / 4  } &P_{i-1,j} + 0_{i,j} + P_{i+1,j}. \tag{Rule\ B.2}\label{eq:ruleB2}
\end{align}
Equivalent reactions govern agent migration and pulling in the other three directions.

\subsection{The Adhesion Model}\label{subsubsec:Adhesion_model}

The Adhesion model consists of adhesive agents that migrate with rate $\rmh$ and perform rules C and D from Figure \ref{fig:ABM_rules}. Suppose an adhesive agent at lattice site $(i,j)$ chooses to move rightwards into site $(i+1,j)$. If the lattice site $(i-1,j)$ is unoccupied, then the agent performs Rule C and moves into site $(i+1,j)$. If the lattice site $(i-1,j)$ is occupied, then the neighboring agent attempts Rule D to adhere to the migrating agent and abort their movement. This event succeeds with probability $\padh$, and neither agent changes its location. This adhesion event fails with probability $1-\padh$, and the migrating agent moves to site $(i+1,j)$ and the neighbor remains at lattice site $(i-1,j)$. These rules can be described by the following trimolecular reaction rates:
\begin{align}
    0_{i-1,j} + H_{i,j} + 0_{i+1,j}& \xrightarrow{\rmh / 4} &0_{i-1,j} + 0_{i,j} + H_{i+1,j}, \tag{Rule\ C} \label{eq:ruleC}\\
    H_{i-1,j} + H_{i,j} + 0_{i+1, j}& \xrightarrow{ (1 -\padh)\rmh / 4  } &H_{i-1,j} + 0_{i,j} + H_{i+1,j}. \tag{Rule\ D} \label{eq:ruleD}
\end{align}

\subsection{The Pulling \& Adhesion Model}\label{subsubsec:PullingAdhesion_model}

The Pulling \& Adhesion model consists of both pulling and adhesive agents. This model implements Rules A-F from Figure \ref{fig:ABM_rules}. Rules A-D are unchanged from their descriptions in Sections \ref{subsubsec:Pulling_model} and \ref{subsubsec:Adhesion_model}. 
If a pulling agent at lattice site $(i,j)$ chooses to move rightwards into site $(i+1,j)$ while an adhesive agent occupies site $(i-1,j)$, then Rule E dictates the agents' attempts to pull and adhere to each other. The migrating pulling agent succeeds with probability $\ppull$ and moves to site $(i+1,j)$ while pulling the neighboring adhesive agent into site $(i,j)$; the neighboring adhesive agent successfully aborts the pulling agent's migration event with probability $\padh$; both agents fail with probability $1-\padh-\ppull$ and the pulling agent moves to site $(i+1,j)$ while the adhesive agent remains at site $(i-1,j)$. Based on our definition of this rule, it is not possible that both the pulling and adhesion events succeed, so the parameters must satisfy $0\le \ppull +\padh \le 1$. Rule E can be described by the following trimolecular reaction rate:
\begin{align}
    H_{i-1,j} + P_{i,j} + 0_{i+1,j}& \xrightarrow{\ppull \rmp / 4 } &0_{i-1,j} + H_{i,j} + P_{i+1,j}, \tag{Rule\ E.1}\label{eq:rulee1} \\
    H_{i-1,j} + P_{i,j} + 0_{i+1,j}& \xrightarrow{(1 - \padh - \ppull)\rmp / 4} &H_{i-1,j} + 0_{i,j} + P_{i+1,j}. \tag{Rule\ E.2}\label{eq:rulee2}
\end{align}

If an adhesive agent at lattice site $(i,j)$ chooses to move rightwards into site $(i+1,j)$ while a pulling agent occupies site $(i-1,j)$, then Rule F dictates that the adhesive agent moves into site $(i+1,j)$ and the pulling agent remains at site $(i-1,j)$. Rule F can be described by the following trimolecular reaction rate:
\begin{align}
    P_{i-1,j} + H_{i,j} + 0_{i+1,j}& \xrightarrow{\rmh / 4 } &P_{i-1,j} + 0_{i,j} + H_{i+1,j}. \tag{Rule\ F}\label{eq:rulef}
\end{align}

\clearpage

\section{ABM implementation}\label{app:ABM_implementation}

Each model is simulated in the spatial domain $(x,y)\in[0,X]\times[0,Y]$. We choose $X=200\text{ and } Y=40$ to represent a thin rectangle where collective migration primarily occurs along the $x$-dimension and is not affected by the boundary in this dimension. We represent this space with a two-dimensional lattice with square lattice sites with length $\Delta=1$ to imitate a typical cell length. The $(i,j)^{\text{th}}$ lattice site is centered at $(x_i,y_j)$, where $x_i = (i-0.5)\Delta,\ i = 1,\dots,X$, and $y_j = (j-0.5)\Delta,\ j = 1,\dots,Y.$ Each model is an exclusion process, meaning that each agent can only occupy one lattice site at a time, and each lattice site is occupied by at most one agent.  The parameter $\alpha\in(0,1)$ denotes the proportion of nonempty lattice sites that are occupied by adhesive agents in the simulation, and $(1-\alpha)$ denotes the proportion of nonempty lattice sites that are occupied by pulling agents in the simulation.

All model simulations are initialized by populating 75\% of the lattice sites in the middle 20\% of columns, e.g., 75\% of the lattice sites in $\{(x_i,y_j)\}_{j=1}^Y$ are initially occupied for $i=80,\dots,120.$ All other columns are initially empty. This initial condition is chosen to reflect a barrier assay \cite{das_ring_2015}. Reflecting boundary conditions are used at the boundaries of lattice to enforce a no-flux condition in the spatial domain. We simulate each ABM using the Gillespie algorithm, which we provide for the Pulling \& Adhesion ABM in Supplementary Algorithm \ref{algo:gillespie} in electronic supplementary material \ref{sec:Gillespie}. All ABMs are simulated until $t=1000$.

\clearpage
\section{Gillespie algorithm}\label{sec:Gillespie}

Our implementation of the Gillespie Algorithm for the Pulling \& Adhesion ABM is provided in Supplementary Algorithm \ref{algo:gillespie}.
\begin{algorithm}
\SetAlgoLined
Create an $X\times Y$ lattice with user-specified placement of agents

Set $t=0$

Set maximum simulation time $t_\text{end}$

Set $P(t)$ and $H(t)$ equal to the number of Pulling and Adhesive agents on the lattice, respectively

\While{$t<t_\text{end}$}{

Calculate the following random variables, uniformly distributed on $[0,1]: \gamma_1,\gamma_2$

%Randomly choose an agent and determine its lattice site;
Calculate the propensity function $a(t)=\rmp P(t) + \rmh H(t)$

Calculate time step $\tau=-\ln(\gamma_1)/a(t)$

$t=t+\tau$

$R=a(t)\gamma_2$

\uIf{$R< \rmp P(t)$}{

Perform Pulling agent migration (Supplementary Algorithm \ref{algo:PullingMigration}) 
}
\uElseIf{$R< \rmp P(t) + \rmh H(t)$}{

Perform Adhesive agent migration (Supplementary Algorithm \ref{algo:AdhesiveMigration}) }
}

\caption{Gillespie algorithm for the Pulling \& Adhesion ABM}\label{algo:gillespie}
\end{algorithm}

\newcommand{\vectwo}[2]{	
				\left( 
				\begin{array}{c}
				{#1}   \\
				{#2}
				\end{array}
				\right)}

\begin{algorithm}
    \SetAlgoLined

    Randomly choose a pulling agent and determine its lattice site index, $\Vec{x} = (i,j)^T$

    Choose one of the four cardinal migration directions, $\Vec{dx}=(dx,dy)^T \in \{ (1,0)^T, (-1,0)^T, (0,1)^T, (0,-1)^T\}$, with equal probability, $1/4$. The neighboring direction is given by $\hat{dx} = -\Vec{dx}$

    \uIf{$\Vec{x} + \Vec{dx}$ is empty}{

        \uIf{$\Vec{x} + \hat{dx}$ is empty}{\tcc{Rule A}
            Move the chosen pulling agent to lattice site $\Vec{x} + \Vec{dx}$
        
        }
        \uElseIf{$\Vec{x} + \hat{dx}$ is occupied by a Pulling agent}{\tcc{Rule B}

            Calculate the random variable, $\gamma_{3}$, uniformly distributed on $[0,1]$

            \uIf{$\gamma_3 \le \ppull$}{

                Move the chosen pulling agent to lattice site $\Vec{x} + \Vec{dx}$
                
                Move the neighboring agent to lattice site $\Vec{x}$
            
            }
            \uElseIf{$\gamma_3 > \ppull$}{
            
                Move the chosen pulling agent to lattice site $\Vec{x} + \Vec{dx}$
            
            }
        }
        \uElseIf{$\Vec{x} + \hat{dx}$ is occupied by an Adhesive agent}{\tcc{Rule E}

            Calculate the random variable, $\gamma_3$, uniformly distributed on $[0,1]$

            \uIf{$\gamma_3 \le \ppull$}{

                Move the chosen pulling agent to lattice site $\Vec{x} + \Vec{dx}$
                
                Move the neighboring agent to lattice site $\Vec{x}$
            
            }
            \uElseIf{$\gamma_3 \le \ppull + 1-\padh$ }{
            
                Move the chosen pulling agent to lattice site $\Vec{x} + \Vec{dx}$
            
            }

        }
    
    }

    \label{algo:PullingMigration}
    \caption{Pulling Agent migration}
\end{algorithm}

\begin{algorithm}
    \SetAlgoLined

    Randomly choose an adhesive agent and determine its lattice site index, $\Vec{x} = (i,j)^T$

    Choose one of the four cardinal migration directions, $\Vec{dx}=(dx,dy)^T \in \{ (1,0)^T, (-1,0)^T, (0,1)^T, (0,-1)^T\}$, with equal probability, $1/4$. The neighboring direction is given by $\hat{dx} = -\Vec{dx}$

    \uIf{$\Vec{x} + \Vec{dx}$ is empty}{

        \uIf{$\Vec{x} + \hat{dx}$ is empty}{\tcc{Rule C}
            Move the chosen adhesive agent to lattice site $\Vec{x} + \Vec{dx}$
        
        }
        \uElseIf{$\Vec{x} + \hat{dx}$ is occupied by an adhesive agent}{\tcc{Rule D}
        
            Calculate the random variable, $\gamma_{3}$, uniformly distributed on $[0,1]$

            \uIf{$\gamma_3 \le (1-\padh)$ }{

                Move the chosen adhesive agent to lattice site $\Vec{x} + \Vec{dx}$
            
            }
        
        }
        \uElseIf{$\Vec{x} + \hat{dx}$ is occupied by a Pulling agent}{\tcc{Rule F}
        
            Move the chosen adhesive agent to lattice site $\Vec{x} + \Vec{dx}$
        
        }

    }

    \label{algo:AdhesiveMigration}
    \caption{Adhesive agent migration}
\end{algorithm}

\clearpage
\section{Coarse-graining ABM rules into PDE models}\label{app:coarse_graining_ABM_Rules}

We will coarse-grain the Pulling, Adhesion, and Pulling \& Adhesion ABMs into their mean-field PDE models. Each ABM consists of a combination of Rules A-F from Figure \ref{fig:ABM_rules}. Each rule updates the occupancies of three consecutive lattice sites, such as $\{(i,j-1), (i,j), (i,j+1)\}$. Let the variables $P_{i,j}(t)$, $H_{i,j}(t)$, and $0_{i,j}(t)$ denote the probabilities that lattice site $(i,j)$ is occupied by a pulling agent, adhesive agent, or empty at time $t$, respectively. To convert each rule into a PDE model, we invoke the \emph{mean-field assumption}, which supposes that all lattice site occupancies are independent of each other. This assumption simplifies model coarse-graining by allowing us to replace the joint probability of three lattice site occupancies with the product of the three individual lattice site occupancy probabilities. For example, under the mean-field assumption, we can write the probability that lattice sites $(i,j-1), (i,j), \text{ and } (i,j+1)$ are all occupied by pulling agents at time $t$ as $P_{i,j-1}(t)P_{i,j}(t)P_{i,j+1}(t)$; otherwise, we must consider the joint occupancy probability for this triplet of lattice sites. Mean-field DE models can poorly predict ABM behavior when the mean-field assumption is violated during ABM simulations, see \cite{baker_correcting_2010, simpson_reliable_2022, nardini_learning_2021} for further details.

\subsection{Coarse-graining the Pulling ABM}\label{subsec:Pulling_coarse_graining}

The Pulling ABM is composed of Rules A and B from Figure \ref{fig:ABM_rules} and Section \ref{subsubsec:Pulling_model}. We begin coarse-graining this ABM into a PDE model by writing the master equation governing how $P_{i,j}(t)$ changes according to these rules:
\begin{equation}
\ddt{P_{i,j}(t)} = K^{\ref{eq:ruleA}} + K^{\ref{eq:ruleB1}} + K^{\ref{eq:ruleB2}}. \label{eq:master_pulling}
\end{equation}

\ref{eq:ruleA} specifies how pulling agents migrate into an empty lattice site 
 with rate $\rmp/4$ when there is no neighboring agent in the lattice site opposite the direction of migration. This rate is divided by four because the agent randomly chooses to attempt to migrate into one of its four neighboring lattice sites. We write this rule in the master equation as:
\begin{align}
K^{\ref{eq:ruleA}}  = &-\dfrac{2\rmp}{4}\left[ 0_{i,j-1}(t) P_{i,j}(t) 0_{i,j+1}(t) + 0_{i-1,j}(t) P_{i,j}(t) 0_{i+1,j}(t) \right]  \nonumber \\ 
&+ \dfrac{\rmp}{4} \left[ 0_{i,j-2}(t) P_{i,j-1}(t) 0_{i,j}(t) +  0_{i,j} P_{i,j+1} 0_{i,j+2} + 0_{i-2,j} P_{i-1,j} 0_{i,j} +  0_{i,j} P_{i+1,j} 0_{i+2,j}\right], \label{eq:PullingKA}
\end{align}
where the first line describes how a pulling agent moves out of lattice site $(i,j)$, and the second line describes how a pulling agent moves into lattice site $(i,j)$. 

\ref{eq:ruleB1} specifies how a pulling agent migrates into an empty neighboring lattice site and pulls its neighbor along with it, which occurs with probability $\ppull$. We write this rule in the master equation as:
\begin{align}
K^{\ref{eq:ruleB1}}  = -\dfrac{\ppull\rmp}{4}\bigg[ &P_{i,j}(t) P_{i,j+1}(t) 0_{i,j+2}(t)  + 0_{i,j-2}(t) P_{i,j-1}(t) P_{i,j}(t)  \nonumber + \\
&P_{i,j}(t) P_{i+1,j}(t) 0_{i+2,j}(t)  + 0_{i-2,j}(t) P_{i-1,j}(t)  P_{i,j}(t) \bigg] \nonumber \\
\dfrac{\ppull\rmp}{4}\bigg[ &P_{i,j-2}(t) P_{i,j-1}(t) 0_{i,j}(t) + 0_{i,j}(t) P_{i,j+1}(t) P_{i,j+2}(t) + \nonumber \\ 
&P_{i-2,j}(t) P_{i-1,j}(t) 0_{i,j}(t) + 0_{i,j}(t) P_{i+1,j}(t) P_{i+2,j}(t) \bigg]. \label{eq:PullingKB1}
\end{align}
\ref{eq:ruleB2} specifies how a pulling agent migrates into an empty neighboring lattice site and fails to pull its neighbor along with it, which occurs with  probability $1-\ppull$. We write this rule in the master equation as:
\begin{align}
K^{\ref{eq:ruleB2}}  = -\dfrac{(1-\ppull)\rmp}{4}\bigg[ &P_{i,j-1}(t) P_{i,j}(t) 0_{i,j+1}(t) + 0_{i,j-1}(t) P_{i,j}(t) P_{i,j+1}(t)  + \nonumber \\
&  P_{i-1,j}(t) P_{i,j}(t) 0_{i+1,j+1}(t) + 0_{i,j-1}(t) P_{i,j}(t) P_{i+1,j}(t) \bigg] \nonumber \\ 
+ \dfrac{(1-\ppull)\rmp}{4}\bigg[ &P_{i,j-2}(t) P_{i,j-1}(t) 0_{i,j}(t) + 0_{i,j}(t) P_{i,j+1}(t) P_{i,j+2}(t) + \nonumber \\ 
&P_{i-2,j}(t) P_{i-1,j}(t) 0_{i,j}(t) + 0_{i,j}(t) P_{i+1,j}(t) P_{i+2,j}(t) \bigg]. \label{eq:PullingKB2}
\end{align}
%Here, the first two lines describe how a pulling agent either is pulled out of lattice site $(i,j)$ when one of its neighboring agents moves and the last two lines describe how a pulling agent moves into lattice site $(i,j)$ and successfully pulls its neighbor during this process.

To obtain the resulting PDE model for the Pulling ABM, we substitute Equations \eqref{eq:PullingKA}, \eqref{eq:PullingKB1}, and \eqref{eq:PullingKB2} into Equation \eqref{eq:master_pulling} and set $0_{i,j}=1-P_{i,j}.$ We replace each term with its Taylor expansion, up to second order:
\begin{align}
P_{i\pm m,j}(t) &= P_{i,j}(t) \pm m\Delta (P_{i,j}(t))_x + \dfrac{m\Delta^2}{2}(P_{i,j}(t))_{xx} + \mathcal{O}(\Delta^3), &m = -2, -1, 0, 1, 2; \nonumber \\ 
P_{i,j\pm n}(t) &= P_{i,j}(t) \pm n\Delta (P_{i,j}(t))_y + \dfrac{n\Delta^2}{2}(P_{i,j}(t))_{yy} + \mathcal{O}(\Delta^3), &n = -2, -1, 0, 1, 2; \label{eq:pulling_taylor}
\end{align}
where subscripts denote differentiation with respect the the shown variable, and $\Delta$ is the length of each lattice site. As shown in the Mathematica notebook \textbf{Pulling\_model\_coarse\_graining.nb}, taking the limit of the resulting expression as $\Delta\rightarrow0$ leads to the mean-field PDE model for the Pulling ABM:
\begin{equation}
    \dfrac{\partial P}{\partial t} =  \nabla \cdot \left( \dfrac{\rmp}{4} \left( 1 + 3\ppull P^2 \right) \nabla P\right),
\end{equation}
where $P = P_{i,j}(t)$.

\subsection{Coarse-graining the Adhesion ABM}\label{subsec:Adhesion_coarse_graining}

The Adhesion ABM is composed of Rules C and D from Figure \ref{fig:ABM_rules} and Section \ref{subsubsec:Adhesion_model}. We begin coarse-graining this ABM into a PDE model by writing the master equation governing how $H_{i,j}(t)$ changes according to these rules:
\begin{equation}
\ddt{H_{i,j}(t)} = K^{\ref{eq:ruleC}} + K^{\ref{eq:ruleD}}. \label{eq:master_adhesion}
\end{equation}

\ref{eq:ruleC} specifies how adhesive agents migrate into an empty lattice site 
 with rate $\rmh/4$ when there is no neighboring agent in the lattice site opposite the direction of migration. We write this rule in the master equation as:
\begin{align}
K^{\ref{eq:ruleC}}  = -\dfrac{2\rmh}{4}\bigg[ &0_{i,j-1}(t) H_{i,j}(t) 0_{i,j+1}(t) + 0_{i-1,j}(t) H_{i,j}(t) 0_{i+1,j}(t) \bigg]  \nonumber \\ 
+ \dfrac{\rmh}{4} \bigg[ &0_{i,j-2}(t) H_{i,j-1}(t) 0_{i,j}(t) +  0_{i,j}(t) H_{i,j+1}(t) 0_{i,j+2}(t) + \nonumber \\
&0_{i-2,j}(t) H_{i-1,j}(t) 0_{i,j}(t) +  0_{i,j}(t) H_{i+1,j}(t) 0_{i+2,j}(t)\bigg], \label{eq:AdhesionKC}
\end{align}
where the first line describes how an adhesive agent moves out of lattice site $(i,j)$, and the second and third lines describe how an adhesive agent moves into lattice site $(i,j)$. 

\ref{eq:ruleD} specifies how adhesive agents migrate into an empty neighboring lattice site when a neighboring adhesive agent is in the lattice site opposite the direction of migration. The neighboring adhesive agent attempts to adhere to the migrating agent and abort the migration event. The adhesion event succeeds with probability $\padh$, and neither agent changes its position. The adhesion event fails with probability $1-\padh$, and the migrating agent shifts into the previously-empty lattice site while the neighboring agent remains in its previous lattice site. We write this rule in the master equation as:
\begin{align}
K^{\ref{eq:ruleD}}  = -\dfrac{(1-\padh)\rmh}{4}\bigg[ &H_{i,j-1}(t) H_{i,j}(t) 0_{i,j+1}(t) + 0_{i,j-1}(t) H_{i,j}(t) H_{i,j+1}(t) + \nonumber \\
&H_{i-1,j}(t) H_{i,j}(t) 0_{i+1,j}(t) + 0_{i-1,j}(t) H_{i,j}(t) H_{i+1,j}(t) \bigg]  \nonumber \\ 
+ \dfrac{(1-\padh)\rmh}{4} \bigg[ &H_{i,j-2}(t) H_{i,j-1}(t) 0_{i,j}(t) +  0_{i,j}(t) H_{i,j+1}(t) H_{i,j+2}(t) + \nonumber \\
&H_{i-2,j}(t) H_{i-1,j}(t) 0_{i,j}(t) +  0_{i,j}(t) H_{i+1,j}(t) H_{i+2,j}(t)\bigg]. \label{eq:AdhesionKD}
\end{align}
%where the first and second lines describes how an adhesive agent moves out of lattice site $(i,j)$, and the third and fourth lines describe how an adhesive agent moves into lattice site $(i,j)$. 

To obtain the resulting PDE model for the Adhesion ABM, we substitute Equations \eqref{eq:AdhesionKC} and \eqref{eq:AdhesionKD} into Equation \eqref{eq:master_adhesion} and set $0_{i,j}=1-H_{i,j}$. We replace each term with its Taylor expansion, up to second order:
\begin{align}
H_{i\pm m,j}(t) &= H_{i,j}(t) \pm m\Delta (H_{i,j}(t))_x + \dfrac{m\Delta^2}{2}(H_{i,j}(t))_{xx} + \mathcal{O}(\Delta^3), &m = -2, -1, 0, 1, 2; \nonumber \\ 
H_{i,j\pm n}(t) &= H_{i,j}(t) \pm n\Delta (H_{i,j}(t))_y + \dfrac{n\Delta^2}{2}(H_{i,j}(t))_{yy} + \mathcal{O}(\Delta^3), &n = -2, -1, 0, 1, 2. \label{eq:adhesion_taylor}
\end{align}
As shown in the Mathematica notebook \textbf{Adhesion\_model\_coarse\_graining.nb}, taking the limit of the resulting expression as $\Delta\rightarrow0$ leads to the mean-field PDE model for the Adhesion ABM:
\begin{equation}
    \dfrac{\partial H}{\partial t} =  \nabla \cdot \left( \dfrac{\rmh}{4} \left( 3\padh \left( H-\dfrac{2}{3}\right)^2 + 1 - \dfrac{4\padh}{3}\right) \nabla H\right)
\end{equation}
where $H = H_{i,j}(t)$.

\subsection{Coarse-graining the Pulling \& Adhesion ABM}\label{subsec:PullingAdhesion_coarse_graining}

The Pulling \& Adhesion ABM is composed of Rules A to F from Figure \ref{fig:ABM_rules} and Sections \ref{subsubsec:Pulling_model}-\ref{subsubsec:PullingAdhesion_model}. We begin coarse-graining this ABM into a PDE model by writing the master system of equations governing how both $P_{i,j}(t)$ and $H_{i,j}(t)$ change according to these rules:
\begin{align}
\ddt{P_{i,j}(t)} &= K^{\ref{eq:ruleA}} + K^{\ref{eq:ruleB1}} + K^{\ref{eq:ruleB2}} + K^{\ref{eq:rulee1}}_P + K^{\ref{eq:rulee2}} \\
\ddt{H_{i,j}(t)} &= K^{\ref{eq:ruleC}} + K^{\ref{eq:ruleD}} + K^{\ref{eq:rulee1}}_H + K^{\ref{eq:rulef}},\label{eq:master_pullingadhesion}
\end{align}
where $K^{\ref{eq:rulee1}}_P$ denotes how $P_{i,j}(t)$ is affected by \ref{eq:rulee1} and $K^{\ref{eq:rulee1}}_H$ denotes how $H_{i,j}(t)$ is affected by \ref{eq:rulee1}. All other rules affect either $P_{i,j}(t)$ or $H_{i,j}(t)$, but not both. Rules A-D are described in Sections \ref{subsec:Pulling_coarse_graining} and \ref{subsec:Adhesion_coarse_graining}, and we do not restate them here. 

Rule E specifies how a pulling agent migrates into an empty neighboring lattice site when a neighboring adhesive agent is present in the lattice site opposite the direction of migration. In \ref{eq:rulee1}, the pulling agent successfully pulls the adhesive agent as it migrates, which occurs with probability $\ppull$. In this scenario, the pulling agent shifts into the previously-empty lattice site and the adhesive agent moves into the site previously occupied by the pulling agent. We write this rule in the master equation for $P_{i,j}(t)$ as:
\begin{align}
K^{\ref{eq:rulee1}}_P  = -\dfrac{\ppull\rmp}{4}\bigg[ &H_{i,j-1}(t) P_{i,j}(t) 0_{i,j+1}(t) + 0_{i,j-1}(t) P_{i,j}(t) H_{i,j+1}(t) + \nonumber \\
&H_{i-1,j}(t) P_{i,j}(t) 0_{i+1,j}(t) + 0_{i-1,j}(t) P_{i,j}(t) H_{i+1,j}(t) \bigg]  \nonumber \\ 
+ \dfrac{\ppull\rmp}{4} \bigg[ &H_{i,j-2}(t) P_{i,j-1}(t) 0_{i,j}(t) +  0_{i,j}(t) P_{i,j+1}(t) H_{i,j+2}(t) + \nonumber \\
&H_{i-2,j}(t) P_{i-1,j}(t) 0_{i,j}(t) +  0_{i,j}(t) P_{i+1,j}(t) H_{i+2,j}(t)\bigg], \label{eq:PullingKE1P}
\end{align}
and in the master equation for $H_{i,j}(t)$ as:
\begin{align}
K^{\ref{eq:rulee1}}_H  = -\dfrac{\ppull\rmp}{4}\bigg[ &0_{i,j-2}(t) P_{i,j-1}(t) H_{i,j}(t) + H_{i,j}(t) P_{i,j+1}(t) 0_{i,j+2}(t) + \nonumber \\
 &0_{i-2,j}(t) P_{i-1,j}(t) H_{i,j}(t) + H_{i,j}(t) P_{i+1,j}(t) 0_{i+2,j}(t)  \bigg]  \nonumber \\ 
+ \dfrac{\ppull\rmp}{4} \bigg[ &H_{i,j-1}(t) P_{i,j}(t) 0_{i,j+1}(t) + 0_{i,j-1}(t) P_{i,j}(t) H_{i,j+1}(t) + \nonumber \\
&H_{i-1,j}(t) P_{i,j}(t) 0_{i+1,j}(t) + 0_{i-1,j}(t) P_{i,j}(t) H_{i+1,j}(t)\bigg]. \label{eq:PullingKE1H}
\end{align}
The neighboring adhesive agent successfully adheres to the migrating pulling agent and aborts its migration event with probability $\padh$. Neither $P_{i,j}(t)$ or $H_{i,j}(t)$ changes in this scenario as no agents change their locations in response to the adhesion event. In \ref{eq:rulee2}, the adhesive agent fails to adhere to the pulling agent and the pulling agent fails to pull the adhesive agent, which occurs with probability $1-\padh-\ppull$. In this scenario, the pulling agent shifts into the previously-empty lattice site while the neighboring adhesive agent remains in its previous lattice site. We write this rule in the master equation as:
\begin{align}
K^{\ref{eq:rulee2}}  = -\dfrac{(1-\padh-\ppull)\rmp}{4}\bigg[ &H_{i,j-1}(t) P_{i,j}(t) 0_{i,j+1}(t) + 0_{i,j-1}(t) P_{i,j}(t) H_{i,j+1}(t) + \nonumber \\
&H_{i-1,j}(t) P_{i,j}(t) 0_{i+1,j}(t) + + 0_{i-1,j}(t) P_{i,j}(t) H_{i+1,j}(t) \bigg]  \nonumber \\ 
+ \dfrac{(1-\padh-\ppull)\rmp}{4} \bigg[ &H_{i,j-2}(t) P_{i,j-1}(t) 0_{i,j}(t) +  0_{i,j}(t) P_{i,j+1}(t) H_{i,j+2}(t) + \nonumber \\
&H_{i-2,j}(t) P_{i-1,j}(t) 0_{i,j}(t) +  0_{i,j}(t) P_{i+1,j}(t) H_{i+2,j}(t)\bigg]. \label{eq:PullingKE2}
\end{align}

Rule F specifies how adhesive agents migrate into an empty neighboring lattice site when a neighboring pulling agent is in the lattice site opposite the direction of migration. The two agents do not interact with each other in this scenario. As such, the adhesive agent migrates into the empty lattice site with rate $\rmh/4$. We write this rule in the master equation as:
\begin{align}
K^{\ref{eq:rulef}}  = -\dfrac{\rmh}{4}\bigg[ &P_{i,j-1}(t) H_{i,j}(t) 0_{i,j+1}(t) + 0_{i,j-1}(t) H_{i,j}(t) P_{i,j+1}(t) + \nonumber \\
&P_{i-1,j}(t) H_{i,j}(t) 0_{i+1,j}(t) + 0_{i-1,j}(t) H_{i,j}(t) P_{i+1,j}(t) \bigg]  \nonumber \\ 
+ \dfrac{\rmh}{4} \bigg[ &P_{i,j-2}(t) H_{i,j-1}(t) 0_{i,j}(t) +  0_{i,j}(t) H_{i,j+1}(t) P_{i,j+2}(t) + \nonumber \\
&P_{i-2,j}(t) H_{i-1,j}(t) 0_{i,j}(t) +  0_{i,j}(t) H_{i+1,j}(t) P_{i+2,j}(t)\bigg]. \label{eq:AdhesionKF}
\end{align}

To obtain the resulting system of differential equations for the Pulling \& Adhesion ABM, we substitute Equations \eqref{eq:PullingKA}, \eqref{eq:PullingKB1}, \eqref{eq:PullingKB2}, \eqref{eq:AdhesionKC}, \eqref{eq:AdhesionKD}, \eqref{eq:PullingKE1P}, \eqref{eq:PullingKE1H}, \eqref{eq:PullingKE2},  and \eqref{eq:AdhesionKF} into Equation \eqref{eq:master_pullingadhesion} and set $0_{i,j}=1-T_{i,j}$, where $T_{i,j}= P_{i,j} + H_{i,j}$. We replace each term with its Taylor expansion, up to second order, from Equations \eqref{eq:pulling_taylor} and \eqref{eq:adhesion_taylor}. As shown in the Mathematica notebook \textbf{Pulling-Adhesion\_coarse\_graining.nb}, taking the limit of the resulting expression as $\Delta\rightarrow0$ leads to the mean-field system of PDEs for the Pulling \& Adhesion ABM:
\begin{align}
    \dfrac{\partial P}{\partial t} = &\dfrac{\rmp}{4} \nabla \cdot \bigg( (1-T)\nabla P + P\nabla T  \bigg) \nonumber \\
    &+ \padh\dfrac{\rmp}{4} \nabla \cdot \bigg( -3 P(1-T) \nabla H - H (1-T)\nabla P - HP \nabla T  \bigg) \nonumber \\
     &+ \ppull\dfrac{\rmp}{4} \nabla \cdot \bigg( 3P^2 \nabla T  \bigg)  \nonumber \\
    \dfrac{\partial H}{\partial t} =  &\dfrac{\rmh}{4} \nabla \cdot \bigg( (1-T)\nabla H + H\nabla T  \bigg) \nonumber \\
    &+ \padh\dfrac{\rmh}{4} \nabla \cdot \bigg( -4 (1-T)H\nabla H - H^2 \nabla T  \bigg) \nonumber \\
    &+ \ppull\dfrac{\rmp}{4} \nabla \cdot \bigg( -(1-T)H\nabla P + (1-T)P\nabla H + 3HP\nabla T  \bigg), \label{eq:PullingAdhesionMF_Supp}
\end{align}
where $P = P_{i,j}(t), H = H_{i,j}(t), \text{ and } T = T_{i,j}(t)$.

\clearpage
\section{BINN implementation and training}\label{app:BINNs}

\subsection{BINNs architecture}\label{subsubsec:BINNsarchitecture}

Following \cite{lagergren_biologically-informed_2020}, we construct $\TMLP(x,t)$ using a fully-connected feed-forward MLP with three hidden layers, which can be written as:
\begin{align}
    z_0 &= [x,t] \nonumber \\
    z_1 &= \sigma\left( z_0W_1 + b_1 \right) \nonumber\\
    z_2 &= \sigma\left( z_1W_2 + b_2 \right) \nonumber\\
    z_3 &= \sigma\left( z_2W_3 + b_3 \right) \nonumber\\
    \TMLP(x,t) &= \psi\left( z_3W_4 + b_4 \right), \label{eq:TMLPArchitecture}
\end{align}
where each $z_k$ denotes the $k^{\text{th}}$ hidden layer for $k=1,2,3$; the $W_k$ matrices and the $b_k$ vectors provide the weights and biases of each hidden layer, respectively; $\sigma$ denotes the sigmoid activation function $\sigma(x)=1/(1+\exp{(-x)})$, and $\psi$ denotes the softplus activation function $\psi(x)=\log(1+\exp(x))$. Each hidden layer in Equation \eqref{eq:TMLPArchitecture} has 128 neurons, meaning that $W_1\in\mathbb{R}^{2\times128};  W_2, W_3\in\mathbb{R}^{128\times128}; W_4\in\mathbb{R}^{128\times1}; b_1, b_2, b_3 \in \mathbb{R}^{128}; \text{ and } b_4 \in \mathbb{R}$.

The architecture of $\DMLP(T)$ is identical to the architecture for $\TMLP$ in Equation \eqref{eq:TMLPArchitecture}, except $\DMLP$ has a one-dimensional input vector, $T$, instead of the two-dimensional input vector, $[x,t]$.

\subsection{Loss Function}\label{subsubsec:loss_function}
BINNs are trained to concurrently fit the given dataset, $\mean{\TABM(x,t)}^{train}$, and solve Equation \eqref{eq:diffusion_framework} by minimizing the following multi-term loss function:
\begin{equation}
    \mathcal{L}_{total} = \mathcal{L}_{WLS} + \epsilon\mathcal{L}_{PDE} +\mathcal{L}_{constr}. \label{eq:Loss_total}
\end{equation}
The $\epsilon$ parameter ensures the terms $\mathcal{L}_{WLS} \text{ and } \mathcal{L}_{PDE}$ are equally weighted because these terms can be of different orders of magnitude; we find good results for $\epsilon=10^4$.

The $\mathcal{L}_{WLS}$ term of Equation \eqref{eq:Loss_total} computes a weighted mean-squared error between $\TMLP(x,t)$ and $\mean{\TABM(x,t)}^{train}$:
\begin{equation}
    \mathcal{L}_{WLS} = \dfrac{1}{\Tftrain X}\sum_{i=1,j=1}^{X,\Tftrain} w_{i,j}\bigg( \TMLP(x_i,t_j) - \mean{\TABM(x_i,t_j)} \bigg)^2. \label{eq:L_ols}
\end{equation}
We set $w_{i,1}=10.0$ for all values of $i$ and all other $w_{i,j}$ values to 1.0 to ensure that $\TMLP$ closely agrees with the ABM's initial data. By minimizing Equation \eqref{eq:L_ols}, we ensure $\TMLP(x,t)$ closely approximates $\mean{\TABM(x,t)}^{train}$.

The $\mathcal{L}_{PDE}$ term of Equation \eqref{eq:Loss_total} quantifies how closely $\TMLP$ and $\DMLP$ follow Equation \eqref{eq:diffusion_framework}. To ensure the MLPs satisfy this PDE framework throughout the ABM's entire spatiotemporal domain, we uniformly sample 10,000 points, $\{ (x_k,t_k) \}_{k=1}^{10,000}$, from $[0,X]\times[0,750]$. For notational convenience, let $\hat{T}_k=\TMLP(x_k,t_k)$ and $\hat{D}_k = \DMLP\big(\TMLP(x_k,t_k)\big)$. We then compute the mean-squared error between the left- and right-hand sides of Equation \eqref{eq:diffusion_framework} at all sampled points:
\begin{equation}
    \mathcal{L}_{PDE} = \dfrac{1}{10,000}\sum_{i=1}^{10,000} \bigg[ \dfrac{\partial}{\partial t}\hat{T}_k - \dfrac{\partial}{\partial x}\bigg( \hat{D}_k \dfrac{\partial}{\partial x}\hat{T}_k \bigg) \bigg]^2, \label{eq:L_pde}
\end{equation}
where differentiation of $\TMLP$ and $\DMLP$ is performed using automatic differentiation. Minimizing Equation \eqref{eq:L_pde} verifies that $\TMLP$ and $\DMLP$ together satisfy Equation \eqref{eq:diffusion_framework}.

The $\mathcal{L}_{constr}$ term of Equation \eqref{eq:Loss_total} incorporates user knowledge into BINNs training. We  penalize $\DMLP$ for outputting values outside of the interval $[D_{\min}, D_{\max}]$. We set $D_{\min}=0$ because Equation \eqref{eq:diffusion_framework} is ill-posed if $\mathcal{D}(u)<0$, and we set $D_{\max}=1.0$ because the mean-field rates of diffusion are below one for all ABM simulations in this study. We compute this term by squaring any values of $\hat{D_i}$ that are not within $[D_{\min}, D_{\max}]$ and weighting these values by $10^{10}$:
\begin{equation}
    \mathcal{L}_{constr} = \dfrac{1}{10,000}\sum_{\substack{k=1 \\ \hat{D}_k \notin [D_{\min},D_{\max}]}}^{10,000} 10^{10} (\hat{D_k})^2. \label{eq:L_constr}
\end{equation}
This term regularizes the BINN training procedure to prevent $\DMLP$ from outputting unrealistic values.

\subsection{BINN Training Procedure}\label{subsubsec:BINNstraining}

For BINN model training, we randomly partition the training ABM dataset into 80\%/20\% BINN training and BINN validation datasets. We train the BINN parameter values (i.e., the weights and biases for $\TMLP$ and $\DMLP$) to minimize a loss function, $\mathcal{L}$, using the gradient-based ADAM optimizer with its default hyperparameter values on the BINN training dataset. For each new set of BINN parameters, we compute $\mathcal{L}$ on the BINN validation dataset and save the BINN parameters if the newly computed $\mathcal{L}$ value achieves a 1\% or greater relative improvement over the previous smallest recorded value. Following \cite{linka_bayesian_2022}, we perform training in a two-step process: in the first step, we train the BINN to match the ABM data by optimizing $\mathcal{L}=\mathcal{L}_{WLS}$ from Equation \eqref{eq:L_ols}; in the second step, we train the BINN on $\mathcal{L}=\mathcal{L}_{total}$ from Equation \eqref{eq:Loss_total}. The first training step is performed for $10^4$ epochs with an early stopping criterion of $10^3$, meaning that training ends early if the smallest-computed $\mathcal{L}$ value on the validation data is unchanged for $10^3$ epochs. The second step is performed for $10^6$ epochs with an early stopping criterion of $10^5$. Each epoch is computed in minibatches of size $10^3$. BINN model training is performed using the PyTorch deep learning library (version 1.7.1).

Following \cite{lagergren_biologically-informed_2020}, we train five separate BINNs for each ABM dataset using different BINN training and validation datasets because the final trained model can be sensitive to which data is included in these two datasets. We compute the five PDE forward simulations from these trained models and select whichever BINN achieves the smallest mean-squared error against the ABM training data as the final selected BINN model.

\subsection{\new{Comments on BINN training convergence}}

\new{We depict the chosen hyperparameter values for BINN model training in Table \ref{tab:BINN_setup}. Many of these values were chosen to follow previous modeling studies \cite{lagergren_biologically-informed_2020, linka_bayesian_2022}. A current challenge in neural network training is determining the optimal choice of such hyperparameter values \cite{kaplarevic-malisic_identifying_2023}. In our work, we found that BINN model training is most sensitive to the $\epsilon$ parameter as well as the number of epochs and early stopping number used during BINN model training (results not shown). If $\epsilon$ is too small, then the BINN will prioritize fitting the ABM data but not satisfying the PDE framework. Conversely, if $\epsilon$ is too large, then the BINN will ensure it satisfies a PDE framework while neglecting the data. We found a good balance between the two loss functions for $\epsilon=1\times10^{-4}$. Training the BINN with a smaller number of epochs, such as $10^5$ with an early stopping criterion of $10^4$ led to a model that had not fully converged to the data and we found better convergence using $10^6$ epochs with an early stopping criterion of $10^5$.}

\begin{table}
    \centering
    \begin{tabular}{|l|c|}
    \hline
    Hyperparameter description & Value  \\ \hline
    Number of hidden layers & 3 \\ \hline
    Number of neurons per hidden layer & 128 \\ \hline
    Weighting between $\mathcal{L}_{WLS}$ and $\mathcal{L}_{PDE}$ ($\epsilon$) & $10^{-4}$ \\ \hline
    Additional initial condition weighting in $\mathcal{L}_{WLS}$ & 10.0  \\ \hline
    Number of collocation points for $\mathcal{L}_{PDE}$ & 10,000 \\ \hline
    Penalty for $\DMLP$ values outside of $[D_{min},D_{max}]$ & $10^{10}$ \\ \hline
    $D_{min}$ & 0.0 \\ \hline
    $D_{max}$ & 1.0 \\ \hline
    ANN epochs & $10^4$ \\ \hline
    ANN early stopping & $10^3$ \\ \hline
    BINN epochs & $10^6$ \\ \hline
    BINN early stopping & $10^5$ \\ \hline
    \end{tabular}
    \caption{\new{Hyperparameter values used to perform BINN model training.}}
    \label{tab:BINN_setup}
\end{table}

\clearpage
\section{Numerical integration of PDEs} \label{sec:numericalIntegration}

When simulating Equation \eqref{eq:diffusion_framework_1d}, we populate the middle 20\% of the spatial dimension with 75\% confluence and zero confluence everywhere else to match the initial ABM configurations and implement no-flux boundary conditions:
\begin{align}
    T(x,0) &= \begin{cases} 0.75, & 80 \le x \le 120 \\ 0, & \text{otherwise}, \end{cases}, \nonumber \\
    \dfrac{\partial T}{\partial x}(0,t) &= \dfrac{\partial u}{\partial x}(X,t) = 0. \label{eq:ICandBC}
\end{align}

Before integration, we discretize the spatial domain as $x_i=i\Delta x$ with $i=0,...,199$ and $\Delta x=1.0$.  For ease of notation, let $T_i(t) = T(x_i,t)$ and $\D_i(t)=\D(T_i(t))$. We then use the method of lines approach to integrate Equation \eqref{eq:diffusion_framework_1d}. To discretize the right hand side of Equation \eqref{eq:diffusion_framework}, we let 
\begin{equation*}
    \dfrac{\partial T_i(t)}{\partial x} \left( \D_i(t) \dfrac{\partial T_i(t)}{\partial x}\right) \approx \dfrac{P_{i+\nicefrac{1}{2}}(t) - P_{i-\nicefrac{1}{2}}(t)}{\Delta x},
\end{equation*}
where $P_{i\pm\nicefrac{1}{2}}(t)$ denotes the right or left flux through location $x_i$, respectively. Following \cite{kurganov_new_2000}, we approximate these fluxes by
\begin{align}
    P_{i+\nicefrac{1}{2}}(t) &= \dfrac{1}{2}\left(\D_i(t)\dfrac{T_{i+1}(t)-T_{i}(t)}{\Delta x} + \D_{i+1}(t)\dfrac{T_{i+1}(t)-T_{i}(t)}{\Delta x}\right) \nonumber\\
    P_{i-\nicefrac{1}{2}}(t) &= \dfrac{1}{2}\left(\D_{i-1}(t)\dfrac{T_{i}(t)-T_{i-1}(t)}{\Delta x} + \D_{i}(t)\dfrac{T_{i}(t)-T_{i-1}(t)}{\Delta x}\right).\label{eq:flux_difference}
\end{align}
To implement the no-flux boundary conditions, we incorporate the ghost points $x_{-1}$ and $x_{200}$ that enforce $u_{-1}(t)=u_1(t)$ and $u_{198}(t)=u_{200}(t)$ into Equation \eqref{eq:flux_difference}. We integrate Equation \eqref{eq:diffusion_framework_1d} using the \texttt{odeint} command from Scipy's integration package (version 1.8.0), which implements the Livermore Solver for Differential Equations (LSODA) method \cite{petzold_automatic_1983}.

\clearpage
\section{Supplementary figures}

\begin{figure}[h!]
    \centering
    \includegraphics[width=\textwidth]{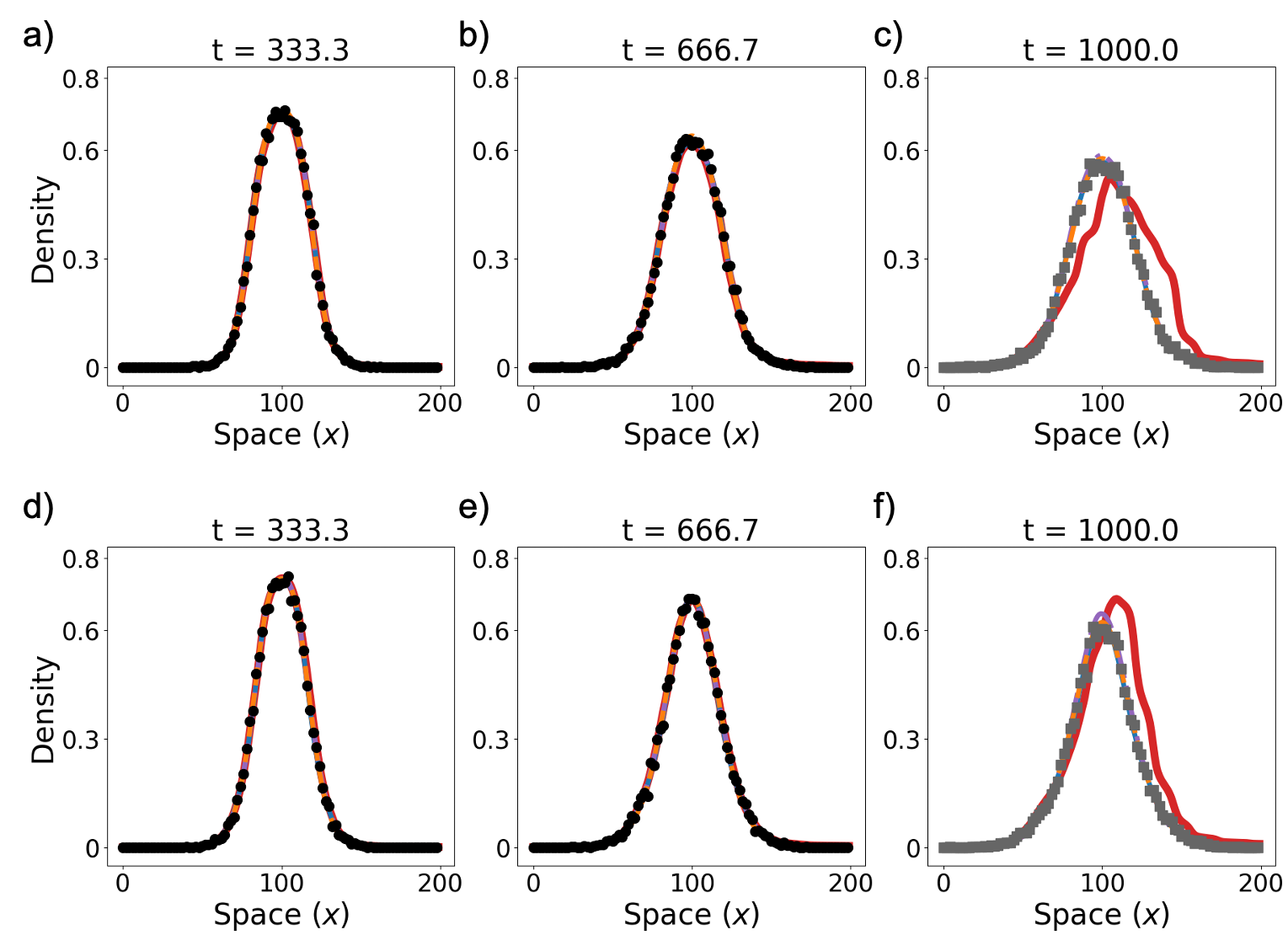}
    \caption{Forecasting ABM data with neural networks and PDEs. ANN and BINN models were trained to fit $\mean{\TABM(x,t)}^{train}$. These two neural networks and the mean-field and BINN-guided PDE simulations were then used to forecast $\mean{\TABM(x,t)}^{train}$ and $\mean{\TABM(x,t)}^{test}$. This was performed for (a-c) the Adhesion ABM with $\Pm = (\rmh,\padh)^T = (1.0, 0.5)^T$ and (d-f) the Pulling \& Adhesion ABM with $\Pm = (\rmp,\rmh,\ppull,\padh,\alpha)^T = (1.0,0.25, 0.33,0.33,0.5)^T$.}

    \label{fig:forecasting_baseline_sims}
\end{figure}

\begin{figure}[h!]
    \centering
    \includegraphics[width=\textwidth]{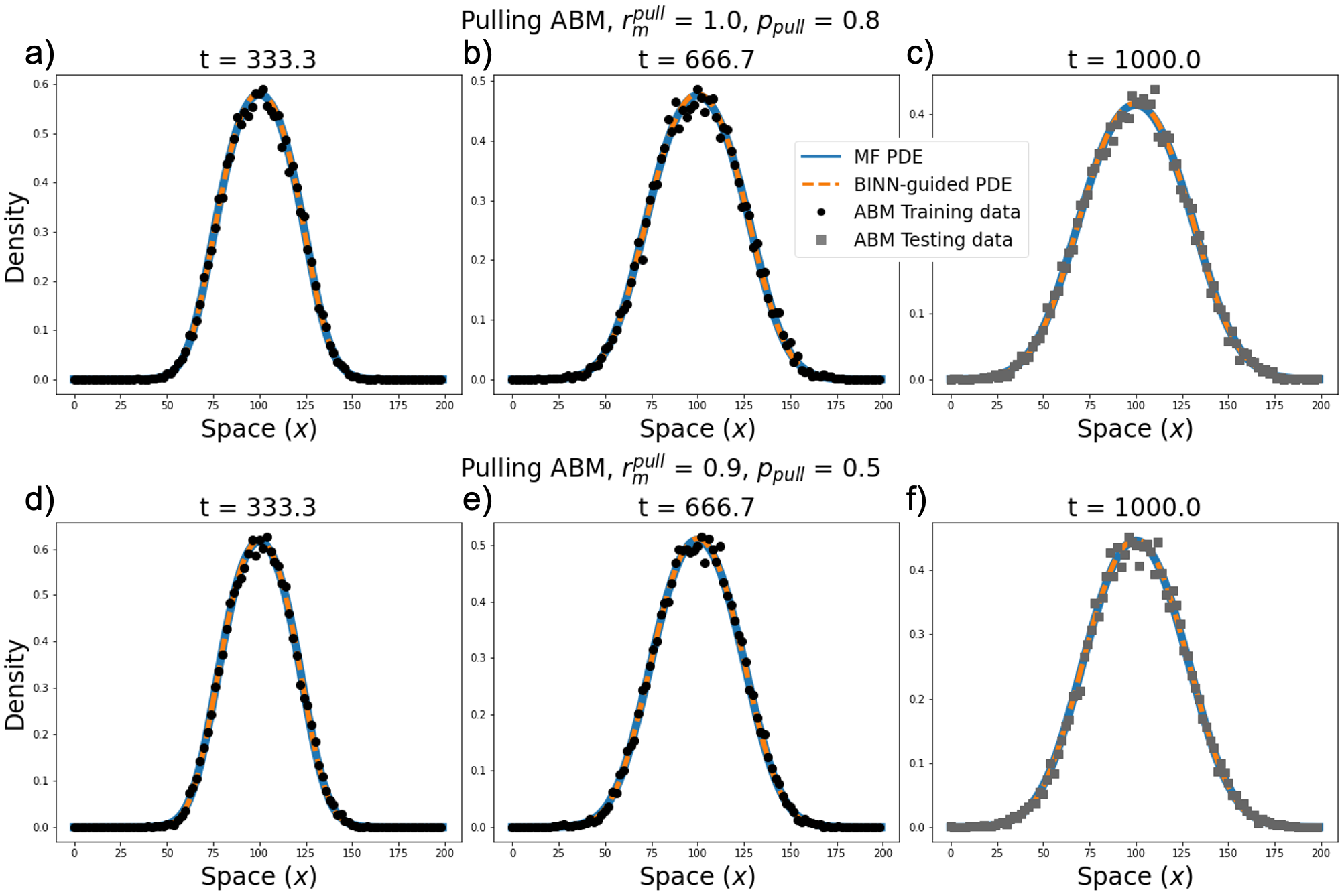}
    \caption{Forecasting Pulling ABM data with mean-field (MF) and BINN-guided PDE models. The mean-field and BINN-guided PDE simulations are used to forecast Pulling ABM data for (a-c) $\rmp=1.0, \ppull=0.8$ (d-f) $\rmp=0.9, \ppull=0.5$.}
    \label{fig:pulling_PDE_simulations}
\end{figure}

\begin{figure}
    \centering
    \includegraphics[width=\textwidth]{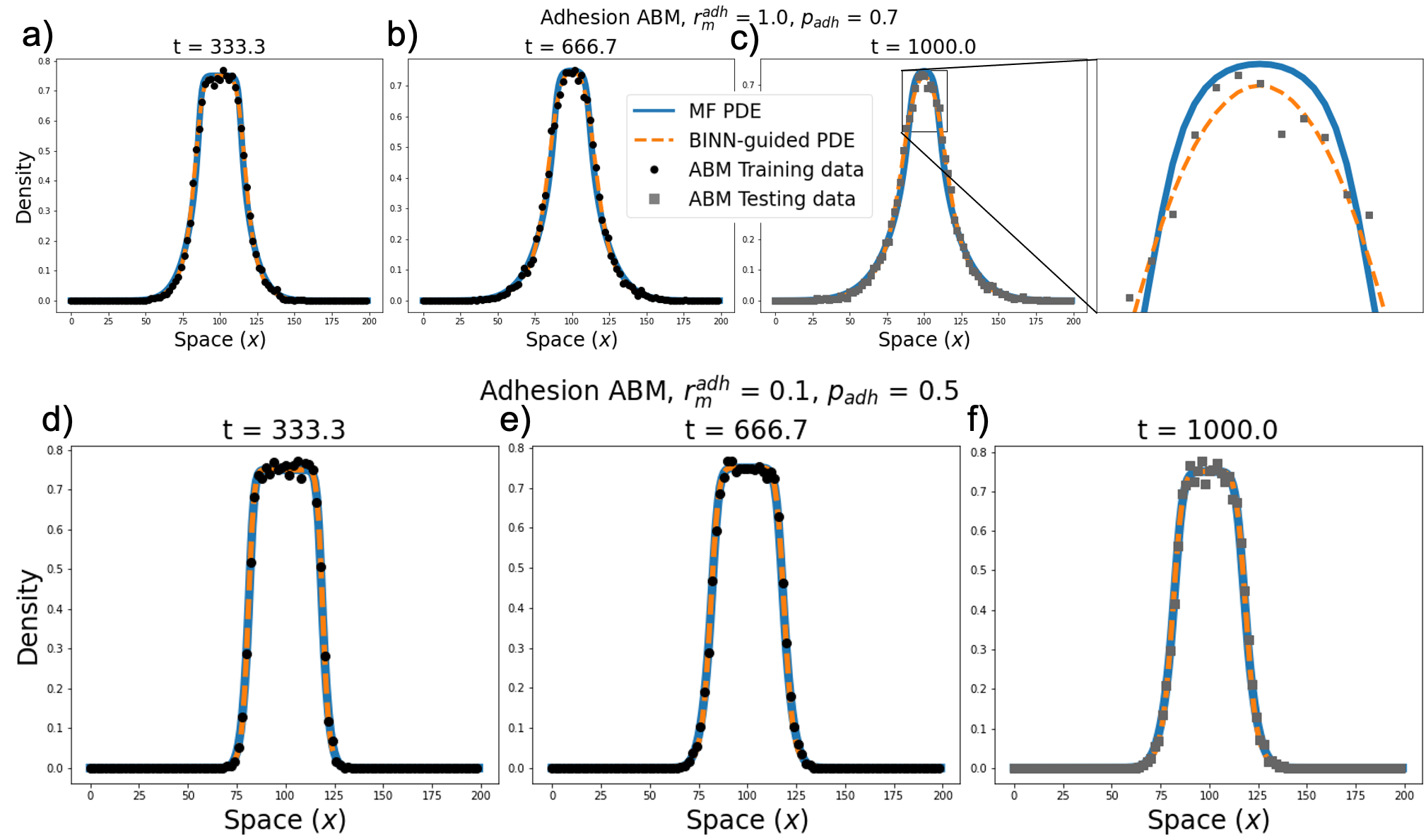}
    \caption{Forecasting Adhesion ABM data with mean-field and BINN-guided PDE models. The mean-field and BINN-guided PDE simulations are used to forecast Adhesion ABM data for (a-c) $\rmh=1.0, \padh=0.7$ (d-f) $\rmh=0.1, \padh=0.5$.}
    \label{fig:adhesion_PDE_simulations}
\end{figure}

\begin{figure}
    \centering
    \includegraphics[width=\textwidth]{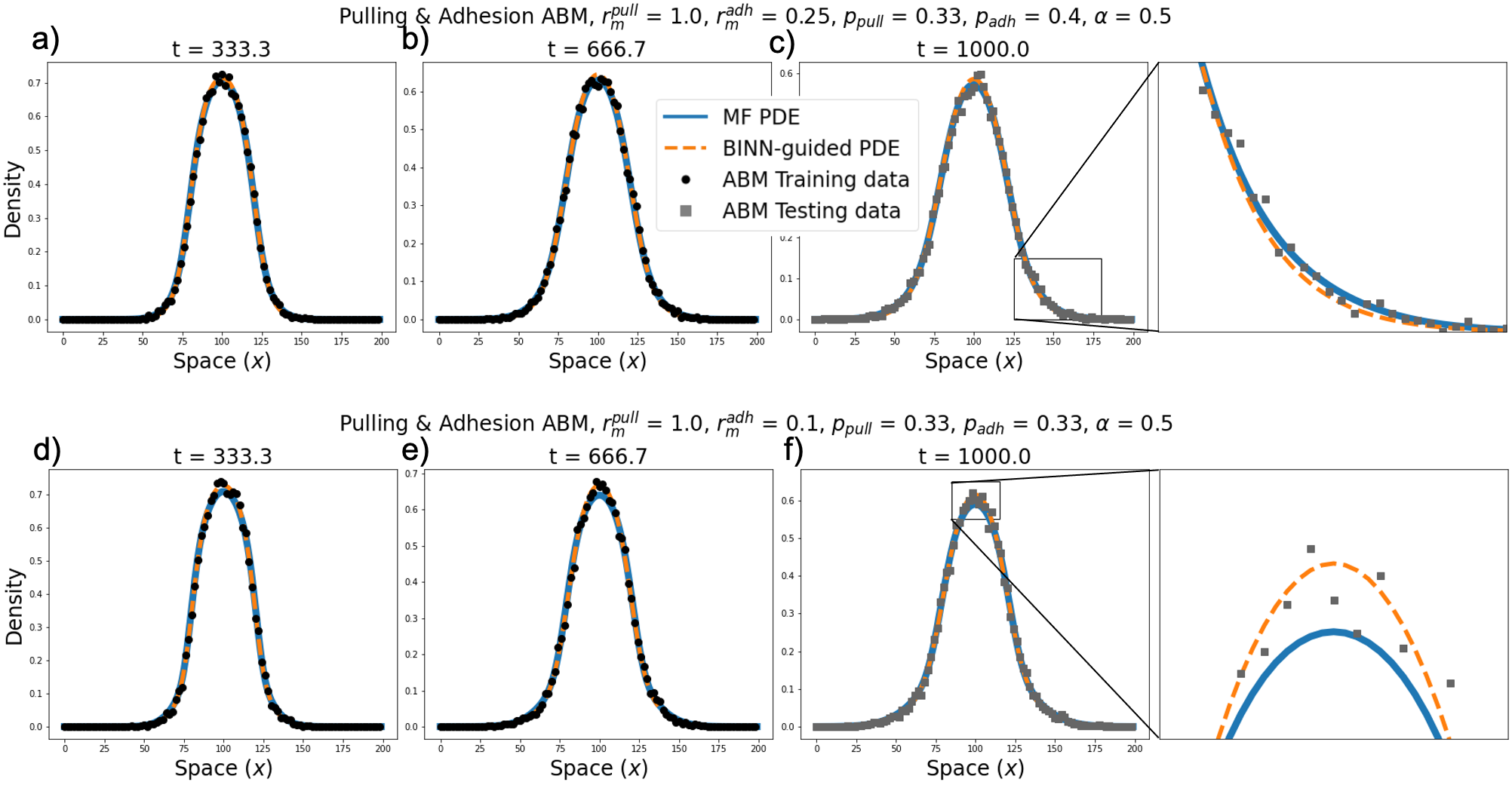}
    \caption{Forecasting Pulling \& Adhesion ABM data with mean-field (MF) and BINN-guided PDE models. The mean-field and BINN-guided PDE simulations are used to forecast Pulling \& Adhesion ABM data for the base parameter values ($\rmp=1.0, \rmh=0.25, \ppull = 0.33, \padh=0.33$, and $\alpha=0.5$), except (a-c) $\padh = 0.4$ (d-f) $\rmh = 0.1.$}
    \label{fig:Pullingadhesion_PDE_simulations}
\end{figure}

\begin{figure}
    \centering
    \includegraphics[width=\textwidth]{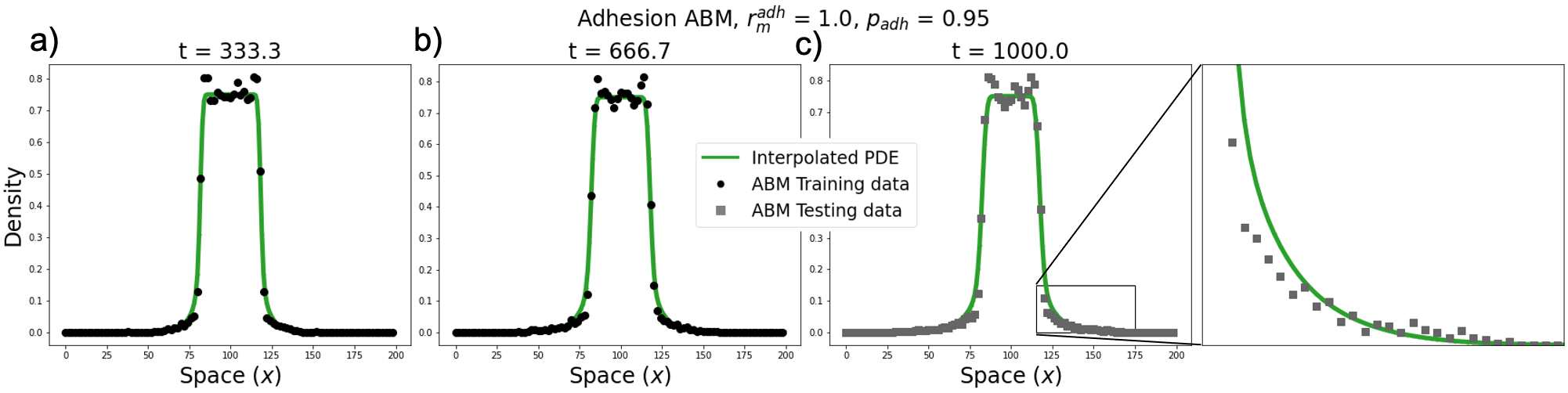}
    \caption{Predicting Adhesion ABM data with the interpolated PDE model. The interpolated PDE model predicts Adhesion ABM data for (a-c) $\rmh=1.0$ and $\padh=0.95$.}
    \label{fig:Adhesion_PDE_predictions_Pm_fixed}
\end{figure}

\begin{table}[]
    \centering
    \centering
    \begin{tabular}{|c|c|}
    \hline
    Sample &   $\Pm = (r_m^{adh},\ {p_{adh}})^T$ \\ \hline
    1 &   (0.145, 0.825)$^T$ \\ \hline
    2 &   (0.505, 0.575)$^T$ \\ \hline
    3 &   (0.415, 0.725)$^T$ \\ \hline
    4 &   (0.865, 0.525)$^T$ \\ \hline
    5 &   (0.955, 0.625)$^T$ \\ \hline
    6 &   (0.235, 0.775)$^T$ \\ \hline
    7 &   (0.685, 0.675)$^T$ \\ \hline
    8 &   (0.325, 0.875)$^T$ \\ \hline
    9 &   (0.775, 0.925)$^T$ \\ \hline
    10 &   (0.595, 0.975)$^T$ \\ \hline
    \end{tabular}
    \caption{Latin hypercube sampling for the Adhesion ABM. The samples from the new parameter dataset for the Adhesion ABM when varying $\rmh$ and $\padh$. The samples are ordered by increasing testing MSE values (see Figure \ref{fig:adhesion_interpolation_rmh_padh}(c)).}
    \label{tab:Adhesion_LHC_sample}
\end{table}

\begin{figure}
    \centering
    \includegraphics[width=\textwidth]{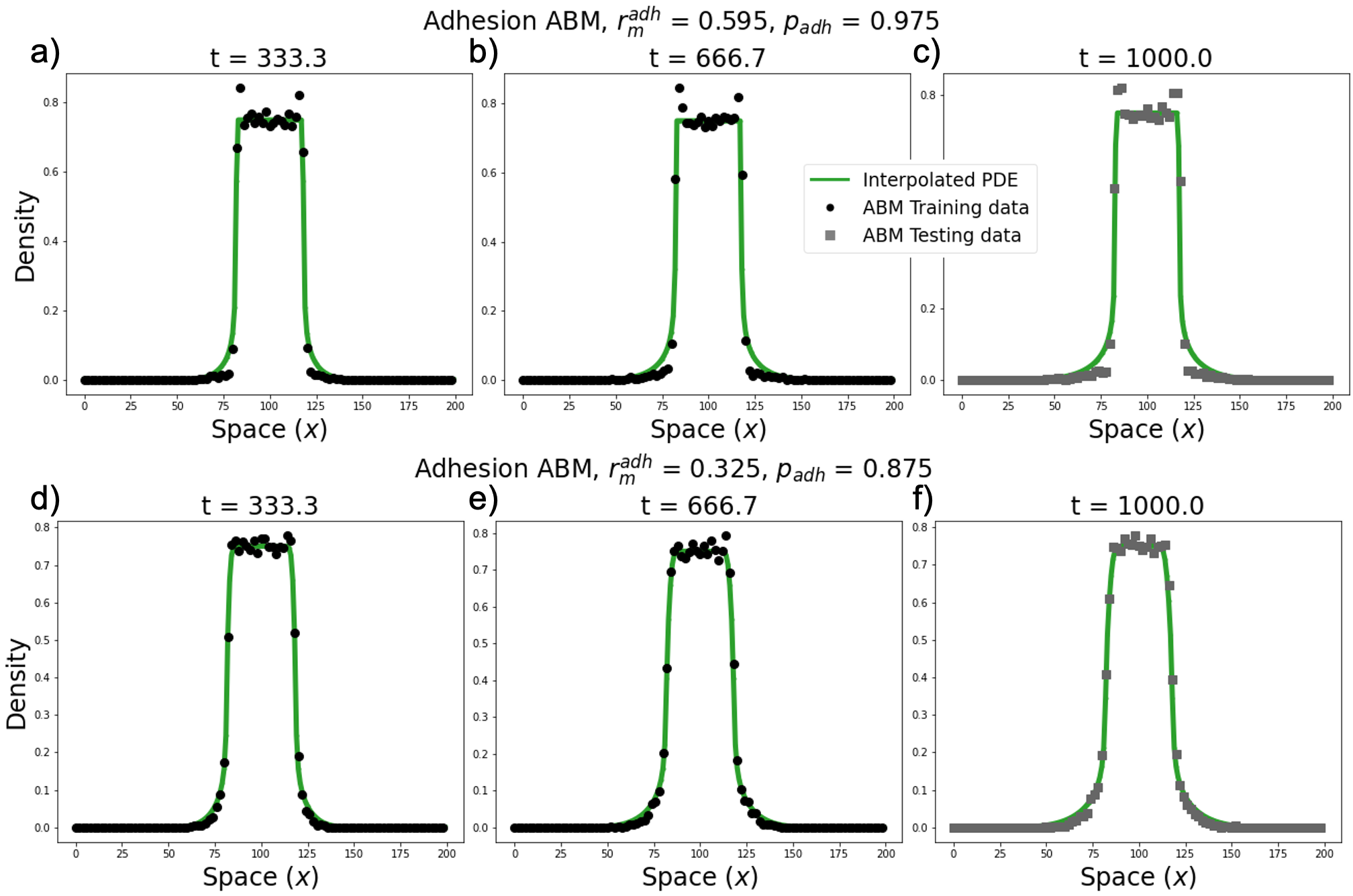}
    \caption{Predicting Adhesion ABM data with the interpolated PDE model. The interpolated PDE model predicts Adhesion ABM data for (a-c) $\rmh=0.595$ and $\padh=0.975$ and (d-f) $\rmh=0.325$ and $\padh=0.875$.}
    \label{fig:Adhesion_PDE_predictions}
\end{figure}

\begin{table}[]
    \centering
    \footnotesize
    \begin{tabular}{|c|c|}
    \hline
    Sample &   $\Pm = \ (r_m^{pull},\ r_m^{adh},\ p_{pull},\ p_{adh},\ \alpha)^T$ \\ \hline
    1 &   (1.0, 0.25, 0.394, 0.578, 0.912)$^T$ \\ \hline
    2 &   (1.0, 0.25, 0.293, 0.528, 0.938)$^T$ \\ \hline
    3 &   (1.0, 0.25, 0.008, 0.226, 0.988)$^T$ \\ \hline
    4 &   (1.0, 0.25, 0.511, 0.477, 0.862)$^T$ \\ \hline
    5 &   (1.0, 0.25, 0.41, 0.109, 0.962)$^T$ \\ \hline
    6 &   (1.0, 0.25, 0.075, 0.595, 0.888)$^T$ \\ \hline
    7 &   (1.0, 0.25, 0.042, 0.544, 0.838)$^T$ \\ \hline
    8 &   (1.0, 0.25, 0.327, 0.059, 0.712)$^T$ \\ \hline
    9 &   (1.0, 0.25, 0.444, 0.31, 0.662)$^T$ \\ \hline
    10 &   (1.0, 0.25, 0.209, 0.209, 0.612)$^T$ \\ \hline
    11 &   (1.0, 0.25, 0.126, 0.41, 0.762)$^T$ \\ \hline
    12 &   (1.0, 0.25, 0.193, 0.042, 0.588)$^T$ \\ \hline
    13 &   (1.0, 0.25, 0.059, 0.561, 0.462)$^T$ \\ \hline
    14 &   (1.0, 0.25, 0.243, 0.26, 0.788)$^T$ \\ \hline
    15 &   (1.0, 0.25, 0.427, 0.494, 0.512)$^T$ \\ \hline
    16 &   (1.0, 0.25, 0.595, 0.327, 0.812)$^T$ \\ \hline
    17 &   (1.0, 0.25, 0.025, 0.461, 0.388)$^T$ \\ \hline
    18 &   (1.0, 0.25, 0.377, 0.176, 0.488)$^T$ \\ \hline
    19 &   (1.0, 0.25, 0.226, 0.645, 0.538)$^T$ \\ \hline
    20 &   (1.0, 0.25, 0.528, 0.126, 0.688)$^T$ \\ \hline
    21 &   (1.0, 0.25, 0.561, 0.075, 0.562)$^T$ \\ \hline
    22 &   (1.0, 0.25, 0.142, 0.193, 0.362)$^T$ \\ \hline
    23 &   (1.0, 0.25, 0.31, 0.092, 0.738)$^T$ \\ \hline
    24 &   (1.0, 0.25, 0.176, 0.662, 0.412)$^T$ \\ \hline
    25 &   (1.0, 0.25, 0.645, 0.008, 0.638)$^T$ \\ \hline
    26 &   (1.0, 0.25, 0.343, 0.293, 0.312)$^T$ \\ \hline
    27 &   (1.0, 0.25, 0.092, 0.611, 0.238)$^T$ \\ \hline
    28 &   (1.0, 0.25, 0.109, 0.628, 0.012)$^T$ \\ \hline
    29 &   (1.0, 0.25, 0.159, 0.343, 0.212)$^T$ \\ \hline
    30 &   (1.0, 0.25, 0.26, 0.142, 0.188)$^T$ \\ \hline
    31 &   (1.0, 0.25, 0.36, 0.377, 0.262)$^T$ \\ \hline
    32 &   (1.0, 0.25, 0.276, 0.36, 0.038)$^T$ \\ \hline
    33 &   (1.0, 0.25, 0.578, 0.243, 0.288)$^T$ \\ \hline
    34 &   (1.0, 0.25, 0.628, 0.159, 0.062)$^T$ \\ \hline
    35 &   (1.0, 0.25, 0.477, 0.511, 0.138)$^T$ \\ \hline
    36 &   (1.0, 0.25, 0.611, 0.276, 0.338)$^T$ \\ \hline
    37 &   (1.0, 0.25, 0.461, 0.444, 0.162)$^T$ \\ \hline
    38 &   (1.0, 0.25, 0.544, 0.427, 0.112)$^T$ \\ \hline
    39 &   (1.0, 0.25, 0.494, 0.394, 0.088)$^T$ \\ \hline
    40 &   (1.0, 0.25, 0.662, 0.025, 0.438)$^T$ \\ \hline
    \end{tabular}
    \caption{Latin hypercube sampling for the Pulling \& Adhesion ABM. The samples from the prior parameter dataset for the Pulling \& Adhesion ABM when varying $\ppull$, $\padh$, and $\alpha$. The samples are ordered by increasing training MSE values.}
    \label{tab:PullingAdhesionLHCprior}
\end{table}

\begin{table}[]
    \centering
    \begin{tabular}{|c|c|}
    \hline
    Sample &   $\Pm = \ (r_m^{pull},\ r_m^{adh},\ p_{pull},\ p_{adh},\ \alpha)^T$ \\ \hline
    1 &   (1.0, 0.25, 0.285, 0.519, 0.775)$^T$ \\ \hline
    2 &   (1.0, 0.25, 0.419, 0.352, 0.875)$^T$ \\ \hline
    3 &   (1.0, 0.25, 0.486, 0.117, 0.525)$^T$ \\ \hline
    4 &   (1.0, 0.25, 0.553, 0.285, 0.375)$^T$ \\ \hline
    5 &   (1.0, 0.25, 0.385, 0.586, 0.475)$^T$ \\ \hline
    6 &   (1.0, 0.25, 0.586, 0.184, 0.175)$^T$ \\ \hline
    7 &   (1.0, 0.25, 0.62, 0.151, 0.325)$^T$ \\ \hline
    8 &   (1.0, 0.25, 0.184, 0.084, 0.625)$^T$ \\ \hline
    9 &   (1.0, 0.25, 0.352, 0.385, 0.925)$^T$ \\ \hline
    10 &   (1.0, 0.25, 0.653, 0.05, 0.275)$^T$ \\ \hline
    11 &   (1.0, 0.25, 0.151, 0.653, 0.075)$^T$ \\ \hline
    12 &   (1.0, 0.25, 0.452, 0.251, 0.125)$^T$ \\ \hline
    13 &   (1.0, 0.25, 0.084, 0.218, 0.225)$^T$ \\ \hline
    14 &   (1.0, 0.25, 0.318, 0.62, 0.725)$^T$ \\ \hline
    15 &   (1.0, 0.25, 0.519, 0.017, 0.825)$^T$ \\ \hline
    16 &   (1.0, 0.25, 0.117, 0.419, 0.425)$^T$ \\ \hline
    17 &   (1.0, 0.25, 0.251, 0.486, 0.975)$^T$ \\ \hline
    18 &   (1.0, 0.25, 0.017, 0.452, 0.025)$^T$ \\ \hline
    19 &   (1.0, 0.25, 0.05, 0.318, 0.575)$^T$ \\ \hline
    20 &   (1.0, 0.25, 0.218, 0.553, 0.675)$^T$ \\ \hline
    \end{tabular}
    \caption{Latin hypercube sampling for the Pulling \& Adhesion ABM. The samples from the new parameter dataset for the Pulling \& Adhesion ABM when varying $\ppull$, $\padh$, and $\alpha$. The samples are ordered by increasing training MSE values.}
    \label{tab:PullingAdhesionLHCnew}
\end{table}

\begin{figure}
    \centering
    \includegraphics[width=\textwidth]{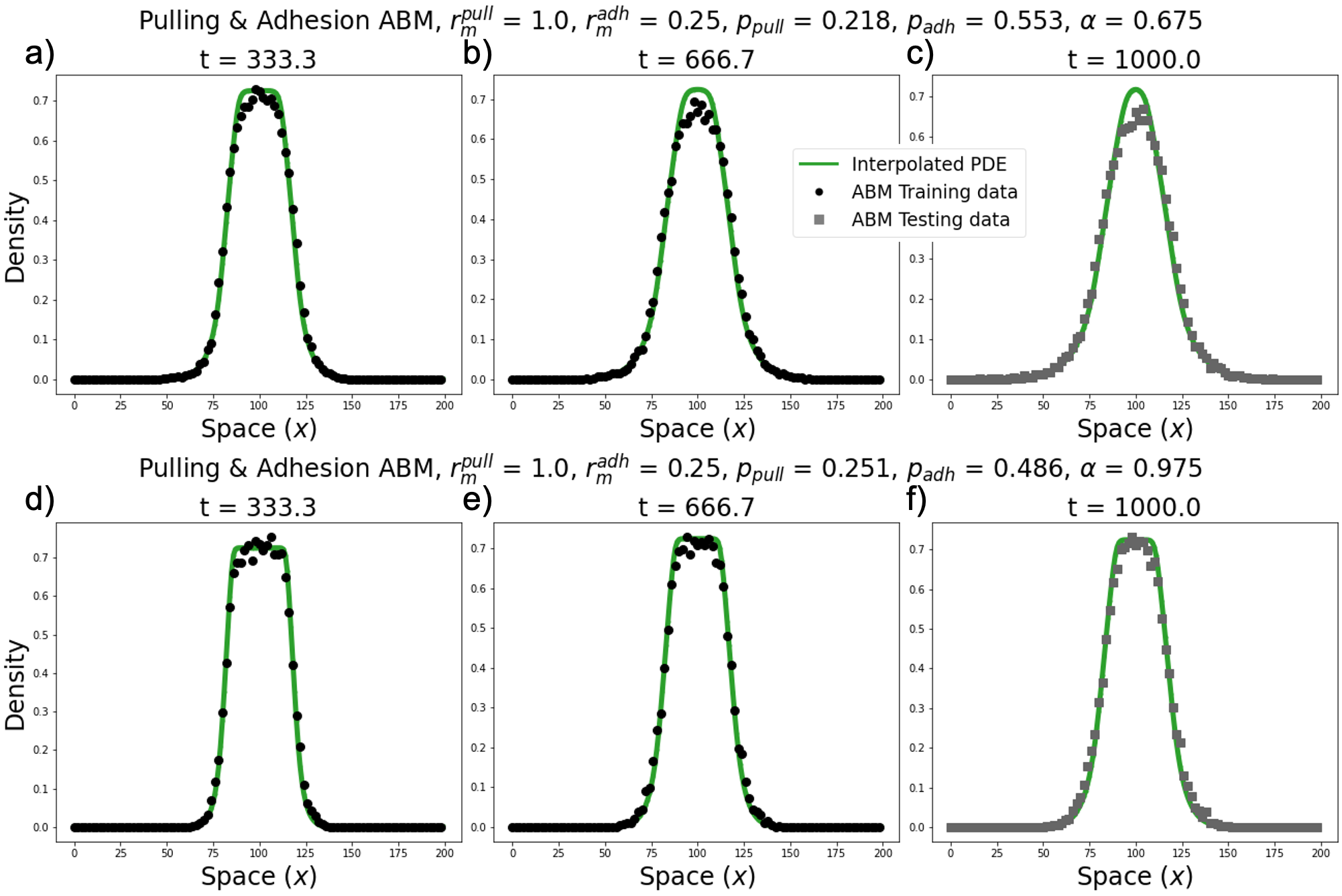}
    \caption{Predicting Pulling \& Adhesion ABM data with the interpolated PDE model. The interpolated PDE model predicts Adhesion ABM data for $\rmp=1.0$, $\rmh=0.25$, and (a-c) $\ppull=0.218$, $\padh=0.553$, and $\alpha=0.675$ (d-f) $\ppull=0.251$, $\padh=0.486$, and $\alpha=0.975$.}
    \label{fig:PullingAdhesion_PDE_predictions}
\end{figure}

\begin{figure}
    \centering
    \includegraphics[width=.9\textwidth]{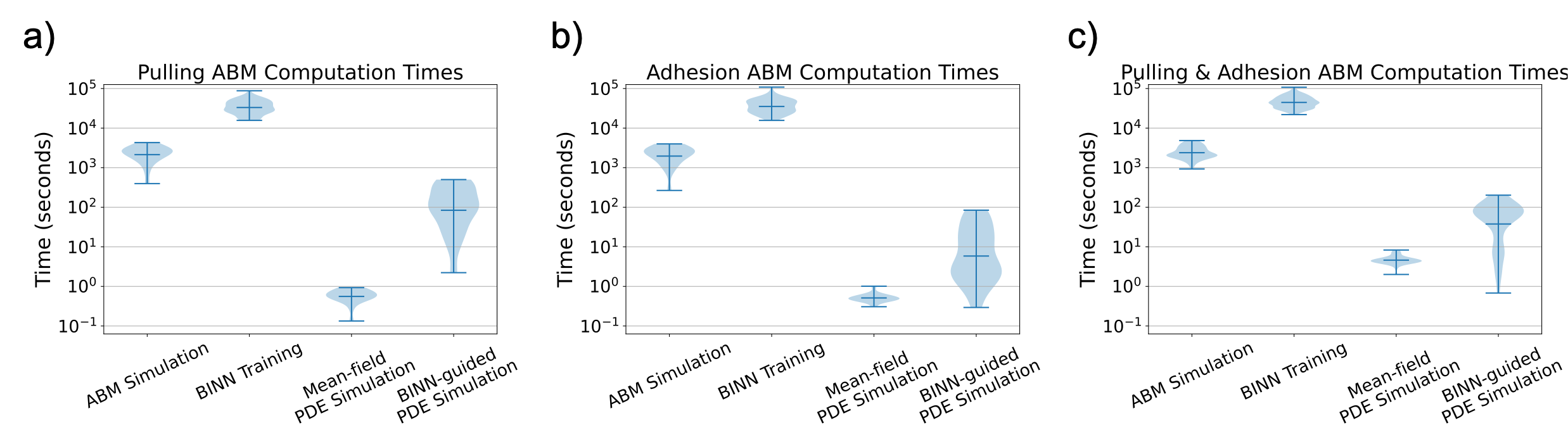}
    \caption{Computational expenses of each modeling approach. Violin plots represent the distribution of wall time computations for ABM simulations, BINN training, mean-field PDE simulations, and BINN-guided PDE simulations for the (a) Pulling ABM, (b) Adhesion ABM, and (c) Pulling \& Adhesion ABM.}
    \label{fig:wall_time_estimation}
\end{figure}

\end{document}